%% file: _draft.tex
\documentclass[12pt,letterpaper]{article}

\makeatletter
\newcommand*{\centerfloat}{%
  \parindent \z@
  \leftskip \z@ \@plus 1fil \@minus \textwidth
  \rightskip\leftskip
  \parfillskip \z@skip}
\makeatother

\usepackage[bottom]{footmisc} 

\usepackage{subcaption} 
\usepackage{amsmath}
\usepackage{mathtools}
\usepackage[longnamesfirst]{natbib}

\usepackage{bm} 
\usepackage{amsthm,amssymb}
\theoremstyle{plain} 
\newtheorem{assumption}{Assumption}
 
\newtheorem{proposition}{Proposition}

\newtheorem{definition}{Definition}

\usepackage{tikz}

\usepackage[labelfont=bf]{caption}
\usepackage{pdflscape}

\usepackage{graphicx,subcaption}
\usepackage{xpatch} 
\xpatchcmd{\proof}{\itshape}{\normalfont\proofnamefont}{}{} 
\newcommand{\proofnamefont}{\bfseries} 

\usepackage[title]{appendix} 
\usepackage{threeparttable, booktabs, siunitx} 
\usepackage{subcaption} 

\usepackage{dcolumn}
\usepackage{enumitem}

\usepackage{lipsum}
\usepackage{setspace}
\usepackage{hyperref}
\definecolor{internationalkleinblue}{rgb}{0.0, 0.18, 0.65}
\definecolor{blue(pigment)}{rgb}{0.2, 0.2, 0.6}
\hypersetup{colorlinks=true, citecolor=internationalkleinblue, linkcolor=internationalkleinblue, urlcolor  = black}
\usepackage[letterpaper, margin=1.2in, top=1.0in, bottom=1.0in]{geometry}  

\usepackage{longtable}

\usepackage{multirow} 
\usepackage{tocloft}
\usepackage{dirtytalk} 
\usepackage{dcolumn}

\usepackage{titlesec}
\titleformat{\section}
  {\sffamily\Large\bfseries}{\thesection.}{1em}{}
\titleformat{\subsection}
  {\sffamily\large\bfseries}{\thesubsection.}{1em}{}
\titleformat{\subsubsection}
  {\sffamily\large\bfseries}{\thesubsubsection.}{1em}{}

\renewcommand{\refname}{References}
\makeatletter
\renewcommand\bibsection{%
  \section*{\sffamily\refname\@mkboth{\refname}{\refname}}%
}%
\makeatother

\title{\vspace{-1.0cm}\textbf{\textsf{Money Creation and Banking: \\Theory and Evidence}}\footnote{This paper is a revised version of Chapter 1 of my PhD dissertation at the University of Missouri. I am deeply indebted to my advisor, Chao Gu, for her guidance at all stages of this research project. I am grateful to my committee members, Joseph Haslag, Aaron Hedlund, and Xuemin (Sterling) Yan for helpful comments. I also thank Lukas Altermatt, Bora{\u{g}}an Aruoba, Lonnie Hofmann, Duhyeong Kim, Sunham Kim, Shawn Ni, Sungmin Park, William Park, and Christian Wipf as well as seminar participants at the 2019 MVEA Annual Meeting, the 2019 KIEA Annual Meeting, the 2020 SEA Annual Meeting, the Bank of Korea, the 2021 WEAI Conference, the 2021 KER International Conference, University of Basel, and University of Mississippi for useful feedback and discussion. The views expressed in this paper are solely those of the author. All remaining errors are mine.}}

\author{\textsc{Heon Lee}\footnote{Contact: \href{mailto:heonlee68@gmail.com}{heonlee68@gmail.com}} \\ University of Missouri}

\date{First Draft: August, 2019 \\ This Draft: June, 2024}

\begin{document}
\maketitle
\onehalfspacing
\begin{abstract}
\noindent 
This paper studies the role of banks' money creation in monetary transmission. I develop a monetary-search model where demand for the monetary base and the money multiplier are endogenously determined through banks' money creation. The model and data show that reserves are not independent of interest rate policy, even with ample reserves. Furthermore, short-term rates and interest on reserves play distinct roles in monetary transmission. I evaluate the theory by matching it with data, and the calibrated model can account for the evolution of reserves, excess reserves, and the money multiplier from 1968 to 2015.
\end{abstract}


\newpage
\spacing{1.1}

\begingroup\begin{quote} [A] model of the banking system in which currency, reserves, and deposits play distinct roles ... seems essential if one wants to consider policies like reserves requirements, interest on deposits, and other measures that affect different components of the money stock differently. \hspace*{4.98cm}\textbf{\cite{lucas2000inflation}}\end{quote}\endgroup

\onehalfspacing
\section{Introduction}
\label{sec:introd}

This paper develops a theory of money and banking that articulates banks' role in money creation and its interaction with credit. The focal points are the endogenous determination of demand for base money through banking, and the central bank's control of base money to peg the short-term interest rate, which in turn influences the bank's money creation activity and other macroeconomic variables.

The central bank conducts monetary policy through interventions in the market for base money. However, most leading models of monetary policy analysis do not consider the transmission mechanism via the market for the monetary base. For instance, New Keynesian models abstract from the money market mechanism. Moreover, many models assert the independence of the quantity of reserves from interest management policy, especially since the Federal Reserve started paying interest on reserves. In the New Monetarist framework, where monetary frictions are explicit, most models focus on monetary aggregates rather than the transmission mechanism from the market for the monetary base to the monetary aggregate through the banking sector. Instead of abstracting from the key mechanism, this paper revisits the issue of money creation to understand the role of banking in monetary transmission and to account for a number of observations within a unified framework.

Specifically, Section \ref{sec:motive} provides four empirical findings from US data, which guide the modeling.
\begin{enumerate}
\setlength\itemsep{-0.5em}
    \item Short-term interest rates are not independent of the amount of reserves, even during 2009-2015. These rates are related to the quantity of reserves and the bank's money creation activity.
    \item The excess reserves to deposit ratio had been close to zero until 2007. The excess reserve ratio skyrocketed as the Fed introduced the interest on reserves. 
    \item The required reserve ratio does not have a negative relationship with the M1 money multiplier, and there were two structural breaks in the evolution of the M1 money multiplier: 1992 and 2008.
    \item Adding unsecured credit into the money demand equation as a regressor recovers the downward-sloping and stable M1 money demand.
\end{enumerate}

The first finding challenges the conventional approach that assumes independence between the quantity of reserves and the interest rate during the 2009-2015 period, known as the zero lower bound period. Contrary to this widely used assumption which has been a cornerstone in many unconventional monetary policy analyses, the finding implies that the quantity of reserves still plays a significant role in interest rate management. The second finding, which was already known to many economists, is worth emphasizing again, especially in light of the first finding. Taken together, the first and second findings suggest that the short-term policy rate and the interest on reserves may have distinct roles to play in interest rate management. 

The third finding contradicts the undergraduate textbook theory of the money multiplier which predicts a negative relationship between the money multiplier and the required reserves ratio. Since the textbook theory assumes zero excess reserves, the absence of a negative relationship may not be a surprising finding during the post-2008 period with excess reserves. However, there is no negative relationship during the pre-2008 period as well. Finally, the last finding suggests that considering the role of unsecured credit could be essential to understand money demand. This paper advances a theory of money and banking, which accounts for all four observations mentioned above as well as the money creation process in the US economy.

This paper builds a monetary banking model based on \cite{lagos2005unified} to understand the monetary transmission. The model features the explicit structure of monetary exchange and the role of financial intermediation in money creation. Agents can trade by using cash, transaction deposits, and unsecured credit. Banks create deposits by making loans, which can be either transaction deposits or non-transaction deposits. However, the creation of transaction deposits is constrained by reserve requirements. Given the monetary policy and credit conditions of the economy, banks determine their holdings of excess reserves and loans, and thus, the money multiplier is determined endogenously.

When banks hold excess reserves, their reserve requirement constraint does not bind, and a change in reserve requirement does not
change the money multiplier. Instead, the money multiplier is determined by the short-term policy rate and the interest on reserves. Lowering the short-term policy rate increases reserves, but the banks do not create deposit money proportionally, which lowers the money multiplier. Higher interest on reserves decreases the money multiplier because banks have more incentive to hold reserves and less incentive to create deposit money. The interest rate on reserves and the short-term rate play distinct roles, and they jointly determine the quantity of reserves. These rates also have different impacts on the lending rate. An increase in the short-term rate raises the lending rate, while an increase in the interest on reserves lowers the lending rate. 

Another ingredient of the model is unsecured credit which can substitute for other means of payment as in \cite{gu2016money}.\footnote{By modeling unsecured credit with an exogenous credit limit, this paper follows \cite{gu2016money}. For other approaches to introducing credit to the monetary economy, see \cite{sanches2010money},  \cite{lotz2016money}, and  \cite{williamson2016scarce}.} Better credit conditions lower transaction deposit balances as credit can substitute for money. This decrease in deposit balances leads to a lower money multiplier, regardless of whether banks hold excess reserves or not. 

Next, I quantify the model by calibrating the model and asking how it accounts for observations in Section \ref{sec:motive} and the money creation process. Given monetary policy behaved as it did, how well can the model account for the behavior of reserves, excess reserves, and money multiplier? The analysis shows the model can explain the historical evolution of the money creation process including all the observations from  Section \ref{sec:motive}. Consistent with data, the model generates zero excess reserves between 1968 and 2007, as well as massive increases in excess reserves after 2008. The model provides the counterfactual reserves to output ratio, which closely tracks its actual behavior from 1968 to 2015. It also generates drops in the money multiplier during the 1990s and 2000s without excess reserves, and its more drastic drops after 2008, which were accompanied by a massive increase in excess reserves.

The quantitative exercise shows that dramatic changes in the money multiplier after 2008 are mainly driven by the introduction of the interest on reserves whereas the decrease in the money multiplier before 2008 is driven by enhanced credit conditions. Contrary to previous approaches that assumed no changes in the short-term policy rate during 2009-2015 (so-called the zero-lower-bound period) and merely focused on the quantity of reserves, this study confirms that the short-term interest rate did change during that period, and its movements were directly reflected in the quantity of reserves. \\

\noindent\textbf{Related Literature } This paper contributes to three strands of literature. First, it contributes to the growing literature on unconventional monetary policy and bank reserves. Many models of unconventional monetary policy take the zero lower bound constraint as a given and focus on the effect of asset purchases by issuing reserves, which is assumed to be independent of short-term interest rates.  (\citealp*{curdia2011central}, \citealp*{gertler2011model} and \citealp*{lee2021quantitative}). By focusing on which assets the central bank purchases, some recent works (e.g., \citealp*{williamson2016scarce} and \citealp*{bhattarai2015time}) study quantitative easing as reducing maturity of government debt. They take the ZLB constraint seriously and treat bank reserves and short-term bonds as perfect substitutes when the interest rate is at the ZLB. Given this ZLB constraint, they focus on the effect of maturity transformation. Whereas these models assume independence of reserves to short-term interest rates, this paper shows that this independence assumption does not hold in the data and provides a model which can account for this observation. In contrast to the conventional approach to unconventional monetary policy, the quantitative analysis shows that the changes in reserves correspond to the changes in short-term interest rates. Rather than relying on the ad hoc assumption of independence, this paper articulates an explicit mechanism that determines the demand for reserves and the money multiplier.

Second, this paper contributes to a large literature on inside money and money creation. Previous works capturing the explicit role of reserve requirements and money creation include \cite{freeman1987reserve}, \cite{haslag1998money}, and \cite{freeman2000monetary}. \cite{freeman1987reserve} and  \cite{haslag1998money} study the impact of money creation and the reserve requirements on seigniorage revenue. \cite{freeman2000monetary}  develop a tractable model of the endogenous money multiplier. They show that the money-output correlation can be explained by the endogenous money supply resulting from households’ choices in response to the business cycle. Recent advances in monetary economics based on a search-theoretic framework provide a deeper understanding of banking and inside money. For example, \cite{gu2013banking} study the environment where banking arises endogenously, and show that banking can improve the economy by facilitating trade using inside money. \cite{andolfatto2020money} integrate a model of bank and financial markets by  \cite{diamond1997liquidity}  with \cite{lagos2005unified} framework and deliver a model where the fractional reserve banking arises in the equilibrium. \cite{altermatt2022inside} studies an economy in which a bank creates inside money by extending loans to entrepreneurs, which can be used for investment. This paper contributes to the literature by constructing a model of money and banking, which establishes the conditions under which banks hold excess reserves, provides an explicit mechanism for determining the demand for reserves and the money multiplier, and explains the money creation process as observed in the data.

Third, this work relates to the literature that studies money and credit explicitly. \cite{gu2016money} show that if money is essential, the credit is irrelevant. Changes in credit conditions only crowd out real balances. This neutrality result can be overturned if one introduces the costly credit (\citealp*{bethune2020frictional}; \citealp*{wang2020sticky}).  In this paper, there are fiat money, deposit money, and unsecured credit, and the neutrality result does not hold. This is because there exists monetary credit and the bank's intermediation of deposit money is costly.     \\

This paper is organized as follows. Section \ref{sec:motive} provides motivating evidence. Section \ref{sec:model} constructs the search-theoretic monetary model of money creation. Section \ref{sec:quant} calibrates the model to quantify the theory. Section \ref{sec:conclusion} concludes.  


\section{Motivating Evidence}
\label{sec:motive}

This section presents a list of observations about money creation and money demand, which motivates the theoretical framework developed in the next sections. \\

\noindent\textbf{Observation 1.} Short-term interest rates are not independent of the amount of reserves, even during 2009-2015. These rates are related to the quantity of reserves and the bank's money creation activity.  \\

\begin{figure}[bp!]
\centerfloat
\includegraphics[width=7.5cm,height=6cm]{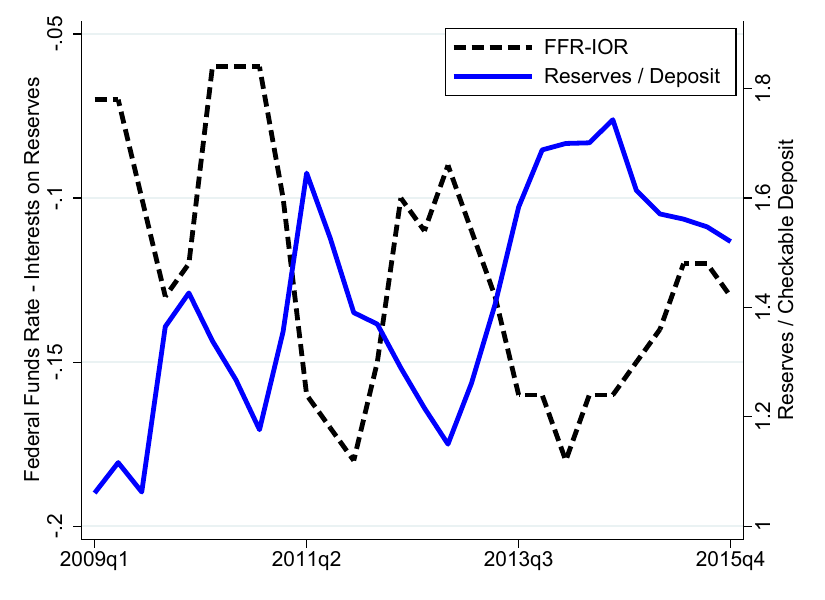}
\includegraphics[width=7.5cm,height=6cm]{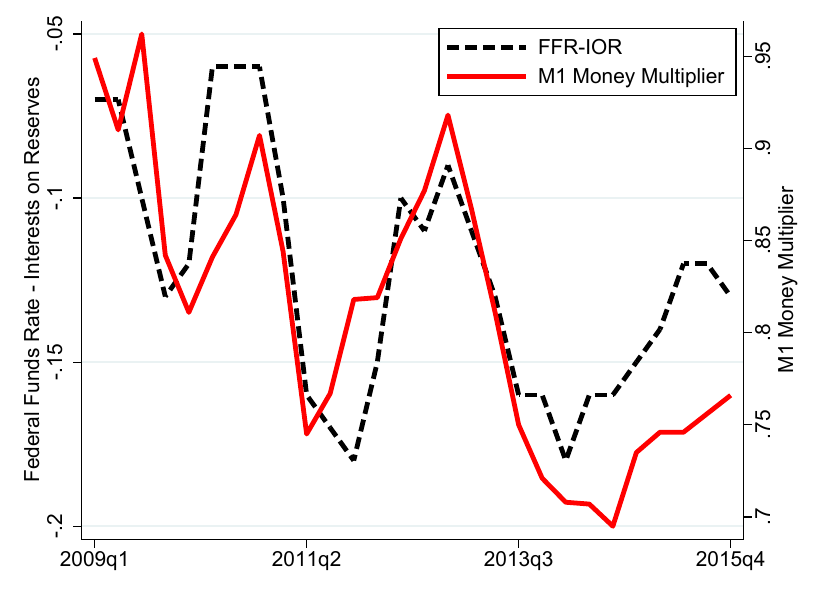}
\caption{US excess reserves and M1 multiplier in the post-2008 period}
\label{fig:motive4}
\end{figure}

The left panel of Figure \ref{fig:motive4} shows the reserves to deposit ratio and the spread between the federal funds rate and the interest on reserves during 2009-2015, which is often called the zero lower bound (ZLB) period. Their opposite movements are evident. It suggests that the quantity of reserves still plays a role in interest rate management. This contradicts many monetary models of unconventional monetary policies in the post-2008 era, which assume the independence of the quantity of reserves from interest rate management in an ample-reserves regime. (e.g., 
\citealp*{bech2011mechanics};
\citealp*{curdia2011central};
\citealp*{kashyap2012optimal}; \citealp*{cochrane2014monetary};    \citealp*{ennis2018simple}; \citealp*{piazzesi2019money}).

The assumption of independence is based on the idea that the interest rate paid on reserves sets a floor for the short-term policy rate. When the target policy rate reaches the interest rate on reserves, money is “divorced” from interest rate management, allowing the central bank to determine the quantity of reserves independently of the interest rate. However, in contrast to this notion, interest on reserves has been the upper bound of target policy rates. Appendix \ref{sec:floor} provides a further discussion on interest on reserves and the overnight reverse repurchase facility.

In addition, the right panel of Figure \ref{fig:motive4} shows that the M1 money multiplier moves together with the interest rate spread. This implies that the monetary policy during the post-2008 period is closely related to the bank's money creation activity, which was not given much attention in many monetary models of the last few decades. \\

\begin{figure}[tp!]
\centerfloat
\includegraphics[width=12cm,height=5.5cm]{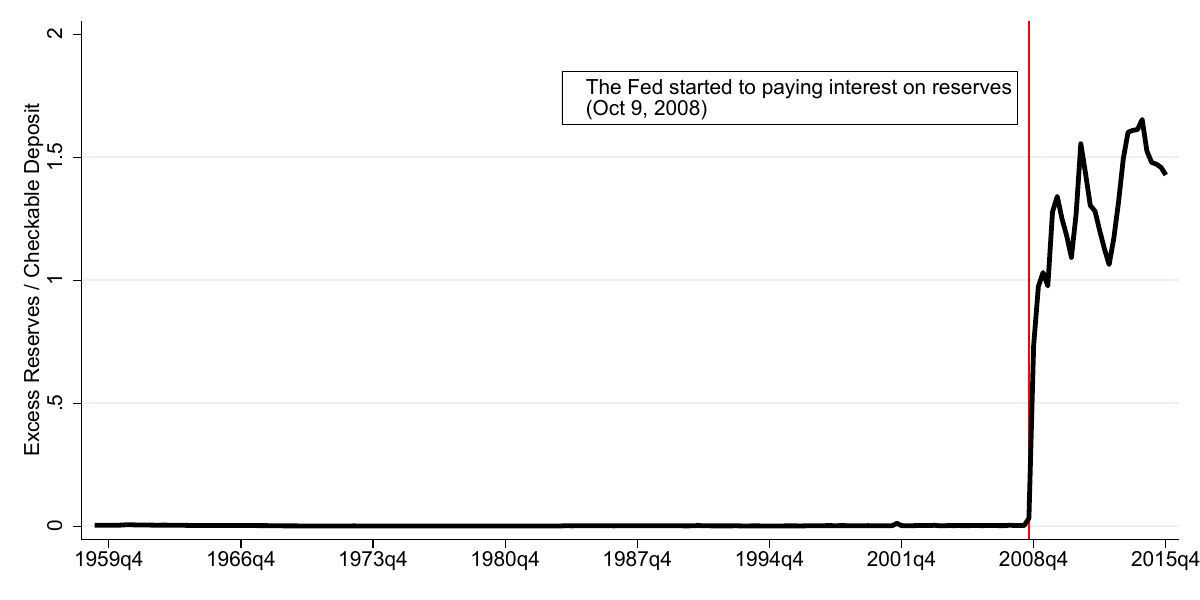}
\caption{Excess reserves ratio}
\label{fig:excess_ratio}
\end{figure}

\noindent\textbf{Observation 2.} The excess reserves to deposit ratio had been zero until 2007. The excess reserve ratio skyrocketed as the Fed introduced the interest on reserve.  \\

Figure \ref{fig:excess_ratio} plots the excess reserves to deposit ratio from 1959Q1 to 2015Q4. Before the 2008 Great Recession, the ratio remained at zero. However, it rose drastically after the recession and exceeded the value of 1, implying that banks have held more excess reserves than the amount of checking account balances they issued. Figure \ref{fig:excess_ratio} also shows that the dramatic increase in excess reserves coincided with the Fed's introduction of the interest on reserves. Taken together, the first and second findings suggest that the short-term policy rate and
interest on reserves may have distinct roles to play in interest rate management.  \\

\noindent\textbf{Observation 3.} The required reserve ratio and the money multiplier do not exhibit a negative correlation, and there were two structural breaks in the evolution of the money multiplier: 1992 and 2008. \\

\begin{figure}[t]
\centerfloat
\includegraphics[width=7cm,height=5.4cm]{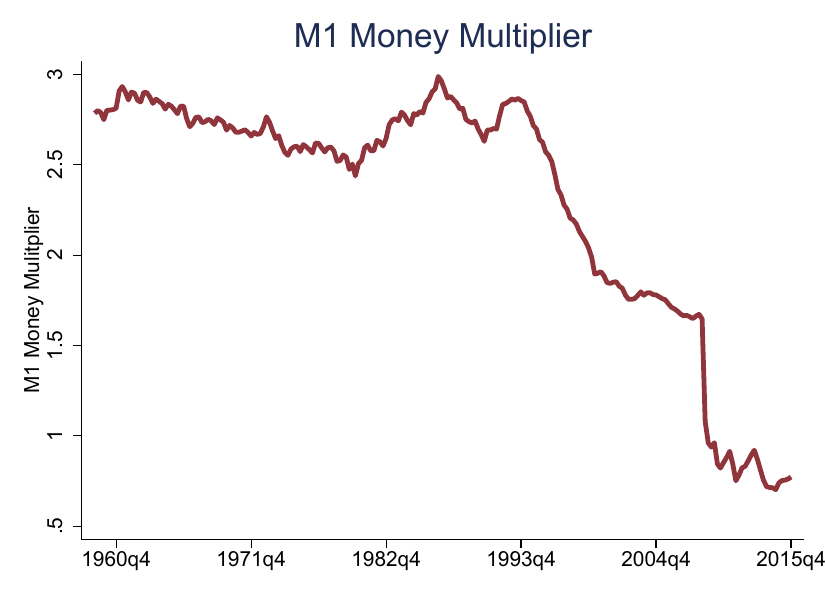}
\includegraphics[width=7cm,height=5.4cm]{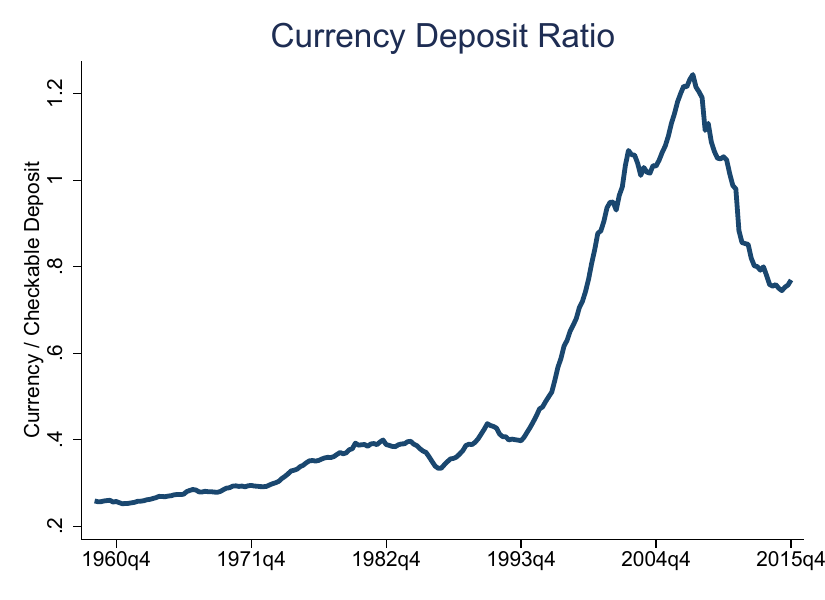}
\includegraphics[width=7cm,height=5.4cm]{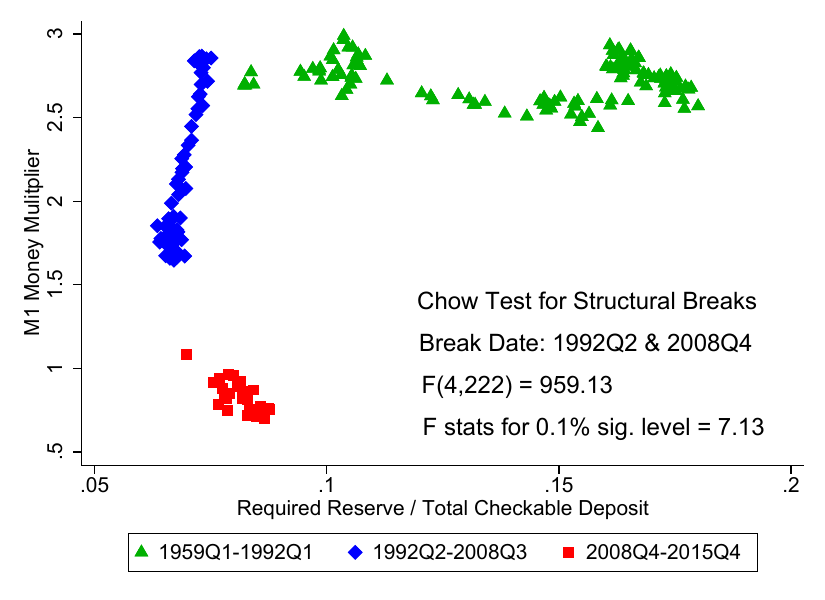} 
\includegraphics[width=7cm,height=5.4cm]{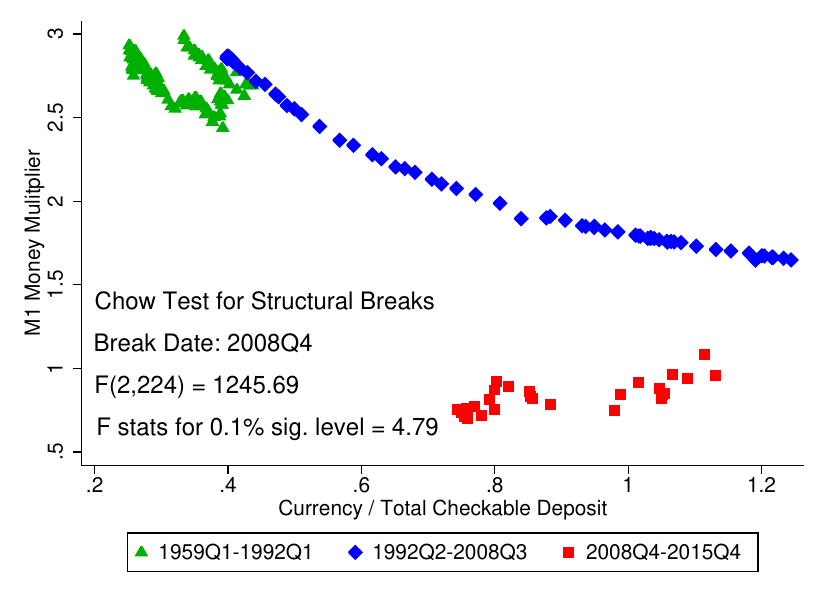}
\caption{Money multiplier, currency/deposit ratio and required reserve ratio}
\label{fig:multip1}
\medskip 
\begin{minipage}{0.97\textwidth} 
{\footnotesize Chow tests for structural breaks are implemented. The bottom-left panel reports a test statistic with the null hypothesis of no structural breaks in 1992Q2 and 2008Q4 and the bottom-right panel reports a test statistic with the null hypothesis of no structural break in 2008Q4. Sample periods are 1959Q1-2015Q4. Appendix \ref{sec:append_1} contains details of the Chow tests. \par}
\end{minipage}
\end{figure}

The top-left panel of Figure \ref{fig:multip1} plots the US M1 multiplier over time. While the money multiplier decreased drastically after 2008, the declining trend had already begun in the early 1990s. An increase in required reserves did not accompany this decrease in the money multiplier. The bottom-left panel of Figure \ref{fig:multip1} plots the  M1 multiplier against the required reserves.\footnote{The required reserve ratio presented in Figure \ref{fig:multip1} is computed by (Required Reserves)/(Total Checkable Deposits). The legal reserve requirement for net transaction accounts was 10\% from April 2, 1992, to March 25, 2020, but some banks are imposed upon by lower requirements or exempt depending on the size of their liabilities. These criteria changed 27 times from the 1st quarter of 1992  to the last quarter of 2019. From March 2020, all the required reserve ratios have become zero. See \cite{feinman1993reserve} and \href{https://www.federalreserve.gov/monetarypolicy/reservereq.htm}{https://www.federalreserve.gov/monetarypolicy/reservereq.htm} for more details on the historical evolution of the reserve requirement policy of the United States.} It shows that the required reserve ratio and the money multiplier do not exhibit a negative correlation. (The correlations are 0.5682 for 1959Q1-2015Q4 and 0.6039 for 1959Q1-2007Q4.)

The left panels of Figure \ref{fig:multip1} show that the US M1 money multiplier has been decreasing since 1992. However, the declining trends before and after 2008 are different.  As the top-right panel of Figure \ref{fig:multip1} shows, the decline during 1992-2007 is accompanied by a huge increase in the ratio of currency to deposit, whereas the decline after 2008 is accompanied by a huge drop in the ratio of currency to deposit. It is also worth noting that the M1 multiplier has been lower than 1 since 2009, which contradicts the textbook theory of money creation.

The difference in the declining trends of the M1 money multiplier is further illustrated through structural breaks. The bottom-left panel of Figure \ref{fig:multip1} displays two structural breaks in the relationship between the M1 multiplier and the required reserve ratio: one in 1992Q3 and another in 2008Q4. In addition, the bottom-right panel of Figure \ref{fig:multip1} shows a structural break in the relationship between the M1 multiplier and the currency deposit ratio, which occurred in 2008Q4. The Chow tests for these breaks are described in more detail in Appendix \ref{sec:append_2}. The structural break of 2008Q4  coincided with the dramatic increase in excess reserves, which is illustrated in Observation 2. 

The absence of a negative correlation suggests that the evolution of the money multiplier can be more complicated than the textbook explanation. Observation 4 suggests that enhanced credit conditions could be a potential explanation.  \\

\noindent\textbf{Observation 4.} Adding unsecured credit into the money demand equation as a regressor recovers the downward-sloping, stable M1 money demand.  \\

\begin{figure}[tp!]
\centerfloat
\includegraphics[width=5.05cm,height=4.5cm]{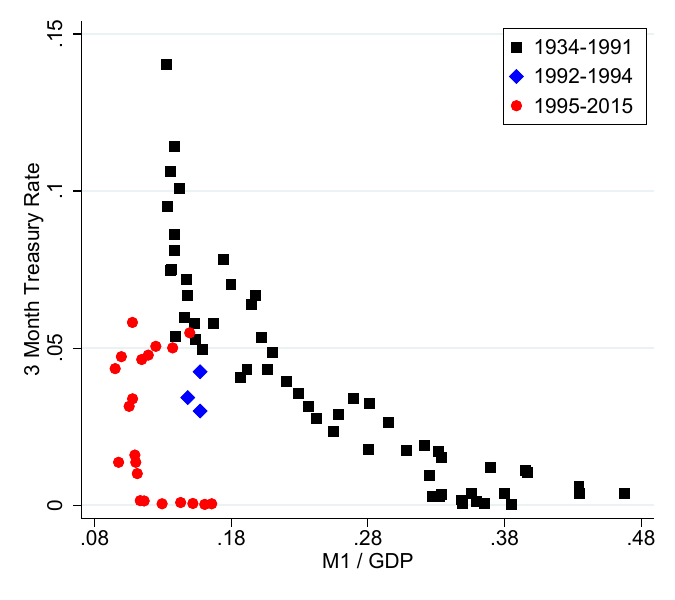}
\includegraphics[width=5.05cm,height=4.5cm]{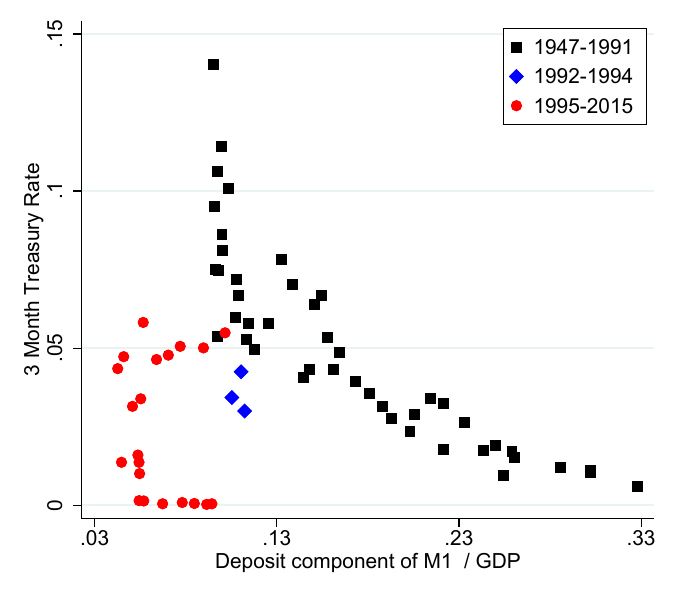}
\includegraphics[width=5.05cm,height=4.5cm]{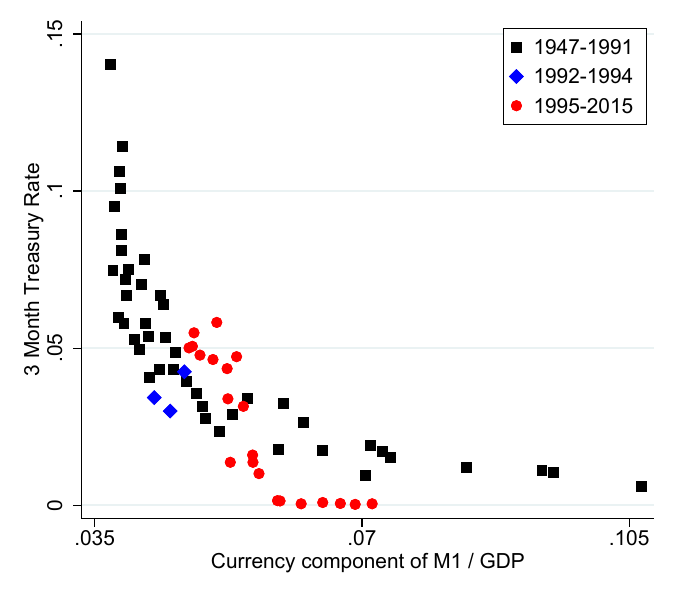}
\caption{US Money demand for M1 and its components}
\label{fig:motive3}
\end{figure}

Observation 3, discussed earlier, identified two structural breaks in 1992 and 2008. To gain a better understanding of the 1992 break, M1 can be decomposed into its deposit and currency components, as illustrated in Figure \ref{fig:motive3}. The figure plots the ratio of M1 and its components to GDP against the 3-month Treasury Bill rate, revealing a breakdown in M1 around 1992 that coincided with the structural break observed in Figure \ref{fig:multip1}.\footnote{One may think this is due to the introduction of retail sweep accounts and Automatic Transfer Service (ATS) in the 1990s. However, ATS was introduced in 1994 and the break of M1 occurred in 1992 which was before 1994. In addition to that, using sweep adjusted M1, \cite{kejriwal2022two} also found a similar structural break. See Appendix \ref{sec:sweep} for more discussion.} As noted by \cite{lucas2015stability}, this breakdown was caused by the deposit component; in contrast, the currency component displays a stable downward-sloping demand.

If an increased availability of consumer credit crowds out deposits but not cash, which implies that once one account for the substitution effect of the newly available consumer credit, there should still be a negative relationship between the real money balance and the interest rate.

\begin{table}[tp!]
\caption{Cointegration regressions and tests}
\label{tab:motive_reg}
\footnotesize
\centering
\begin{threeparttable}
\begin{tabular}{lD{.}{.}{4}D{.}{.}{4}D{.}{.}{4}cD{.}{.}{4}D{.}{.}{4}D{.}{.}{4}} 
\toprule \hline
Dependent Variable:  & 
\multicolumn{3}{c}{$ln(m_t)$}& & \multicolumn{3}{c}{$ln(d_t)$} \\ \cline{2-4}  \cline{6-8} 
&\multicolumn{1}{c}{OLS}&\multicolumn{1}{c}{OLS}&\multicolumn{1}{c}{CCR} &  & \multicolumn{1}{c}{OLS}& \multicolumn{1}{c}{OLS}&  \multicolumn{1}{c}{CCR} \\ 
&\multicolumn{1}{c}{(1)}&\multicolumn{1}{c}{(2)}&\multicolumn{1}{c}{(3)} & & \multicolumn{1}{c}{(4)} &\multicolumn{1}{c}{(5)}      & \multicolumn{1}{c}{(6)}    \\ \hline
$r_t$     &1.600^{***} & -2.298^{***}  & -2.755^{**} & & 4.928^{***}& -3.259^{**} & -4.990^{*}  \\ 
          &  (0.432)   &  (0.740) & (1.171)    & & (0.943) &  (1.481)  &  (2.647)        \\
$ln(uc_t)$&            & -0.322^{***}  & -0.282^{***} & &           & -0.677^{***} & -0.583^{***}   \\
          &            &(0.056) &  (0.098)       & &           & (0.110) &  (0.124)     \\ \hline
$adj R^2$ &  0.109     & 0.416 &   0.981       & & 0.230      & 0.520  &  0.959       \\  
Observation & \multicolumn{1}{c}{112} & \multicolumn{1}{c}{112}  & \multicolumn{1}{c}{112}   & &  \multicolumn{1}{c}{112} & \multicolumn{1}{c}{112} &  \multicolumn{1}{c}{112}    \\  \hline
Johansen $r=0$              & 15.031 & \multicolumn{2}{c}{50.151}  & & 10.348 & \multicolumn{2}{c}{53.881}       \\
Johansen $r=1$              & 0.034  & \multicolumn{2}{c}{9.842}  & & 0.563  &  \multicolumn{2}{c}{10.538}       \\
\hline \bottomrule 
\end{tabular}
\end{threeparttable}
\begin{minipage}{0.88\textwidth} 
{\scriptsize  Notes: Columns (1)-(2) and (4)-(5) report OLS estimates and columns (3) and (6) report the canonical cointegrating regression (CCR) estimates.  First-stage long-run variance estimation for CCR is based on Bartlett kernel and lag 1.  For (1)-(2) and (4)-(5) Newey-West standard errors with lag 1 are reported in parentheses. Intercepts are included but not reported. ***, **, and * denote significance at the 1, 5, and 10 percent levels, respectively. Johansen cointegration test results are reported. Appendix \ref{sec:append_2} contains unit root tests for each series. The data are quarterly from 1980Q1 to 2007Q4.  \par}
\end{minipage}
\label{tab:reg1}
\end{table}
Following \cite{cagan1956monetary}, and \cite{ireland2009welfare}, I relate the natural logarithm of $m$, the ratio of money balances to income, to the short-term nominal interest rate, denoted by $r$. I also regress $r$ on the natural logarithm of $d$, the ratio of deposit balances to income.
$$ln(m_t)=\beta_0+\beta_1 r_t+\epsilon_t, \quad ln(d_t)=\beta_0+\beta_1 r_t+\epsilon_t $$
In addition to the above specifications, to capture the impact of the improved availability of consumer credit that can substitute the deposit, I add a logarithm of $uc$, the ratio of unsecured credit to income as another regressor as follows.\footnote{Following to \cite{krueger2006does}, I use revolving consumer credit.}
$$ln(m_t)=\beta_0+\beta_1 r_t+\beta_2 ln(uc_t)+\epsilon_t, \quad ln(d_t)=\beta_0+\beta_1 r_t+\beta_2 ln(uc_t)+\epsilon_t
$$

I focus on the post-1980 period, until the arrival of the Great Recession. In Table \ref{tab:reg1},  columns (1) and (4) report the estimates without unsecured credit, and columns (2)-(3) and (5)-(6) report the estimates with unsecured credit. The Johansen tests in columns (1) and (4) fail to reject the null hypothesis of no cointegration, which confirms the apparent breakdowns from Figure \ref{fig:motive3}, and ordinary least squares (OLS) estimates from columns (1) and (4) both report positive coefficients on $r_t$ that contradict the conventional notion of money demand: the stable downward-sloping relationship
between real balances and the interest rate. 

In columns (2)-(3) and (5)-(6), however, the Johansen tests reject their null hypothesis of no cointegration at a 99 percent confidence level, suggesting that there exists a stable relationship among real money balances, the interest rate, and unsecured credit. To estimate the cointegration relationship, I implement the canonical cointegrating regression, proposed by \cite{park1992canonical}, in columns (3) and (6).\footnote{The OLS estimates would likely be biased given the non-stationarity of the data and long-run variances from the cointegration relationship. Columns (2) and (5) report OLS estimates for the same variables just for comparison.}  The estimated coefficients on $r_t$ and $ln(uc_t)$ both are negative and significantly different from zero. Thus, using the cointegrating regressions and tests, I document the evidence that once one accounts for the substitution effect of consumer credit, there still exists a stable negative relationship between real money balances and the interest rate. This substitution effect is a potential explanation for the decline of the money multiplier during 1992Q2-2008Q3. 

These findings suggest that a desirable monetary model for studying monetary transmission should have the following properties. First, the model should be capable of explaining how the amount of reserves is linked to interest rate management, regardless of whether banks hold excess reserves or not. Second, the model should feature the distinct roles of interest on reserves and nominal interest rates. Third, the model should be able to answer why banks are holding excess reserves, whereas they did not before 2008. Lastly, the model needs to capture the interaction between money and credit. In the following sections, 
by incorporating these four properties, I develop a theoretical model of the money creation process which is consistent with the above four observations, both qualitatively and quantitatively.

\section{Model}
\label{sec:model}
The model constructed here extends the standard monetary search model \citep{lagos2005unified} by introducing fractional reserve banking, unsecured credit, and capital accumulation. Time is discrete and two markets convene sequentially in each time period: (1) a decentralized market (DM, hereafter), where buyers and sellers meet and trade bilaterally, followed by (2) a frictionless centralized market (CM, hereafter), where agents work, consume, and adjust their balances. The DM trade features imperfect record-keeping and limited commitment. Due to these two frictions, some means of payment are needed in DM trades. 

There is a unit mass of households who discount their utility each period by $\beta$. The preferences of the households for each period are
$$\mathcal{U}= U(C)-\zeta H+u(q)-c(q)$$
where $C$ is the CM consumption, $\zeta> 0$ is a parameter, $H$ is the CM disutility from production, and $q$ is the DM consumption. As standard, assume $U'$, $u'$, $c'>0$, $U''$, $u''<0$, $c''\geq0$, and $u(0) = c(0)=0$. Furthermore, I assume that the CM utility has a constant relative risk aversion equal to 1, i.e., $-C\frac{ U''(C)}{U'(C)}=1$. Consumption goods are perishable.  The efficient quantity of DM consumption in the DM is denoted by $q^*$ which solves $u'(q^*)=c'(q^*)$. At the beginning of the DM, each household receives a preference shock such that they can be either  buyer or a seller. A household will be a buyer with probability $\nu$, while with probability $1-\nu$ a household is a seller.

There is a representative firm. During the CM young firms are born, and they become old and die in the next CM. Firms maximize their profit by producing CM consumption goods by hiring labor and using capital as inputs. The production technology is given by $F(K,N)$ where $K$ is the capital input and $N$ is the labor input. The production function $F(K,N)$ satisfies $F_K,F_N>0$, $F_{KK},F_{NN}<0$ and constant returns to scale. Given a constant returns to scale technology, we can define $f(k)\equiv F(k,1)=F(K,N)/N$ where $k=K/N$. 
Capital is depreciated by $\delta$ every period. Assume that firms are anonymous in the CM, and they cannot commit to honour intertemporal promises. Therefore, a firm needs a medium of exchange to purchase capital goods. To finance investments, firms can borrow from banks. There exists a capital producer whose technology  can
transform CM consumption goods into capital goods with cost $\Phi(\cdot)$. 

There are measure $n$ of active banks that will be endogenously determined by a free entry condition in equilibrium. In the CM, banks make portfolio choices for reserves, loans, transaction deposits, and non-transaction deposits. Banks extend loans to firms by creating deposits that households can use as a means of payment to trade goods in the DM. Loans are paid back with interest $i_\ell$.  Enforcing repayment is costly. The cost function is described by $\eta(\tilde{\ell})$, where $\tilde{\ell}$ is the amount of loans in real terms, $\eta',\eta''> 0$, and $\eta(0)=0$. Managing reserve balances also incurs a cost. The cost is represented by a cost function $\gamma(\tilde{r})$, where $\tilde{r}$ is the amount of reserves in real terms, $\gamma',\gamma''>0$, and $\gamma(0)=0$. Each unit of reserves earns a nominal interest rate of $i_r$.  Banks are subject to a reserve requirement: for transaction deposits recorded as liabilities on their balance sheet, a fraction must be held as reserves. Specifically, a bank must hold at least $\chi\Tilde{d}$ as reserves, where $\chi$ is the reserve requirement and $\tilde{d}$ is the real balance of transaction deposits. 


There are three types of DM meetings depending on payment methods that sellers accept: DM1, DM2, and DM3. In DM1, there is no record-keeping device, and the seller can only recognize cash. In DM2, the seller  accepts transaction deposits and unsecured credit. In DM3, she accepts cash, transaction deposits and unsecured credit. The buyer can trade using unsecured credit with credit limit $\bar{b}$ as the trading is monitored imperfectly.\footnote{The acceptance of different means of payment can be endogenized as in \cite{lester2012information} or \cite{lotz2016money} but here we assume the types of meetings are exogenously given. \cite{lester2012information} endogenize the meeting types by allowing sellers' costly \textit{ex ante} choice to acquire the technology for recognizing the certain type of assets. Similarly, \cite{lotz2016money} study the environment with costly record-keeping technology where sellers must invest in a record-keeping technology to accept credit.}  The buyer's probability of a type $j$ meeting is $\sigma_j$, while the seller's probability of a type $j$ meeting is $\frac{\sigma_{j}\nu}{1-\nu}$.

The central bank controls the base money supply $B_t$ in the CM. Let $\mu$ denote the base money growth rate. Then, changes in the quantity of base money can be written as 
$$\mu_{t+1} B_t=B_{t+1}-B_t,$$
The base money is held in two ways: (1) $M_t$ as currency in circulation, i.e., outside money held by households; (2) $R_t$ as reserves held by banks. Thus,
$$ B_t=M_{t}+ R_{t}. $$
The central bank can control the base money supply in two ways. First, it can conduct a lump-sum transfer or collect a lump-sum tax in the CM. Second, it can increase the money supply by paying interest on reserves, $i_r$. Let $T$ represents a lump-sum transfer (or tax if it is negative). The central bank's constraint is
$$\mu_{t+1}\phi_{t+1} B_{t}=\phi_{t+1} (B_{t+1}-B_{t})=T_{t+1}+  i_{r,t+1}\phi_{t+1} R_{t+1}, $$ where $\phi$ is the price of money in terms of the CM consumption good.

\subsection{The CM Problem}
Let $W(m,d,s,b)$ denote the CM value function where $m$ is the cash holding, $d$ is the transaction deposit balance, $s$ is the non-transaction deposit balance, and $b>0$ is the unsecured debt owed to the seller from the previous DM (or unsecured loans to the buyer from the previous DM, if $b<0$). All the state variables are
in unit of the current CM consumption good. The CM problem is 
\begin{align*}
W(m_t,d_t,s_t,b_t)  = \max_{\substack{C_t,H_t,  \hat{m}_{t+1}, \hat{d}_{t+1}, \hat{s}_{t+1}} } &U(C_t)-\zeta H_t+\beta V(\hat{m}_{t+1}, \hat{d}_{t+1}, \hat{s}_{t+1})  \\
\text{  s.t.  }  \frac{\phi_t}{\phi_{t+1}}(\hat{m}_{t+1}+\hat{d}_{t+1}+\hat{s}_{t+1})+C_t & =w_tH_t+ m_t+(1+i_{d,t})d_t+(1+i_{s,t})s_t-b_t+T_t 
\end{align*}
where $\hat{m}_{t+1}$, $\hat{d}_{t+1}$, and $\hat{s}_{t+1}$  is the cash holding, transaction deposit balance, and non-transaction deposit balance, respectively, carried to the next DM, and $w$ is the real wage. The first-order conditions (FOCs) are
\begin{align} 
-\frac{\phi_t}{\phi_{t+1}}\frac{\zeta}{w_t}&+\beta V_m(\hat{m}_{t+1},\hat{d}_{t+1},\hat{s}_{t+1}) \leq 0, \text{} = \text{ if }
\hat{m}_{t+1}>0  \label{eq:gcmfoc1} \\ 
-\frac{\phi_t}{\phi_{t+1}}\frac{\zeta}{w_t}&+\beta V_d(\hat{m}_{t+1},\hat{d}_{t+1},\hat{s}_{t+1})\leq 0, \text{} = \text{ if } \hat{d}_{t+1}>0  \label{eq:gcmfoc2} \\ 
-\frac{\phi_t}{\phi_{t+1}}\frac{\zeta}{w_t}&+\beta V_s(\hat{m}_{t+1},\hat{d}_{t+1},\hat{s}_{t+1})  \leq 0, \text{} = \text{ if } \hat{s}_{t+1}>0 
\label{eq:gcmfoc3}   \\
-\frac{\zeta}{w_t}&+U'(C_t)=0.
\label{eq:gcmfoc4}
\end{align}
The first term on the left-hand side (LHS) of equation (\ref{eq:gcmfoc1}) is the marginal cost of acquiring cash. The second term is the discounted marginal value of carrying cash to the following DM. Therefore, the choice of $\hat{m}_{t+1}>0$ equates the marginal cost and the marginal return on cash. A similar interpretation applies to equations (\ref{eq:gcmfoc2}) and (\ref{eq:gcmfoc3})  for the decision on deposits. The envelope conditions for $W(m,d,s,b)$ are
\begin{align*}
& W_d(m_t,d_t,s_t,b_t)=(1+i_{d,t})\frac{\zeta}{w_t},\quad W_m(m_t,d_t,s_t,b_t)=\frac{\zeta}{w_t}  \\
& W_s(m_t,d_t,s_t,b_t)=(1+i_{s,t})\frac{\zeta}{w_t},\quad W_b(m_t,d_t,s_t,b_t)=-\frac{\zeta}{w_t}    
\end{align*}
which implies $W(m_t,d_t,s_t,b_t)$ is linear in $m$, $d$, $s$, and $b$. This linearity allows us to write
$$W(m_t,d_t,s_t,b_t)=\frac{\zeta}{w_t}\left\{m_t+(1+i_{d,t})d_t+(1+i_{s,t})s_t \right\}-\frac{\zeta}{w_t}b_t+W(0,0,0,0,0).$$

\subsection{The DM Problem}
In the DM, the buyer and seller trade bilaterally. Let $q_j$ and  $p_j$  be the DM consumption and payment in a type-$j$ DM meeting. The bilateral trade is characterized by $(p_j,q_j)$. This trade is subject to $p_j \leq z_j$ where $z_j$ is the total liquidity of the buyer in a type-$j$ meeting. The liquidity position for each type of buyer is 
\begin{align}
z_1&= m  \\ 
z_2&= d(1+i_d)+\bar{b}  \\
z_3&= m+ d(1+i_d)+\bar{b}  
\end{align}
The DM terms of trade are determined by \cite{kalai1977proportional}'s proportional bargaining. Kalai bargaining solves the following problem:

$$\max_{p,q} u(q)-\frac{\zeta }{w}p \quad \textit{s.t} \quad u(q)-\frac{\zeta }{w}p=\theta\left[ u(q)-c(q)\right] \text{ and } p_j \leq z_j$$
where $\theta \in [0,1]$ denotes the buyers' bargaining power. The payment, $p$, can be expressed as $p= v(q)w/\zeta=\{(1-\theta)u(q)+\theta c(q)\}w/\zeta$.
Define \textit{liquidity premium}, $\lambda(q)$, as follows:
\begin{equation*}\lambda(q)=\frac{u'(q)}{v'(q)}-1=\frac{\theta [u'(q)-c'(q)]}{(1-\theta) u'(q)+\theta c'(q)}
\end{equation*} 
where $\lambda(q)>0$ and $\lambda'(q)<0$ for $q<q^*$ and $\lambda(q^*)=0$. When $z_j\geq p^*$, the buyer has sufficient liquidity to purchase efficient DM output $q^*$. In this case, the payment to the seller is $p^*=v(q^*)w/\zeta$. 

The value function of a household at the beginning of DM is
$$V(m,d,s)=\nu V^B(m,d,s)+(1-\nu) V^S(m,d,s)$$
where $V^B(m,d,s)$ and $V^S(m,d,s)$ denote the value function for a buyer and a seller, respectively. By using the linearity of $W$, we can write a DM value function for a seller as follows: 
$$ V^S(m,d,s)= \sum_{j=1}^3 \left\{ \frac{\sigma_{j}\nu}{1-\nu} \left[\frac{\zeta }{w}p_j-c(q_j)\right]\right\}+W(m,d,s,0) $$
and the value function of a buyer in the DM is
$$ V^B(m,d,s)= \sum_{j=1}^3 \left\{\sigma_{j} \left[u(q_j)-\frac{\zeta }{w}p_j\right]\right\}+W(m,d,s,0) $$
where $p_j \leq z_j$. The second term on the right-hand side (RHS) is the continuation value when there is no trade. The rest of the RHS is the surplus from the DM trade. The DM payments are constrained by $p_j\leq z_j$. For compact notation, define inflation rate as $\pi_{t+1}\equiv\phi_t/\phi_{t+1}-1$. Assuming interior, differentiating $V$ and substituting its derivatives into the FOCs from the CM problem yields 
\begin{align}
(1+\pi_{t+1})U'(C_t)&=\beta U'(C_{t+1})[1+\nu\sigma_1\lambda(q_{1,t+1})+\nu\sigma_3\lambda(q_{3,t+1})] \label{eq:agent_ps} \\ 
(1+\pi_{t+1})U'(C_t)&=\beta  U'(C_{t+1})[1+\nu\sigma_2\lambda(q_{2,t+1})+\nu\sigma_3\lambda(q_{3,t+1})](1+i_{d,t+1})
\label{eq:agent_pd} \\
(1+\pi_{t+1})U'(C_t)&=\beta U'(C_{t+1})(1+i_{s,t+1})
\label{eq:agent_pe}
\end{align}
where $q_{j,t+1}=\min\{q^*,v^{-1}(\zeta z_{j,t+1}/w_{t+1})\}$ and  $\lambda(q^*)=0$. 

\subsection{The Bank's Problem}
The banking sector is perfectly competitive and banks take the interest rates as given: lending rate $i_{\ell,t}$, transaction deposit rate $i_{d,t}$, non-transaction deposit rate $i_{s,t}$ and interest on reserves $i_{r,t}$. The bank maximizes its profit by choosing $\{\tilde{\ell}_t,\tilde{r}_t, \tilde{d}_t,\tilde{s}_t\}$ subject to its balance sheet identity constraint and reserve requirement constraint, where $\tilde{\ell}_t$ is lending,  $\tilde{r}_t$ is reserve balance, $\tilde{d}_t$ the transaction deposit issuance, and $\tilde{s}_t$ the non-transaction deposit issuance, respectively, denoted in real terms: 
\begin{equation}
\begin{split}
\max_{\Tilde{r}_t,\Tilde{d}_t,\Tilde{\ell}_t,\Tilde{s}_t} \quad  (1+i_{\ell,t})\Tilde{\ell}_t+(1+i_{r,t})& \Tilde{r}_t-(1+i_{d,t})\Tilde{d}_t-(1+i_{s,t})\Tilde{s}_t-\gamma(\Tilde{r}_t) -\eta(\Tilde{\ell}_t)   \\
\text{ subject to } \quad &\Tilde{\ell}_t+\Tilde{r}_t= \Tilde{d}_t+\Tilde{s}_t \quad \text{ and } \quad  \Tilde{r}_t\geq \chi\Tilde{d}_t
\end{split}
\end{equation}
In the first constraint, balance sheet identity, the LHS represents the value of assets such as reserves and loans, and the RHS represents the value of liabilities such as transaction deposits and non-transaction deposits. The second constraint is the reserve requirement. Let $\lambda_{\chi,t}$ denote the Lagrange multiplier for the reserve requirement constraint. Assuming interior, the FOCs for the bank's problem can be written as 
\begin{align}
&0=i_{\ell,t}-i_{s,t}-\eta'(\ell_t) \label{eq:bank_foc1} \\
&0=i_{r,t}-i_{s,t}-\gamma'(\Tilde{r}_t)+\lambda_{\chi,t} \label{eq:bank_foc2}  \\
&0=i_{s,t}-i_{d,t}-\chi\lambda_{\chi,t}. \label{eq:bank_foc3} 
\end{align}
The bank's \textit{ex post} profit equals to the entry cost, $\kappa$
\begin{equation}
\label{eq:bank_entry} 
(1+i_{\ell,t})\Tilde{\ell}_t+(1+i_{r,t}) \Tilde{r}_t-(1+i_{d,t})\Tilde{d}_t-(1+i_{s,t})\Tilde{s}_t-\gamma(\Tilde{r}_t) -\eta(\Tilde{\ell}_t)=\kappa.   
\end{equation}
Suppose there are active banks i.e., $n>0$. Consider two cases. In the first case, the reserve requirement constraint is binding, i.e., $\lambda_{\chi,t}>0$. In the second case, the reserve requirement constraint is loose, i.e., $\lambda_{\chi,t}=0$.  We call the first case a \say{scarce-reserves case,} and the second an \say{ample-reserves case.}  \\

\noindent\textbf{The Scarce-Reserves Case } Consider the case where the bank does not have enough reserves. It needs to acquire reserves to issue more transaction deposits, which implies a binding constraint. With $\lambda_{\chi,t}>0$, the bank's FOCs (\ref{eq:bank_foc1})-(\ref{eq:bank_foc3}) give
\begin{align}
i_{d,t}&=(1-\chi) i_{s,t}+\chi i_{r,t}-\chi\gamma'(\tilde{r}_t) \label{eq:bankfoc_sca1} \\
i_{\ell,t} &=  i_{s,t}+\eta'(\tilde{\ell}_t). \label{eq:bankfoc_sca2}
\end{align}

\noindent\textbf{The Ample-Reserves Case } Consider the case where the bank has sufficient reserves. Its reserve requirement constraint does not bind, $\lambda_{\chi,t}=0$. Then the three FOCs for the bank's problem become
\begin{align}
i_{d,t}&=i_{s,t} \label{eq:bankfoc_amp1} \\
i_{r,t} &= i_{s,t}+\gamma'(\Tilde{r}_t)
\label{eq:bankfoc_amp2}    \\
i_{\ell,t} &=  i_{s,t}+\eta'(\tilde{\ell}_t). \label{eq:bankfoc_amp3}
\end{align}

The key difference between these two cases is that in the scarce-reserve case, banks only hold required reserves as the constraint binds. In contrast, in the ample-reserve case, banks can hold excess reserves in addition to required reserves because the reserve requirement constraint is no longer binding. 

\subsection{The Firm and Capital Producer}
A representative firm maximizes its profit by producing CM consumption goods and using its capital $K_t$ and hiring labor $N_t$ as inputs. In the CM of $t-1$,  the firm borrows funds $L_t$ from banks, and purchases capital goods $K_t$ using the funds. A firm purchases capital from a perfectly competitive capital producing firm at the end of period $t-1$. This capital is used in production at $t$ and its undepreciated $(1-\delta)K_t$ part is resold to a capital producer once the production is over. The firm's problem can be written as follows:
\begin{align*}
\max_{N_{t},K_{t},L_{t}} L_t-Q_{t-1} K_t+\beta\left[F(K_t,N_t)-w_tN_t+Q_t(1-\delta)K_t-(1+i_{\ell, t})\frac{L_t}{1+\pi_t} \right]
\end{align*}
subject to $L_{t}=Q_{t-1} K_{t}$, where $Q_t$ is price of capital in terms of CM consumption good at period $t$. The firm’s problem gives
\begin{equation}
\label{eq:firm_foc}
F_N(K_t,N_t)=w_t, \quad F_K(K_{t},N_{t})=Q_{t-1}\frac{1+i_{\ell, t}}{1+\pi_{t}}-Q_{t}(1-\delta).
\end{equation}
The capital law of motion is given as: 
$$ K_{t+1}=(1-\delta)K_t+I_t.$$
A capital producer can transform CM consumption goods into capital goods with cost $\Phi(I_t)$. Formally, a capital producer solves the following profit-maximization problem:
$$\max_{I_t}Q_t I_t-\Phi(I_t)$$
which gives $Q_t=\Phi'(I_t)$. Assuming a linear cost function, $\Phi(I_t)=I_t$, gives $Q_t=1$ for all $t$. Given above results, define the real lending rate:
$$\rho_{t}\equiv\frac{1+i_{\ell,t}}{1+\pi_{t}}-1$$
Then we have the following equilibrium condition for the real lending rate and the marginal product of capital as follows:
\begin{equation}\label{eq:mpk}
F_K(K_{t},N_{t})=\rho_{t}+\delta
\end{equation}

\subsection{Equilibrium}
In the equilibrium, 
the resource constraint for CM consumption goods and labor market clearing condition are satisfied.
\begin{equation}
C_t+K_{t+1} =F(K_t,N_t)+(1-\delta)K_t,\text{ and } N_t=H_t\label{eq:clear11}
\end{equation}
The money market clearing conditions are given as below
\begin{equation}
\phi_{t+1} M_{t+1}= m_{t+1},\quad 
\phi_{t+1} R_{t+1}=n_{t+1}\Tilde{r}_{t+1}, \quad \text{and} \quad  B_{t+1}   =M_{t+1}+R_{t+1} \label{eq:clear12}
\end{equation}
and market clearing condition for lending and deposits are satisfied.  
\begin{equation}
L_{t+1} =  n_{t+1}\Tilde{\ell}_{t+1}, \quad \text{and} \quad    d_{t+1}=n_{t+1}\Tilde{d}_{t+1} 
\label{eq:clear13}
\end{equation}
Given agents' optimal choices and market clearing conditions, we define a monetary equilibrium as follows:
\singlespacing
\begin{definition}\label{def:me} Given a sequence of monetary policy $\{B_t,i_{rt},\chi_t\}^{\infty}_{t=1}$ and credit condition  $\{\bar{b}_t\}^{\infty}_{t=1}$, and initial conditions $(K_0,B_0)$, a monetary equilibrium is a sequence of quantities $\{K_t,N_t,m_t,d_t,s_t,r_t\}^\infty_{t=1}$, prices $\{\phi_t,i_{\ell t},i_{d,t},w_t,\rho_t\}^\infty_{t=1}$, and measures of active banks, $\{n_t\}^\infty_{t=1}$ that satisfies: 
\begin{enumerate}
\item The Euler equations (\ref{eq:agent_ps})-(\ref{eq:agent_pe}), and transversality conditions:
$$\lim_{t\rightarrow\infty}\beta^t K_t=\lim_{t\rightarrow\infty}\beta^tU'(C_t)\phi_t m_t=\lim_{t\rightarrow\infty}\beta^tU'(C_t)\phi_t d_t=\lim_{t\rightarrow\infty}\beta^tU'(C_t)\phi_t s_t=0$$
\item Optimality conditions of banks and the firm, (\ref{eq:bank_foc1})-(\ref{eq:bank_entry}) and (\ref{eq:firm_foc});
\item  Market clearing  (\ref{eq:clear11})-(\ref{eq:clear13}), and $\phi_t B_t>0$.
\end{enumerate}
\end{definition}
\onehalfspacing

As standard, the short-term policy rate $i_{t}$ is given by the Fisher equation
\begin{equation}\label{equation:peg}
1+i_{t}=(1+\pi_{t+1})\frac{U'(C_t)}{\beta U'(C_{t+1})}
\end{equation}
implying $i_{t}=i_{s,t+1}$. 

Similar to \cite{sargent1975rational} and \cite{gu2020effects}, a central bank can peg $i_{t}$ by letting $B_{t+1}$ evolve endogenously as long as the base money is valued, $\phi_{t+1} B_{t+1}>0$. Here, monetary policy implementation is different from what is assumed in the previous literature. For example, New Keynesian models simply assume that the central bank can determine interest rates. In the other monetary models with fiat money, the central bank implements the monetary policy by controlling the aggregate money supply, or by controlling the growth rate of aggregate money supply. In this model, the central bank can set interest rate by controlling the supply of base money which eventually influences the supply of monetary aggregate, and other macroeconomic variables. Here, for the central bank's  monetary policy implementation, it is crucial  to have monopoly power over the supply of base money which is the sum of reserves and currency in circulation.

Given this environment, we have the following results:


\begin{proposition}\label{prop:pre1}
When the central bank does not pay interest on reserves i.e., $i_{r,t+1}=0$, banks do not hold excess reserves i.e., $\Tilde{r}_{t+1}=\chi\Tilde{d}_{t+1}$.  When banks hold excess reserves, $\partial \Tilde{\ell}_{t+1}/\partial i_{t}>0$, $\partial \Tilde{\ell}_{t+1}/\partial i_{r,t+1}<0$, $\partial \Tilde{r}_{t+1}/\partial i_{t}<0$, $\partial \Tilde{r}_{t+1}/\partial i_{r,t+1}>0$, $\partial i_{\ell,t+1}/\partial i_{t}>0$, and $\partial i_{\ell,t+1}/\partial i_{r,t+1}<0$.
\end{proposition}

Proposition \ref{prop:pre1} says that paying interest on reserves is necessary to have banks hold excess reserves. It also states that changes in the short-term policy rate and the changes in interest on reserves have different effects on the banking sector when banks are holding excess reserves. For example, an increase in the short-term policy rate reduces each bank's reserve balances, while an increase in the interest on reserves raises the bank's reserve balances. This is because 
an increase in the short-term policy rate reduces returns on reserves while an increase in the interest on reserves raises returns on reserves. Also, an increase in the short-term policy rate raises the lending rate, whereas an increase in the interest on reserves lowers the lending rate. \\

\noindent\textbf{Stationary Monetary Equilibrium } In the remaining of this section, I focus on a symmetric stationary monetary equilibrium in which the agents make the same decisions and all real variables are constant over.  Given that $\phi_{t}/\phi_{t+1}=B_{t+1}/B_{t}=M_{t+1}/M_{t}=1+\mu$, the net inflation rate, $\pi$, is equal to the currency growth rate, $\mu$, in the stationary monetary equilibrium. By the Fisher equation, $1+i=(1+\mu)/\beta$.\footnote{Note that $i\geq0$ is necessary for the existence of equilibrium. Whereas the lower bound of the nominal interest rate is zero in this setting, one can relax this by introducing liquid assets or threats of theft. For details on liquid assets, see \cite{rocheteau2018open} and \cite{lee2016money}; for threats of theft, see \cite{kim2024negative}.} This leads to the following
definitions:

\begin{definition}[\textbf{Stationary Monetary Equilibrium}]\label{def:sme}
Given monetary policy, $i$, $i_r$, and $\chi$ and credit limit $\bar{b}$, a stationary monetary equilibrium consists of real balances, $(m,r,d,s,\ell)$, allocation $(q_1,q_2,q_3,C,K,N)$, the measure of banks $n$, and prices $(i_d,i_\ell)$, satisfying Definition \ref{def:me} except for initial conditions.

\end{definition}

\begin{definition}\label{def:type} 
The stationary monetary equilibrium is a scarce-reserves equilibrium when $\tilde{r}=\chi\tilde{d}$ and an ample-reserves equilibrium, when $\tilde{r}>\chi\tilde{d}$, respectively. 
\end{definition}

Given the above definitions, we have the following result. 

\setlength{\abovedisplayskip}{3pt}
\setlength{\belowdisplayskip}{3pt}

\begin{proposition}\label{prop:threshold}
Given $(i, \chi, \bar{b})$: (i)
$\exists !$  ample-reserves equilibrium if and only if $i_r\in(\bar{\iota}_r,\bar\Delta+i)$;
(ii) $\exists$ scarce-reserves equilibrium if and only if $i_r\leq \bar{\iota}_r$; and the thresholds satisfy $\bar\Delta=\gamma'(\underline{r})$ and $\bar{\iota}_r=\gamma'(\bar{r})+i$ where $\underline{r}$ solves $\kappa=\gamma'(\underline{r})\underline{r}-\gamma(\underline{r})$ and $(\bar{r},\bar{\ell},\bar{K},\bar{N},\bar{C},\bar{n})$ solves
$$F_K(\bar{K},\bar{N})=\frac{1}{\beta}-1+\delta+\frac{\eta'(\bar{\ell})}{\beta(1+i)}, \qquad \max\left\{0,\frac{\chi\left\{v(q^*)F_N(\bar{K},\bar{N})/\zeta-\bar{b}\right\}}{\bar{n}}\right\}=(1+i)\bar{r},$$
$\eta'(\bar{\ell})\bar{\ell}+\gamma'(\bar{r})\bar{r}-\gamma(\bar{r})-\eta(\bar{\ell})=\kappa$, $\bar{C}+\delta \bar{K}=F(\bar{K},\bar{N})$, $U'(\bar{C})=\zeta/F_N(\bar{K},\bar{N})$, and $\bar{K}=\bar{n}\bar{\ell}$.
\end{proposition}

When the central bank does not pays interest on reserves or pays low interest such that $i_{r}\leq \bar{\iota}_r$, banks only hold required reserves because the opportunity cost of holding reserves is higher than its benefit. When the central bank pays interest on reserves such that $i_{r}\in(\bar{\iota}_r,\bar{\Delta}+i)$, banks always hold positive amount of excess reserve balances that satisfies $\gamma'(\Tilde{r})=i_r-i$ and $\Tilde{r}>\chi\tilde{d}$.\footnote{When $i_r>\bar{\Delta}+i$, no equilibrium exists with active banks ($n>0$), but an equilibrium without active banks ($n=0$) exists. We'll focus solely on the equilibrium with an active banking sector.} In this case, banks hold a large amount of reserves because the benefit of holding reserves outweighs its opportunity cost of holding reserves.

Having solved the stationary monetary equilibrium, I proceed to establish the results on the money multiplier. Define the monetary aggregate as $\mathcal{M}\equiv m+d$ and the money multiplier as $\xi\equiv \mathcal{M}/(\phi B)$, then we have the following results.
\begin{proposition}\label{prop:compastatics1}
In the ample-reserves equilibrium, the money multiplier is increasing in $i$ and decreasing in $i_r$, $i.e,$ 
$\partial\xi/\partial i>0$ and  $\partial\xi/\partial i_r<0.$
\end{proposition}

Proposition \ref{prop:compastatics1} shows that the money multiplier is increasing in short-term policy rate while it is decreasing in interest on reserves, which is consistent with the observation illustrated in Figure \ref{fig:motive4}. This result is intuitive. In the ample-reserves equilibrium, banks hold reserves because holding reserves itself is profitable regardless of the reserve requirement. The higher interest on reserves decreases the money multiplier because the banks have more incentive to hold reserves and less incentive to create transaction deposits. Increasing the short-term policy rate decreases reserves, but the banks do not create transaction deposits proportionally, which lowers the money multiplier. These are new findings compared to the literature.

To get more analytical results, I assume that the following restriction is satisfied in any equilibrium\footnote{This assumption implies the convex cost function $\gamma(\cdot)$ is not too convex.}:
\begin{assumption}
$1>\gamma'(\tilde{r})+\gamma''(\tilde{r})\tilde{r}$. 
\end{assumption}

\begin{figure}[tp!]
\centerfloat
\includegraphics[width=7cm,height=6cm]{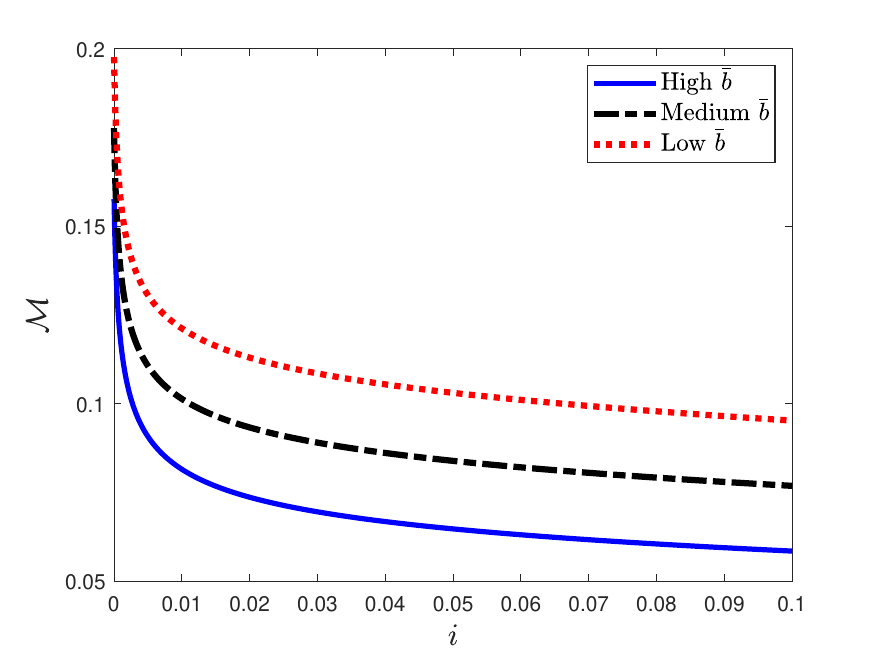}
\includegraphics[width=7cm,height=6cm]{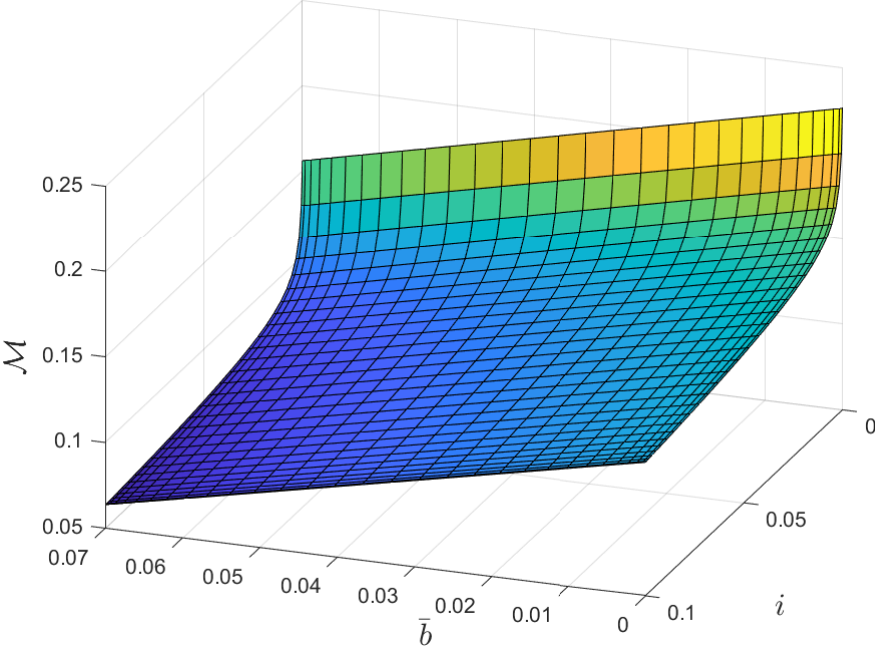}
\caption{Money Demand under Different Credit Limit}
\label{fig:md_dcl}
\end{figure} 
For each type of equilibrium, the following results are proved in Appendix \ref{sec:proof}.
\begin{proposition}\label{prop:compastatics2}
Suppose $p^*>\bar{b}$. In the ample-reserves equilibrium, $\partial i_d/\partial \bar{b}=0$ and $\partial  d/\partial \bar{b}<0$.
In the scarce-reserves equilibrium, when the equilibrium is unique, $\partial i_d/\partial \bar{b}>0$ and  $\partial  d/\partial \bar{b}<0$ if $\bar{b}$ or $\sigma_3$ is small.
\end{proposition}
Proposition \ref{prop:compastatics2} shows two results: (1) a better credit condition increases the deposit rate in the scarce-reserve equilibrium but not in the ample-reserve equilibrium; and (2) an increase in $\bar{b}$ can reduce $d$ in both types of equilibria. One implication for the deposit rate changes is that neutrality between money and credit does not hold in the scarce-reserve equilibrium.\footnote{For more discussion on the neutrality of money and credit, see \cite{gu2016money} and \cite{wang2020sticky}.} This is because the use of transaction deposit money incurs costs for operating banks. An increase in the credit limit lowers the cost of operating deposits by reducing real balances, which increases DM2 and DM3 trades through the increase in the deposit rate. In contrast to the scarce-reserve equilibrium, changes in credit conditions do not impact the deposit rate in the ample-reserve equilibrium; they merely crowd out real balances. Consistent with observation 4 from Section \ref{sec:motive}, the second result implies that an increase in $\bar{b}$ can reduce $d$, which can eventually reduce the real balances of the monetary aggregate, $\mathcal{M}$. Figure \ref{fig:md_dcl} shows some examples.

As can be seen from the results above, the model can successfully address the mechanism illustrated in Section \ref{sec:introd} and \ref{sec:motive}. We can interpret the breakdown of the money demand relationship in the pre-2008 economy as a
result of improved availability of consumer credit under the scarce-reserve equilibrium.
For the post-2008 period, after the Fed started paying interest on reserves, the economy shifted to the ample-reserves equilibrium. The model suggests that the changes in the money multiplier and the excess reserves during the post-2008 period are the results of the Fed’s management of two interest rates: the nominal policy rate and the interest on reserves.  

We now look into DM trades more closely. From (\ref{eq:agent_ps}) and (\ref{eq:agent_pd}), we have 
\begin{align}
\frac{i_t}{\nu}&=\sigma_1\lambda(q_{1})+\sigma_3\lambda(q_{3}) 
\label{eq:DM_trade} \\
\left\{ \frac{1+i}{1+i_d}-1\right\}\frac{1}{\nu}&=\sigma_2\lambda(q_{2})+\sigma_3\lambda(q_{3}).  \label{eq:DM_trade2}
\end{align}
where $i\geq i_d$ and $i\geq 0$. It is straightforward to see that DM1 and DM3 consumptions are efficient, $q_1=q_3=q^*$, when $i_t=0$, i.e. the Friedman rule applies. However, if the central bank pays sufficient interest on reserves, we have $i_d=i$ which gives efficient consumptions in DM3 as well as in DM2 even when $i>0$.  This result can be formally summarized in the following proposition.
\setlength{\abovedisplayskip}{3pt}
\setlength{\belowdisplayskip}{3pt}
\begin{proposition}\label{prop:dmt}
Let the short-term policy rate be positive $i>0$. Then the DM consumptions in  DM2 and DM3 are efficient $q_{2}=q_{3}=q^*$  when $i_{r}\geq\bar{\iota}_r$.
\end{proposition}


The intuition behind the efficient DM consumptions when $i_r>\bar{\iota}_r$  is straightforward. In many monetary models, a higher inflation or interest rate increases the opportunity cost of holding money. In the environment where money is valued as a medium of exchange, having less liquidity in the economy because of the opportunity cost of holding money is inefficient. However, the interest on reserves provides a proportional return. If this return is properly distributed across households, it eliminates the inefficiency arising from the opportunity cost of holding money, leading to efficient consumption in DM2 and DM3. Therefore, when the central bank pays sufficient interest on reserves, the DM2 and DM3 meeting consumptions can be efficient even though the economy is not under the Friedman rule.

In addition to that, if credit limit $\bar{b}$ is sufficiently high, DM2 and DM3 consumptions also can be efficient even though $i>0$.  Appendix \ref{sec:proof} verifies the following:  
\begin{proposition}\label{prop:dmt2}
The threshold $\bar{\iota}_r$ is decreasing in $\bar{b}$. When $\bar{b}\geq p^*$,  $\bar{\iota}_r=i$, $d=0$ and $q_2=q_3=q^*$.  
\end{proposition}

Proposition \ref{prop:dmt2} simply states that if the credit limit is high enough, it results in efficient consumption in both DM2 and DM3 trades. As $\bar{b}\rightarrow p^{*-}$, the household's transaction deposit balance $d$ converges to 0. This is reminiscent of a result by \cite{gu2016money}: if credit
is easy, money is irrelevant; if credit is tight, money is essential, but credit becomes irrelevant. One difference is that even though credit is easy (i.e., $\bar{b}\geq p^*$) the household always holds cash $m>0$ as long as $i<\nu\sigma_1\lambda(0)$ because the household only can trade using cash in the DM1 meeting. Since the better credit condition lowers transaction deposit balance, required reserves also shrink accordingly, leading to decreases in $\bar{\iota}_r$.     

One can also check the interest rate pass-through of the monetary policy. Its pass-through depends on the type of equilibrium. 

\begin{proposition}
\label{prop:compastatics4} 
(i) In the ample-reserve equilibrium, $\partial i_\ell/\partial i>0$, $\partial i_\ell/\partial i_r<0$, $\partial i_d/\partial i=1>0$, $\partial i_d/\partial i_r=0$, and $\partial \rho /\partial i_r<0$ but $\partial\rho/\partial i$ is ambiguous. (ii) In the scarce-reserve equilibrium, when the stationary monetary equilibrium is unique, $\partial i_\ell/\partial i_r<0$, and $\partial \rho /\partial i_r<0$ if $\bar{b}$ or $\sigma_3$ is small. When $\sigma_3$ is small,  $\partial i_\ell/\partial i>0$, and $\partial i_d/\partial i>0$   but $\partial\rho/\partial i$ is ambiguous.
\end{proposition}  

Proposition \ref{prop:compastatics4} tells us that the monetary policy rates pass through the lending rate and deposit rate.
Similar to Proposition \ref{prop:pre1}, changes in the short-term policy rate and the changes in interest on reserves have different effects. In both types of equilibrium, the lending rate is strictly increasing in the short-term policy rate but is strictly decreasing in interest on reserves. In the scarce-reserves equilibrium, the deposit rate is strictly increasing in the short-term policy rate and the interest on reserves. The pass-through from the short-term policy to the real lending rate is ambiguous. However, the real lending rate is strictly decreasing in interest on reserves. As the marginal product of capital is determined by the real lending rate, $F_K(K,N)=\rho+\delta$, the central bank can stimulate the economy by exploiting the pass-through of monetary policy to interest rates. \\

\noindent\textbf{Monetary Transmission and Breaking Neoclassical Dichotomy } In the classical frictionless monetary models, output and the real interest rate are determined independently of monetary policy. In other words, monetary policy is neutral with respect to those real variables. Here, monetary policy could influence the real variable such as investment. It is worth discussing the difference in monetary transmission channels from the previous literature. In the textbook by \cite{gali2015monetary}, Chapter 1  shows that monetary policy is neutral in the classical frictionless monetary models, and  Chapter 3, discusses how the presence of sticky prices makes monetary policy non-neutral.\footnote{The term monetary neutrality is often used differently in the literature. In New Keynesian literature, monetary neutrality implies that changes in short-term policy rates and money supply both do not have impacts on real variables. In contrast, as discussed in \cite{head2012sticky}, in the New Monetarist models, although money is not superneutral, since real effects result from changes in nominal interest rates, inflation, or money growth rate, money is neutral because changes in aggregate money supply do not have real effects on the real variable.}\textsuperscript{,}\footnote{It is worth mentioning that presence of sticky prices does not necessarily make monetary policy non-neutral. \cite{head2012sticky} provides a monetary search model where the price stickiness emerges endogenously in contrast to the models imposing price stickiness exogenously. While \cite{head2012sticky} explains price stickiness and micro-level price level changes which can match with microdata, money is neutral in their model.}  In contrast to those approaches, in this model, monetary policy could influence the real variable without the presence of sticky prices. 

To inspect the mechanism, recall the Equation (\ref{eq:mpk}), $F_K(K_{t+1},N_{t+1})=\rho_{t+1}+\delta$. A decrease in the real lending rate unambiguously lowers the marginal production of capital. To see the anatomy of what constitutes marginal production of capital, rewrite the marginal product of capital as below:
\begin{equation}
\underbrace{F_K(K_{t+1},N_{t+1})}_{\text{marginal product of capital}}=\underbrace{\frac{U'(C_t)}{\beta U'(C_{t+1})}-1+\delta}_{\text{standard neoclassical term}}\quad +\underbrace{\frac{\eta'(\Tilde{\ell}_{t+1})}{1+\pi_{t+1}}}_{\text{bank's marginal cost of lending}}
\end{equation}
What distinguishes this model from the neoclassical growth model is the last term of the above equation. The banks' enforcement cost provides a wedge to finance the investments. For example, as shown in Proposition 1, an increase in interest on reserves lowers $\Tilde{\ell}_{t+1}$ when the banks hold excess reserves. This reduces the marginal cost of financing investment and influences the real lending rate through general equilibrium impact. 

This channel is different from other micro-founded monetary models with capital. \cite{aruoba2003search} is the one of first papers that introduced the neoclassical growth model to the \cite{lagos2005unified} environment. As pointed out by \cite{waller2003comment}, it features a strong neoclassical dichotomy, meaning the outcomes in the DM and the CM can be solved independently. Later \cite{aruoba2011money} and \cite{waller2011random} break this dichotomy by introducing the role of capital in the DM where capital accumulation lowers the cost of producing DM goods. The other way to break down the dichotomy is to introduce pledgeable capital, allowing more credit trade across agents by holding more capital. (e.g.,\citealp*{venkateswaran2013pledgability} and \citealp*{gu2019credit}) Here, what breaks down the neoclassical dichotomy is the limited commitment problem between firms and households.

\section{Quantitative Analysis}
\label{sec:quant}
The above section has developed a model of money creation analyzing monetary transmission. The model is tractable, and analytical results can be established. In this section, I calibrate the model and evaluate the model quantitatively. 

\subsection{Calibration}
The model period length is set to one year. 
The utility functions for the DM and the CM are $u(q)=B[(q+\varsigma)^{1-\varphi}-\varsigma^{1-\varphi}]/(1-\varphi)$ and $U(C)=\log(C)$. The cost function for the DM is $c(q)=q$. The production function takes the form of a standard Cobb-Douglas function, $F(K,N)=K^\alpha H^{1-\alpha}$. The enforcement cost for lending is assumed to be quadratic, $\eta(\tilde{\ell})=\Psi\tilde{\ell}^2$, and the balance sheet cost for managing reserves balances takes the form, $\gamma(\tilde{r})=G\tilde{r}^{g}$.

The calibration period is 1968–2007. The benchmark nominal interest rate is $i=0.0593$, the average 3-month treasury rate, and the benchmark required reserve ratio is $\chi=0.1111$, the average of the ratio between required reserves and total checkable deposits. Since the Federal Reserve did not pay interest on reserve before October 2008, I set $i_r=0$ as the benchmark.

\begin{figure}[tp!]
\centerfloat
\includegraphics[width=7cm,height=8cm]{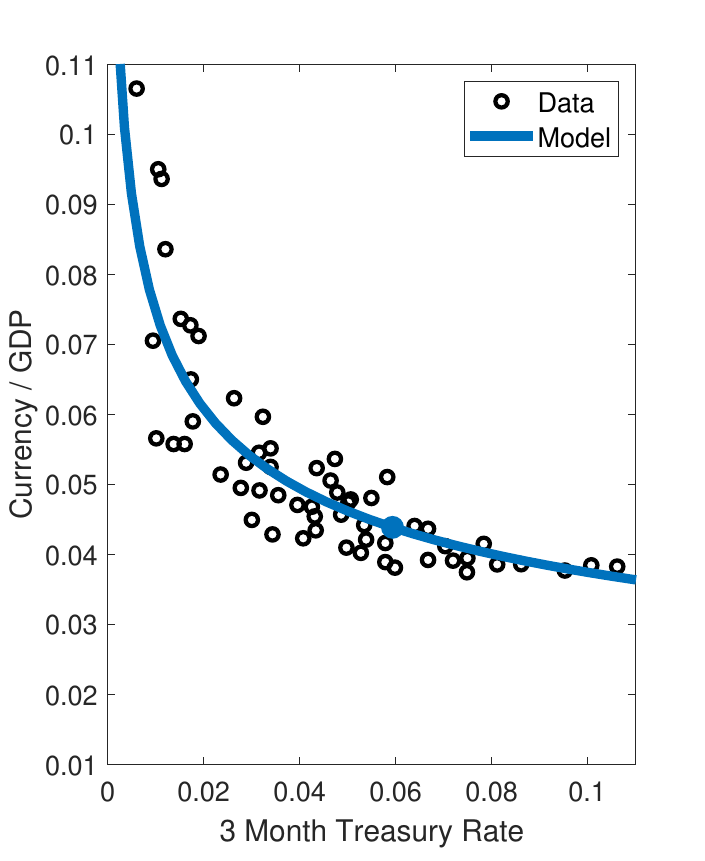}
\includegraphics[width=7cm,height=8cm]{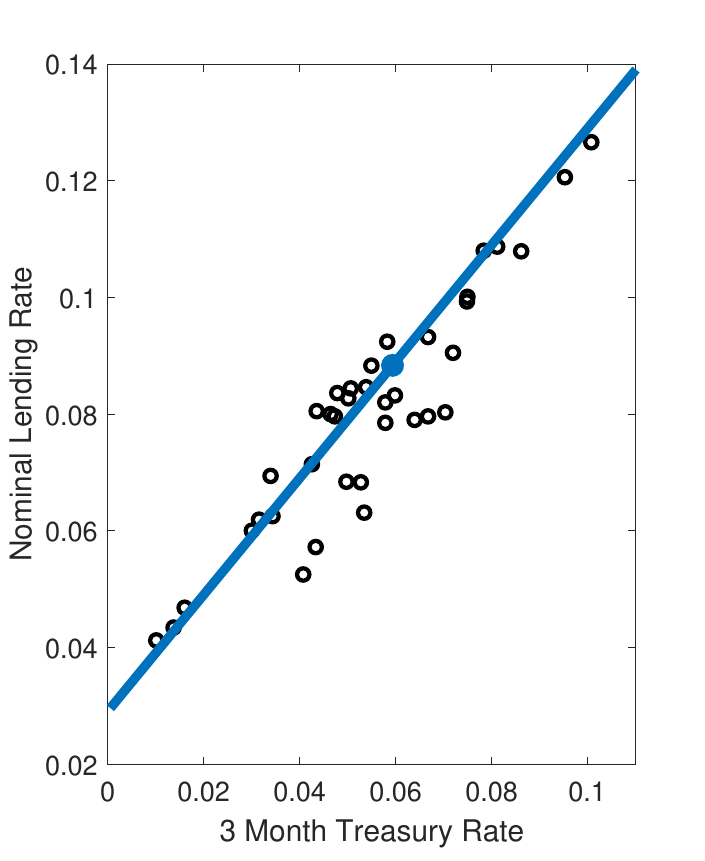}
\caption{Money Demand for Currency and Interest rate Pass-through} 
\label{fig:cd}
\end{figure}

Some parameters are directly pinned down. The discount factor $\beta$ is set to match a $3\%$ real interest rate. The capital share in CM output is set to $\alpha=1/3$ as the standard, and the capital depreciation rate is matched with $I/K=\delta=0.0825$. To ensure that changes in credit conditions do not affect currency holdings, $\sigma_3$ has been set to 0, based on the stable downward sloping currency demand illustrated in Figure \ref{fig:cd}. The fraction of DM2 meetings, $\sigma_2$, has been set to 0.689 so that the equilibrium percentage of unsecured credit users matches 68.9\%, which is the average percentage of US households holding at least one credit card from 1970 to 2007,  based on the Survey of Consumer Finances. For simplicity, the bargaining power has been set to $\theta=1$, which implies  the buyer makes a take-it-or-leave-it offer to the seller in the DM. The probability of being a buyer has been normalized to $\nu=0.5$. The parameter $\varsigma$ has been introduced in $u(q)$ merely to ensure that $u(0)=0$, and has been set to $\varsigma=0.0001$, as in \cite{aruoba2011money}. The curvature parameter of $\gamma(\cdot)$ has been set to $g=1.5$ to ensure that $\gamma(\cdot)$ is less convex than $\eta(\cdot)$.

\begin{table}[tp!]
\footnotesize 
\caption{Model parametrization}
\vspace{-0.2cm}
\centerfloat 
\begin{threeparttable}
\begin{tabular}{llllD{.}{.}{3}D{.}{.}{3}}
\toprule \hline
\textbf{Parameter}      &    \textbf{Description} &    \textbf{Value}  & \textbf{Target Description}  & \multicolumn{1}{r}{\textbf{Target}} & \multicolumn{1}{r}{\textbf{Model}}  \\ \hline
\multicolumn{6}{c}{\textbf{External Parameters}}      \\
$\delta$ &  depreciation rate   & 0.0825 &  investment/capital, $I/K$ &     &   \\
$\alpha$ &  capital share in $F$   & 0.3333 &  labor’s share of income, $2/3$&     &  \\
$\beta$  &  discount factor  & 0.9709 &   real interest rate, $3\%$  &    &   \\
$\sigma_2$ & DM2 matching prob.  & 0.6890 & share of credit meeting  &  &     \\
$\varsigma$  &  parameter of $u(\cdot)$   & 0.0001 & \cite{aruoba2011money}       &  &   \\ 
$\theta$  &  bargaining power   & 1&   take-it-or-leave-it offer   &  &   \\ 
$\nu$  &  prob. of being a buyer  & 0.5 &   normalization   &     &  \\
\multicolumn{6}{c}{\textbf{Internal Parameters}}      \\
$\zeta$ & coeff. on labor supply   & 2.4949 & labor supply, $H$       & 0.3333 & 0.3321     \\
$\sigma_1$ & DM1 matching prob.    & 0.0006 & currency/output, $M/PY$ & 0.0443 & 0.0438   \\
$G$ & parameter of $\gamma(\cdot)$ & 0.0018 & reserves/output, $R/PY$ & 0.0121 & 0.0117    \\
$\bar{b}$ & credit condition       & 0.0555 &  unsecured credit/output& 0.0347 & 0.0347   \\
$\kappa$ & bank entry cost         & 0.0122 & lending rate, $i_\ell$  & 0.0862 & 0.0883    \\
 $B$ & parameter of $u(\cdot)$     & 0.0156 & capital/output, $K/Y$   & 2.1896 & 2.1957   \\
$\varphi$ & parameter of $u(\cdot)$& 3.3145 &  elast. of $M/PY$ to $i$ & -0.1948 & -0.2893  \\
$\Psi$ & parameter of $\eta(\cdot)$& 0.0173 & semi-elast. of $i_\ell$ to $i$ & 11.3673 & 11.2972  \\
\hline \bottomrule
\end{tabular}
\end{threeparttable}
\label{tab:param}
\end{table}

The remaining 8 parameters $(\zeta,\kappa,\sigma_1, \bar{b}, G, B,\varphi,\Psi)$ are set to match the following 8 targets: (i) the standard measure of work as a fraction of discretionary time, $H=1/3$; (ii) the average nominal lending rate, $i_\ell=0.0862$; (iii) the currency output ratio, $M/PY=0.0443$; (iv) the reserves output ratio, $R/PY=0.0121$; (v) the unsecured credit output ratio, $0.0347$; (vi) the capital output ratio, $K/Y=2.1896$; (vii) the elasticity of currency demand to the nominal interest rate, $-0.1948$; (viii) the semi-elasticity of the nominal lending rate to the nominal interest rate, $11.3673$. The targets are computed based on 1968-2007 data.

All of the targets in the model, except the elasticity of currency demand and the semi-elasticity of lending rate, are  directly computed using straightforward formulas given the benchmark nominal interest rate and the required reserve ratio. Similar to  \cite{aruoba2011money}, the elasticity of currency demand is computed using changes in money demand when the interest rate changes from $i-0.05$ to $i+0.05$. The semi-elasticity of lending rate is also  computed using changes in the nominal lending rate when the interest rate changes from $i-0.05$ to $i+0.05$. The calibrated parameters and the targets are summarized in Table \ref{tab:param}, and the calibrated money demand of currency and lending rate pass-through are shown in Figure \ref{fig:cd}. 

\subsection{Results}
\label{sec:quan_result}

\begin{figure}[tp!]
\begin{subfigure}[b]{1\textwidth}
\centerfloat
\includegraphics[width=4.8cm,height=4.8cm]{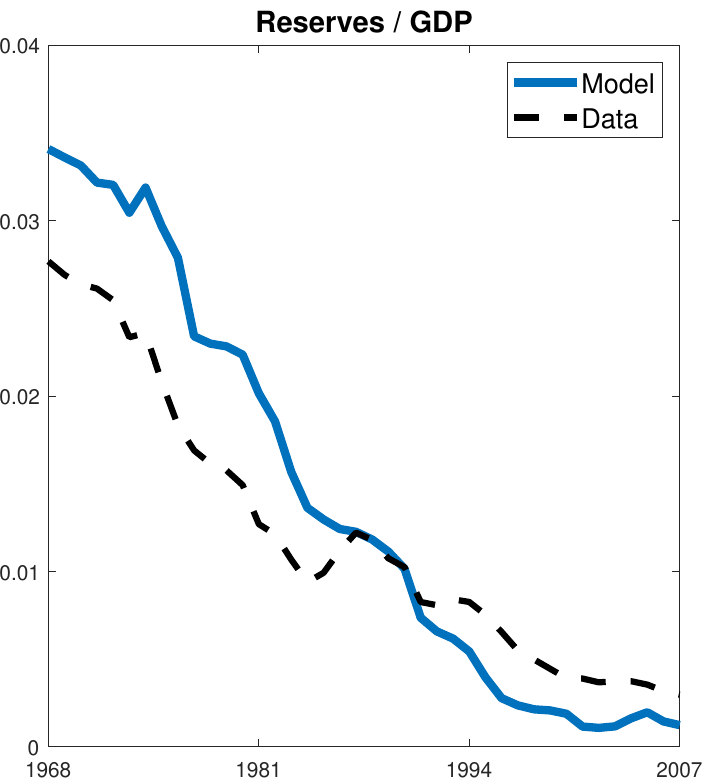}
\includegraphics[width=4.8cm,height=4.8cm]{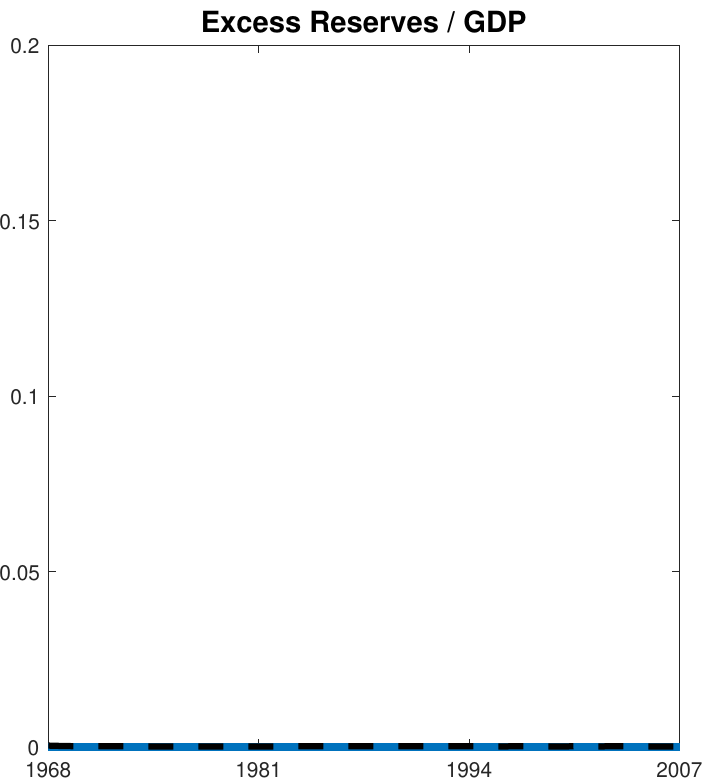}
\includegraphics[width=4.8cm,height=4.8cm]{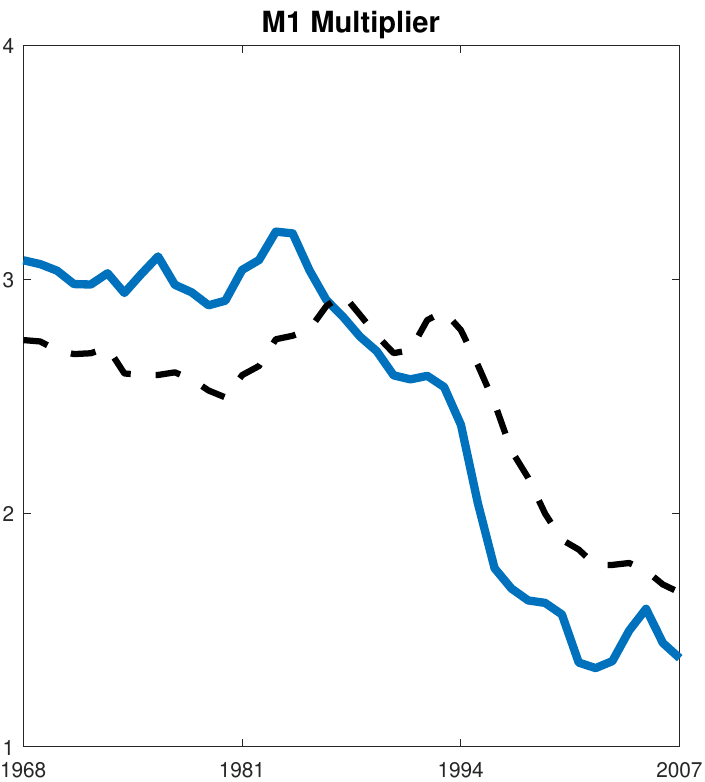}
\end{subfigure}
\begin{subfigure}[b]{1\textwidth}
\centerfloat
\includegraphics[width=4.8cm,height=4.8cm]{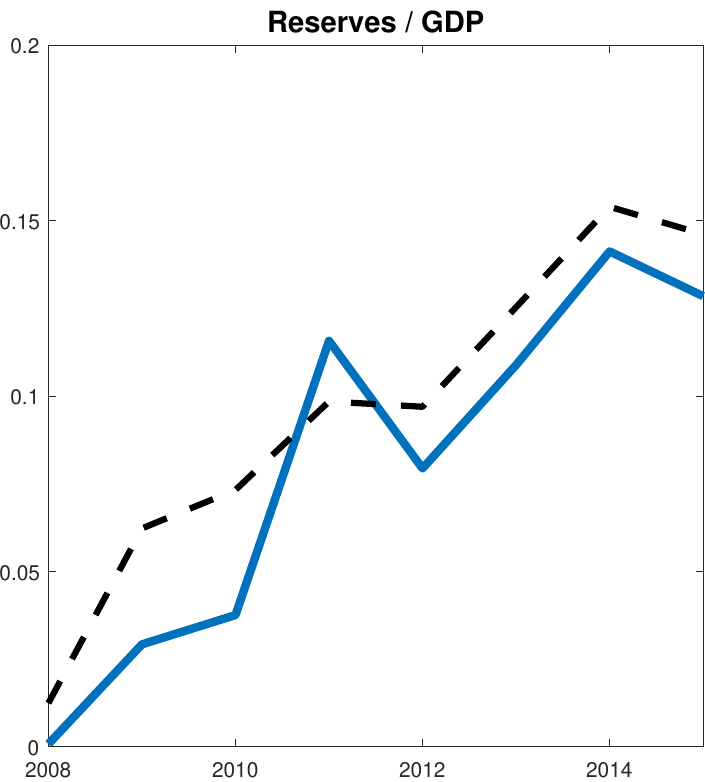}
\includegraphics[width=4.8cm,height=4.8cm]{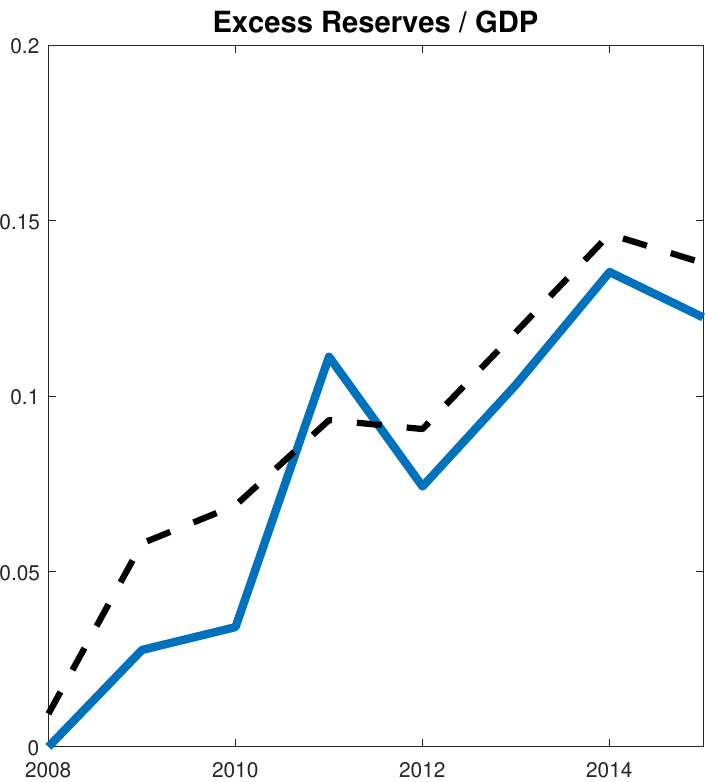}
\includegraphics[width=4.8cm,height=4.8cm]{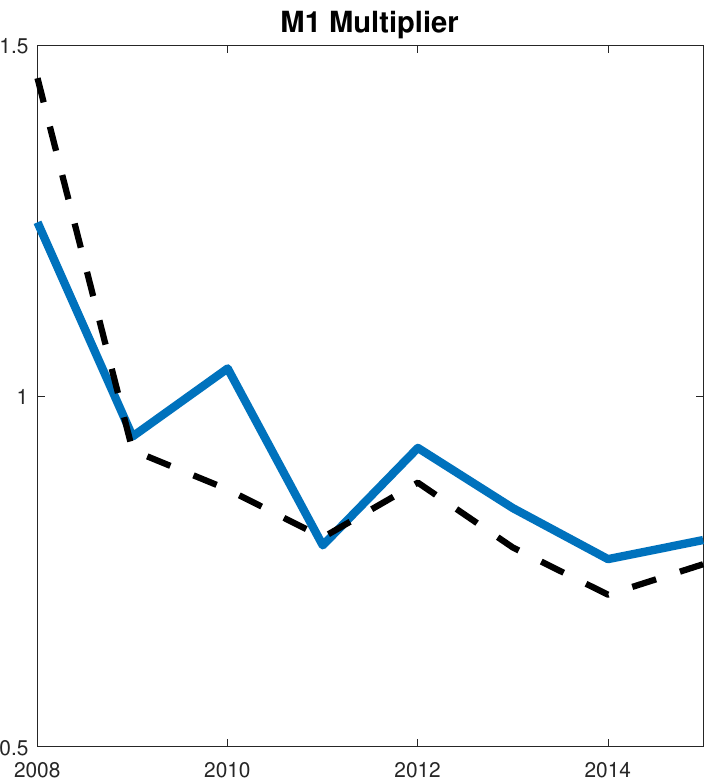}
\end{subfigure}
\caption{Model vs. Data} 
\label{fig:model_fit1}
\end{figure}

This section explores how well the model can account for the low-frequency behavior of reserves and the money creation process, assuming the only driving forces are monetary policy, $(i,i_r,\chi)$ and credit conditions, $\bar{b}$. The unsecured credit limit $\bar{b}$ is computed using the unsecured credit to output ratio. Using the calibrated parameters, I compute the model equilibrium for given  $(i,i_r,\chi,\bar{b})$. For monetary policy variables, I use 3 month treasury rate and interest on excess reserves,\footnote{Whereas the Federal Reserve had announced interest on excess reserves and interest on required reserves separately, the Fed had paid the same interest to both. As of March 2020, the Fed unified two interests on reserves as ``Interest Rate on Reserve Balances" since the Board reduced reserve requirement to zero.} and required reserves to total checkable deposit ratio.  \\ 

\noindent\textbf{Reserves and Money Multiplier } Figure \ref{fig:model_fit1} compares the model and data from 1968 to 2015. The top-left panel and the bottom-left panel shows reserves as a fraction of output from 1968 to 2007 and from 2008 to 2015, respectively. The model generates the quantity of reserves which can match the movement in the data. The top-middle panel and the bottom-middle panel of Figure shows excess reserves as fraction of output from 1968 to 2007 and from 2008 to 2015, respectively. The model also successfully generate the zero excess reserves during 1968-2007 and huge increase during 2008-2015. The top-right panel and the bottom-right panel of Figure \ref{fig:model_fit1} shows M1 money multiplier. While the model does not match all of the movement in the data, there is a very similar basic pattern. During 1968-2015 money multiplier consistently decreased, whereas there is no excess reserves. During 2008-2015, money multiplier decreased lower than one and reflected the behavior of excess reserves.  \\

\noindent\textbf{Matching Observations } We now examine the model’s ability to reproduce the observed patterns discussed in Section 2.

\begin{figure}[tp!]
\centerfloat
\includegraphics[width=7.5cm,height=5.6cm]{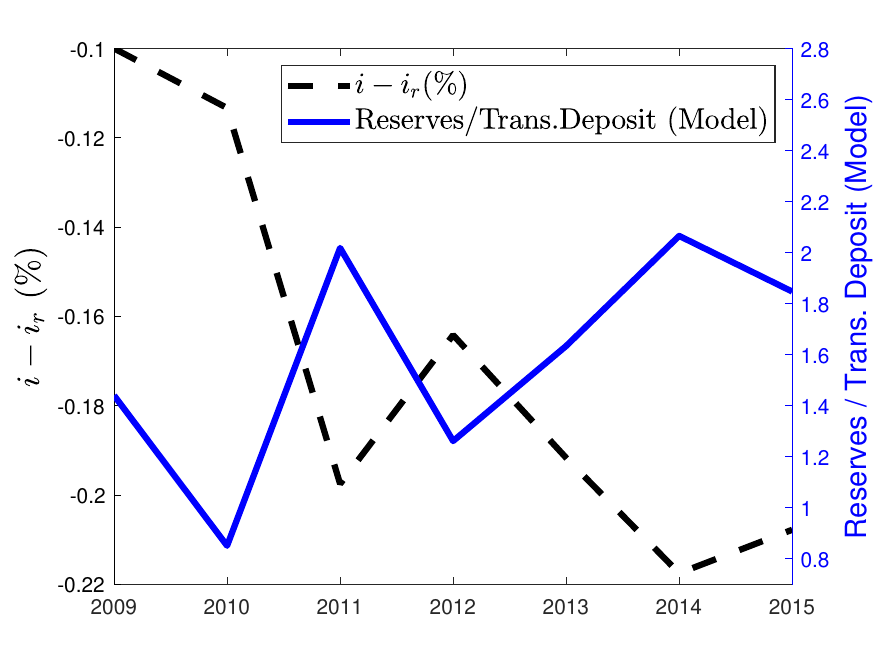}
\includegraphics[width=7.5cm,height=5.6cm]{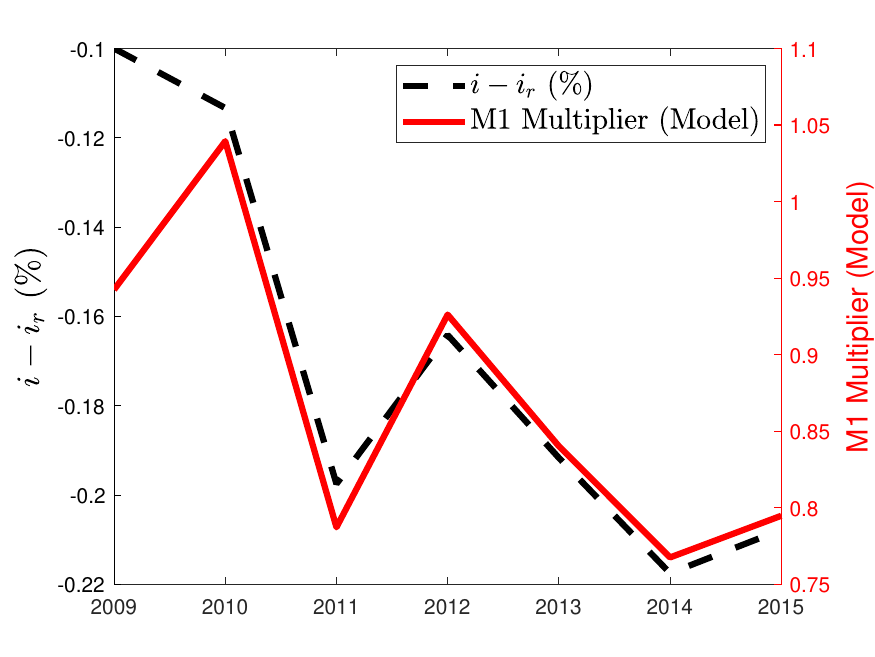}
\caption{Reserves Deposit ratio and M1 multiplier: Model}
\label{fig:mo1}
\end{figure}

Observation 1 shows that the quantity of reserves is not independent of the short-term policy rate. 
Figure \ref{fig:mo1} displays the counterfactual data from the model. As in Figure \ref{fig:motive4}, Figure \ref{fig:mo1} show the evident opposite movements of the reserves to deposit ratio with respect to the spread between short-term policy and interest on reserves. Also, the counterfactual M1 money multiplier moves together with the spread as in Figure \ref{fig:motive4}. As already shown in Proposition \ref{prop:compastatics1}, a higher interest on reserves increases the incentive hold to reserves because the banks earn interest just by holding reserves. However, banks do not create transaction deposits proportional to the increases in reserves which eventually lowers the money multiplier. In contrast, a higher short-term policy rate lowers reserve balances by disincentivizing the bank to hold reserves which eventually increases money multipliers. 

\begin{figure}[tp!]
\centerfloat
\includegraphics[width=12cm,height=5cm]{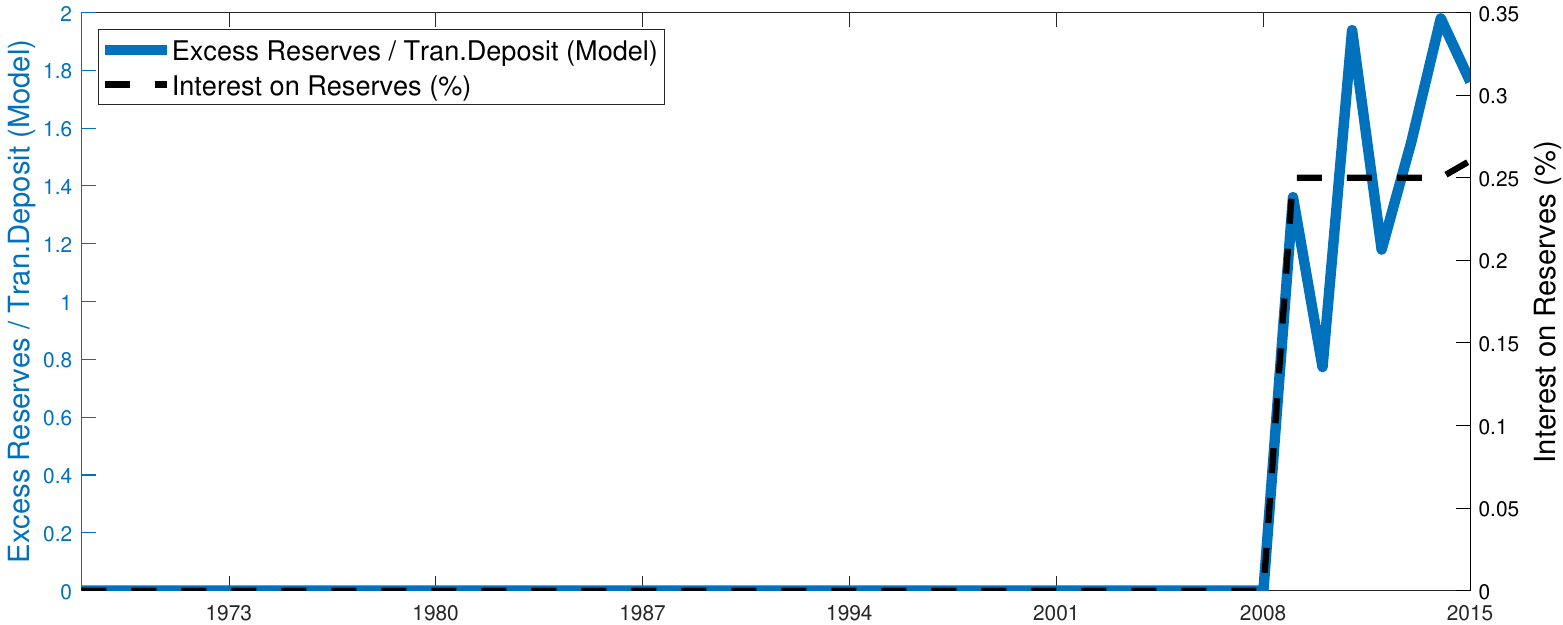}
\caption{Excess reserves ratio (Model)}
\label{fig:excess_ratio_ms}
\end{figure}

Observation 2 shows that the excess reserves deposit ratio remained at zero until the introduction of interest on reserves, after which they skyrocketed. Similar to Figure \ref{fig:excess_ratio}, Figure \ref{fig:excess_ratio_ms} shows the excess reserves deposit ratio remains zero before the introduction of interest on reserves as well as a huge increase in excess reserves. Paying interest reserves gives more incentive to hold reserves, resulting in banks holding more reserves than the required level.

\begin{figure}[tp]
\centerfloat
\includegraphics[width=7cm,height=5.5cm]{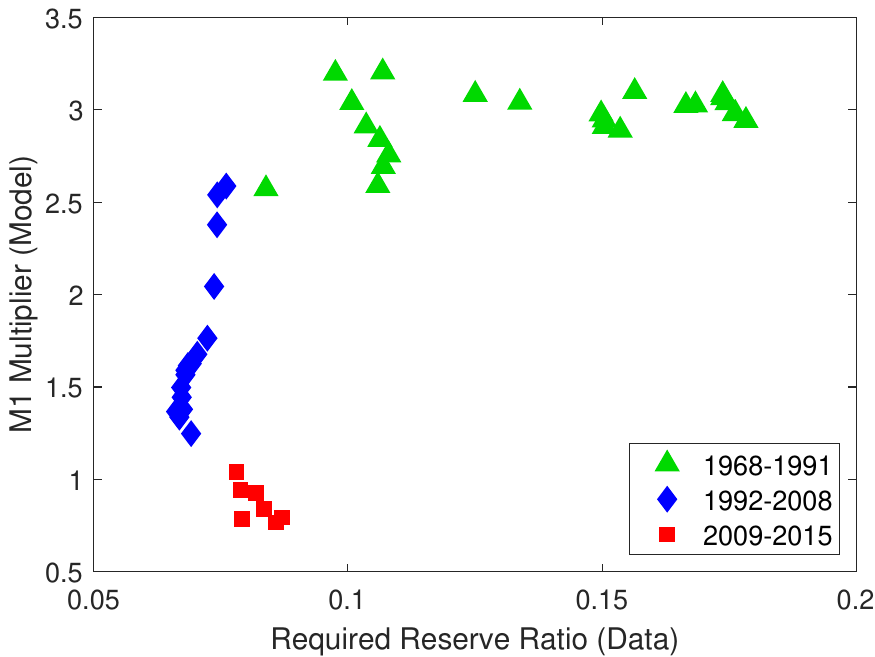}
\includegraphics[width=7cm,height=5.5cm]{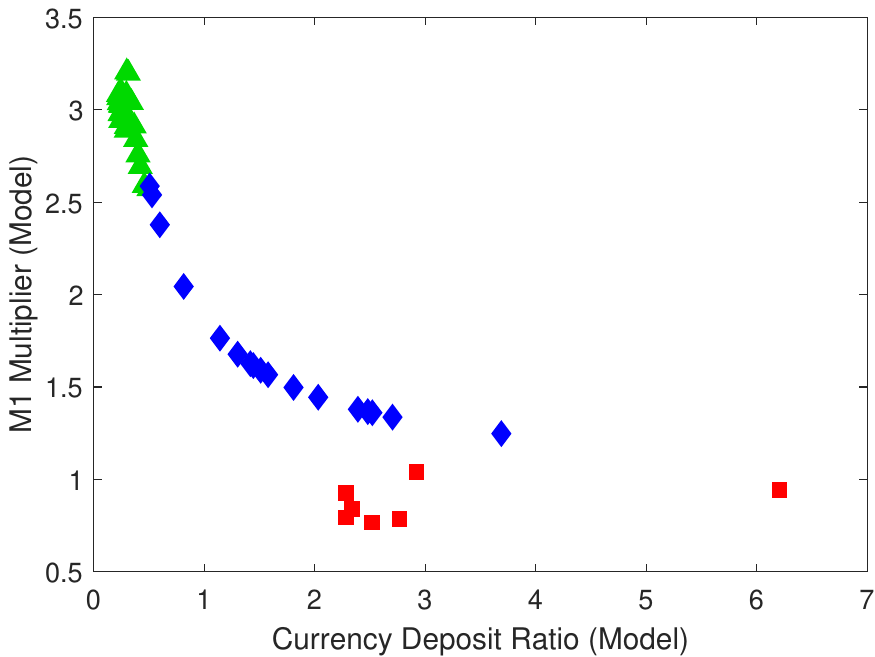}
\caption{Money multiplier, currency/deposit ratio and required reserve ratio: Model}
\label{fig:mo3}
\end{figure}

Observation 3 shows that there is no negative relationship between the required reserve ratio and the money multiplier, and it also shows two structural breaks, one in 1992 and another in 2008. As shown in Figure \ref{fig:multip1}, there has been a consistent decrease in the money multiplier since 1992, which was not accompanied by a decrease in the required reserve ratio but rather by an increase in the currency deposit ratio. Notably, the decrease in the money multiplier after 2008 was not accompanied by an increase in the currency deposit ratio. In Figure \ref{fig:mo3}, the model successfully reproduces this observation, showing a persistent decrease in the money multiplier since 1992 that was not accompanied by a decrease in the required reserve ratio but rather by an increase in the currency deposit ratio. Additionally, the model also reproduces a decrease in the money multiplier did not accompany an increase in the currency deposit ratio after 2008.

\begin{table}[tp!]
\caption{Money demand and Credit: Model vs. Data}
\label{tab:reg_compare}
\footnotesize
\centering
\begin{threeparttable}
\begin{tabular}{lD{.}{.}{5}cD{.}{.}{5}ccD{.}{.}{5}D{.}{.}{5}} 
\toprule \hline
Dependent Variable: $ln(m_t)$ & 
\multicolumn{4}{c}{OLS}& & \multicolumn{2}{c}{CCR}  \\ 
\cline{2-5}  \cline{7-8} 
&\multicolumn{1}{c}{Data}&\multicolumn{1}{c}{Model}&\multicolumn{1}{c}{Data}&\multicolumn{1}{c}{Model}  & &\multicolumn{1}{c}{Data}&  \multicolumn{1}{c}{Model} \\ 
&\multicolumn{1}{c}{(1)}&\multicolumn{1}{c}{(2)}&\multicolumn{1}{c}{(3)}&\multicolumn{1}{c}{(4)}&  & \multicolumn{1}{c}{(5)}       & \multicolumn{1}{c}{(6)}    \\ \hline
$r_t$     & 1.600^{***} & 10.462 & -2.298^{***} & -1.807 & &   -2.755^{**} & -1.635  \\ 
          & (0.432)     &        & (0.740)      &        & &  (1.171)      &        \\
$ln(uc_t)$&             &        & -0.322^{***} & -1.014 & &  -0.282^{***} & -1.005  \\
          &             &        & (0.056)      &        & &   (0.098)    &       \\ \hline
$adj R^2$ &  0.109      &  0.567 & 0.416 & 0.919& &  0.981    &    0.948    \\   
\hline \bottomrule 
\end{tabular}
\end{threeparttable}
\begin{minipage}{0.85\textwidth} 
{\scriptsize  Notes: Columns (1) and (3) report OLS estimates and columns (2) and (4) report the canonical cointegrating regression (CCR) estimates.  First-stage long-run variance estimation for CCR is based on Bartlett kernel and lag 1.  For (1) and (2) Newey-West standard errors with lag 1 are reported in parentheses. Intercepts are included but not reported.\par}
\end{minipage}
\label{tab:mo3}
\end{table}

Observation 4 shows that consideration of unsecured credit recovers downward-sloping M1 money demand. The model can reproduce this result as well. Using the counterfactual data generated from the model, Table \ref{tab:reg_compare} shows that, in the model, unsecured credit consideration recovers the downward-sloping money demand, as in the data.\footnote{To compare the result from Table \ref{tab:motive_reg}, the model parts of Table \ref{tab:reg_compare} use annualized quarterly series instead of annual data. } Column (2) shows that the regressing only on interest rate gives positive estimates. The columns (4) and (6) show that adding unsecured credit to output ratio to the money demand equation recovers downward-sloping money demand, as in the data.   \\

\begin{figure}[tp!]
\centerfloat
\includegraphics[width=7.3cm,height=4.9cm]{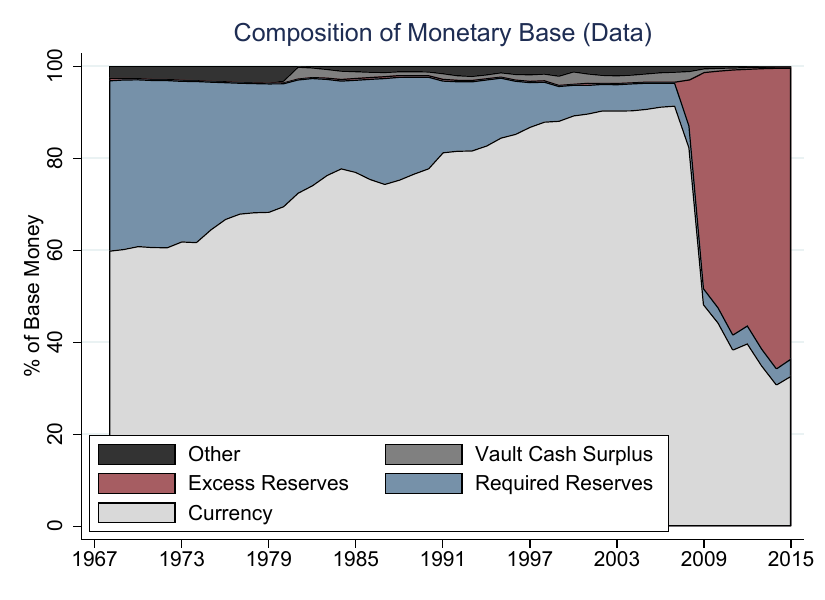}
\includegraphics[width=7.3cm,height=4.9cm]{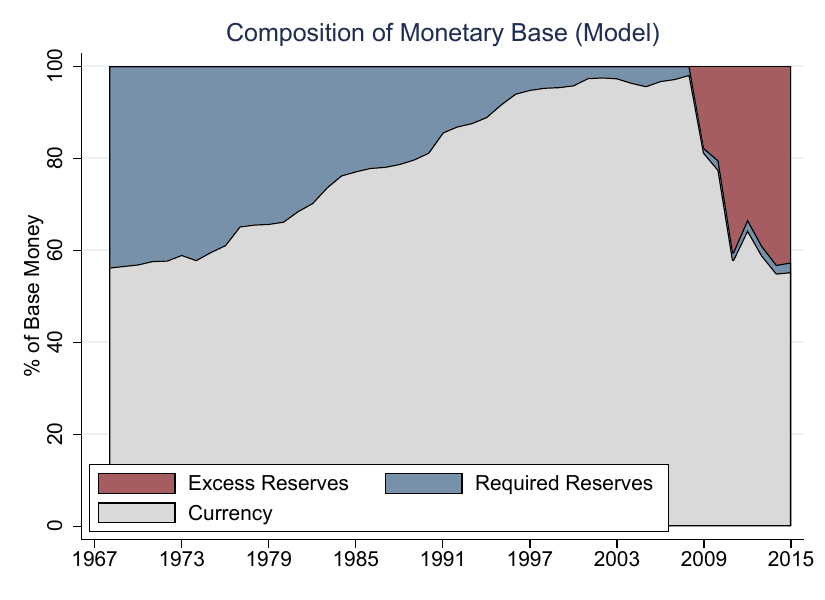}
\caption{Composition of monetary base: data vs. model\protect\footnotemark}
\label{fig:portf}
\end{figure} 
\footnotetext{The monetary base and currency are sourced from `H.6 Money Stock Measures', published by the Federal Reserve. These data can be found in the Federal Reserve Economic Data (FRED) database under the series BOGMBASE and CURRSL. Additionally, data on required reserves, excess reserves, and vault cash surplus are obtained from `H.3 Aggregate Reserves of Depository Institutions and the Monetary Base', also published by the Federal Reserve. These figures are compiled in the FRED under the series EXCRESNS, EXCSRESNS REQRESNS, and VAULTSUR. The category `Other' represents the monetary base that does not fall into any of these specified categories. }

\noindent\textbf{Composition of Monetary Base }  The model also generates the composition of the monetary base over time. Figure \ref{fig:portf} compares the composition of the monetary base between the data and the model. The model successfully captures the changes in each component of the monetary base - currency, required reserves, and excess reserves - both before and after 2008. The currency portion consistently increased from 1968 to 2007 and then drastically decreased as the portion of excess reserves significantly increased. The portion of required reserves has consistently decreased. It is worth emphasizing that the share of currency in the monetary base has been substantial. Until 2008, its share had increased from 60\% to 90\%. After 2008, its share has hovered around 40\%. Even though the share of currency has been reduced, it still accounts for a large portion of the central bank's balance sheet. This implies that if one wants to consider the central bank's balance sheet as an important channel of monetary policy, incorporating the currency in the analysis might be necessary.

\subsection{Reserve Demand and Interest on Reserves}

\begin{figure}[tp!]
\centerfloat 
    \includegraphics[width=7cm,height=6cm]{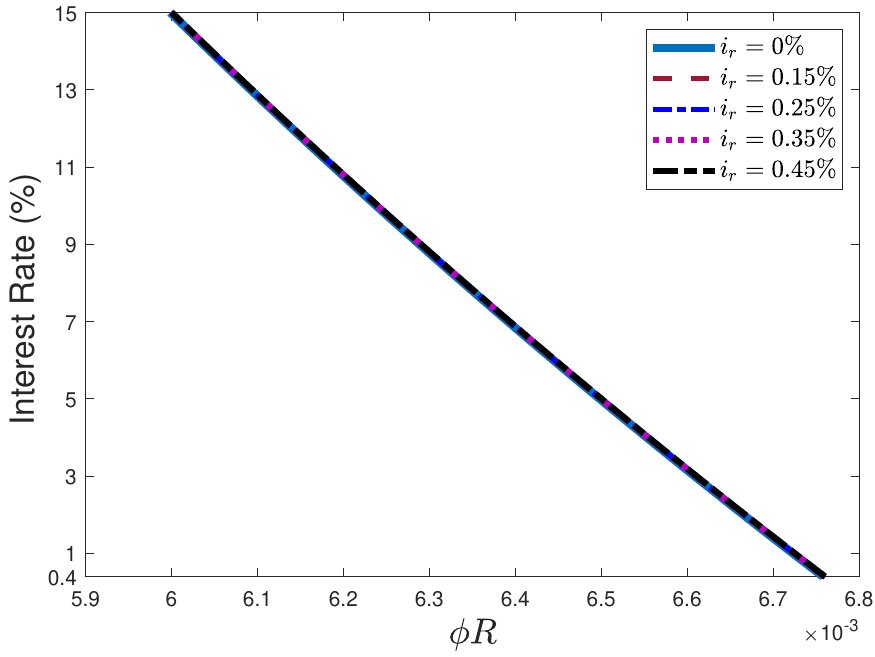}
    \includegraphics[width=7cm,height=6cm]{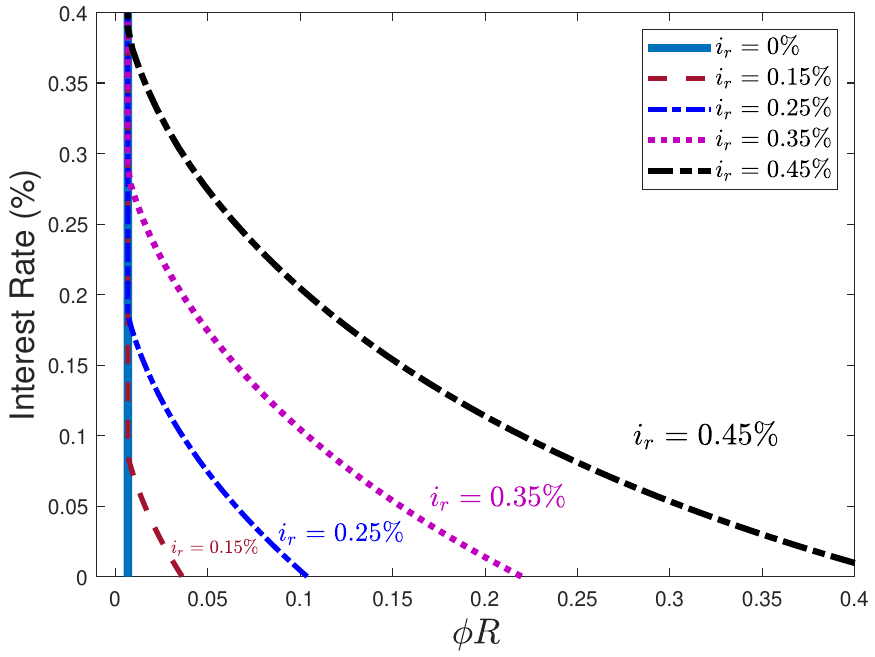}  
    \caption{Reserves Demand under different Interest on Reserves} \label{fig:rd}
\end{figure} 
This section reviews the reserve demand from the model under calibrated parameters to illustrate the underlying mechanism of some quantitative results. Figure \ref{fig:rd} plots reserve demand under different interest rates on reserves. For readability, I divide the plot of reserve demand by interest rate into two ranges: 0.4\% to 15\% and 0\% to 0.4\%, as the reserve demands exhibit drastic changes with respect to interest on reserves.

The left panel of Figure \ref{fig:rd} plots reserve demand for interest rates ranging from 0.4\% to 15\%, while the right panel shows the reserve demand for interest rates ranging from 0\% to 0.4\%.
The left panel demonstrates that interest on reserves virtually does not change the reserve demand under moderate or high $i$. In contrast, the right panel reveals that when the central bank pays interest on reserves, the lower part of the reserve demand curve becomes flatter, but not completely flat. This flattening occurs because the corresponding equilibrium is the ample reserve equilibrium. In the ample reserve regime, banks hold reserves not because of regulatory requirements but because it is profitable to do so.

As both $i$ and $i_r$ are more closely linked to the banks' profit through the reserve balances in the ample-reserve regime, the banks' reserve demand becomes more elastic with respect to changes in $i$ and $i_r$. The right panel of Figure \ref{fig:rd} illustrates that in the low interest rate region, given the same $i$, the reserve demand increases drastically as interest on reserves rises. Thus, paying interest on reserves and lowering the short-term policy rate can lead the economy to an ample reserve regime. This channel drives a huge increase in excess reserves, as shown in Figures \ref{fig:excess_ratio} and \ref{fig:excess_ratio_ms}.

The reserve demand in Figure \ref{fig:rd} also provides a clear explanation for the opposite movements of the reserves to deposit ratio with respect to the spread between the short-term policy rate and interest on reserves, as presented in Figures \ref{fig:motive4} and \ref{fig:mo1}. When the economy is in the ample reserve regime, the quantity of reserves decreases as the short-term policy rate rises and increases when interest on reserves goes up. A reduction in the short-term policy rate leads to an increase in reserves; however, banks do not create deposit money in proportion to this increase, resulting in a lower money multiplier. Similarly, a higher interest rate on reserves reduces the money multiplier, as banks are more motivated to hold reserves and less inclined to create deposit money. Interest on reserves and the short-term rate play distinct roles, and they jointly determine the quantity of reserves.

\subsection{Implication for Monetary Transmission }
In the model economy, the central bank sets interest rates by controlling the monetary base. This is consistent with the conventional notion of monetary policy. Recall \cite{romer2000keynesian}: 
\begingroup\begin{quote} 
[The] appropriate concept of money is unambiguously high-powered money. Here $M$ is not a variable the central bank is targeting, but rather one it is manipulating to make interest rates behave in the way it desires. This is an excellent description of high-powered money. Moreover, for high-powered money, the assumption that the opportunity cost of holding money is the nominal rate is appropriate. In addition, the assumption that the central bank can control the money stock is a much better approximation for high-powered money than for broader measures of the money stock. 
\end{quote}\endgroup

However, the recent literature does not seem to align in this understanding of monetary policy. Many macroeconomic models got rid of this monetary transmission, especially after the Fed introduced the interest on reserves. For example, \cite{goodfriend2011central} put it as “Interest on reserves frees monetary policy to fund credit policy independently of interest rate
policy.” and \cite{cochrane2014monetary} put as “[The] Fed can separate interest rate changes ... from balance sheet policy” \cite{cochrane2014monetary} goes even further by putting “Federal Reserve policy in the future goes so far past “monetary" that the label will no longer be appropriate.” The result in this paper is the opposite. This paper shows that paying interest on reserves does not make the quantity of reserves independent of interest rate policy. 

The quantitative analysis shows that, given credit conditions, there exists a one-to-one mapping from a set of monetary policy variables to the quantity of currency, the quantity of reserves, the quantity of excess reserves, and the money multiplier. In this approach, one does not need to rely on the assumption that the central bank directly controls monetary aggregates or the even more abstract assumption that the central bank controls interest rates out of nothing. Instead, the model gives an explicit process of monetary policy from setting short-term policy rates by controlling the monetary base to its influence on the monetary aggregate and other macroeconomic variables. 

It not only provides an explicit monetary transmission mechanism but also gives different views on monetary policy. For example, unconventional monetary policies, such as quantitative easing, are not asset purchases independent of the target level of the short-term policy rate. Rather, they can be viewed as the uses of two different interest rates: interest on reserves and short-term policy rates. Changes in reserves reflect the changes in short-term policy rates and vice versa, implying we also can interpret them as the uses of interest on reserves and quantity of monetary base (or reserves) as well.\footnote{It is worth noting that the Fed currently supplies and absorbs reserves using standing facilities and their administered rates, such as the overnight reverse repo (repurchase agreement), standing repo facility, and discount windows and their rates. The Fed adopted this approach to control the target interest rate because the previous system was operationally more difficult than the current approach as it ``requires forecasting the many exogenous factors that affect the amount of bank reserves outstanding, and then engaging in open market operations on a near-daily basis to keep reserves at a level consistent with the FOMC's target range" \citep{dudley2018important}.} The model does not require the assumption of independence between equilibrium prices (short-term policy rates) and quantities (monetary base or reserves).

\section{Concluding Remarks}
\label{sec:conclusion}

This paper develops a monetary-search model with fractional reserve banking and unsecured credit, and studies the role of money creation in monetary transmission. In the fractional reserve banking system, money is created when banks make loans. The bank's inside money creation, however, can be constrained by the reserve requirement and the reserves. Whether the reserve requirement constraint binds or not is endogenously determined by the banks' profit maximization. 

Banks hold excess reserves when the central bank pays sufficiently high interest on reserves. In this case, the money multiplier and the quantity of the reserve depend on the short-term policy rate and the interest on reserves rather than the reserve requirement. In contrast to previous works, these two interest rates play distinct roles, and the quantity of reserves is not independent of interest rate management. Furthermore, their impact on the lending rate are different: the lending rate increases with the short-term interest rate, while it decreases with interest on reserves. Quantitative analysis can generate simulated data that resemble the actual data. This paper provides evidence from the model and the data that suggests that the dramatic changes in the money multiplier after 2008 are mainly driven by the introduction of the interest on reserves. 

This work can be extended in various ways. Although I focus on the centralized market for the reserves with homogeneous banks, in reality, the market for reserves is a decentralized interbank market and banks have different portfolios. Therefore, one can further investigate how much the market structure and heterogeneity matter for the transmission of monetary policy (e.g., \citealp*{afonso2015trade}; \citealp*{armenter2017excess}; \citealp*{afonso2019model}). Second, I assume that bank assets are composed of loans and reserves. But commercial banks' assets are composed of securities, loans, and reserves. Extending the model to incorporate banks' portfolio choices and analyzing the role of investment, financial regulation, and monetary policy can open up other
research avenues. (e.g., \citealp*{rocheteau2018corporate}).


\newpage

\bibliographystyle{aer} 
\bibliography{bibi}

\clearpage
\begin{center}
\textsc{\LARGE{Appendix For Online Publication}}
\end{center}
\spacing{1.15}

\begin{appendices}
\section{Interest on reserves and floor system?}
\label{sec:floor}

\begin{figure}[h!]
\centerfloat
\includegraphics[width=9cm,height=5cm]{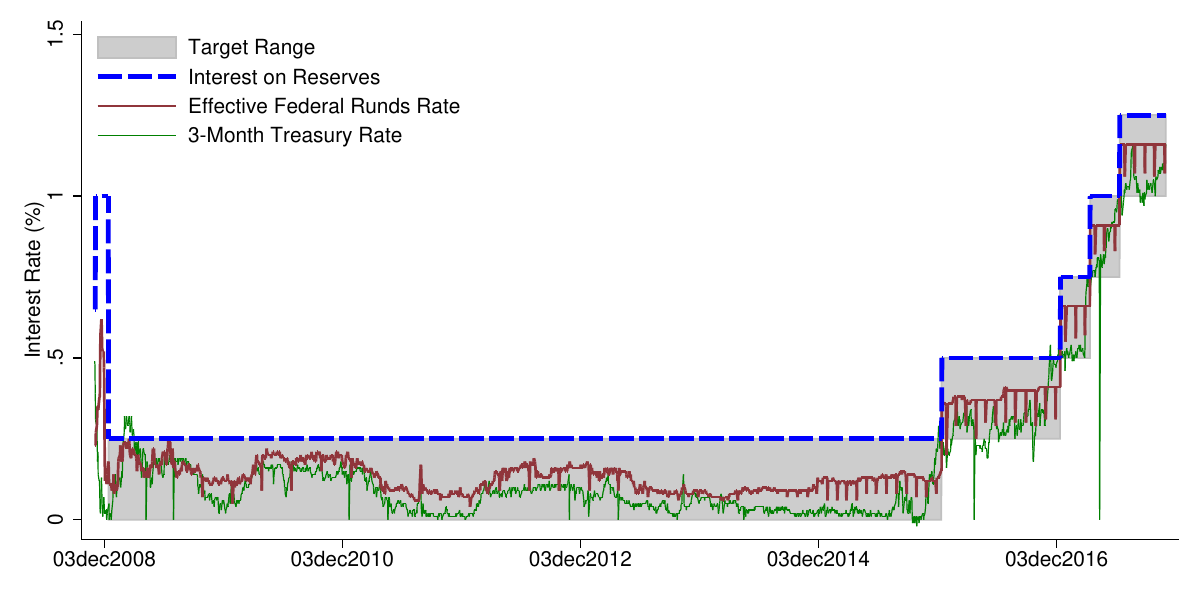}
\includegraphics[width=8cm,height=5cm]{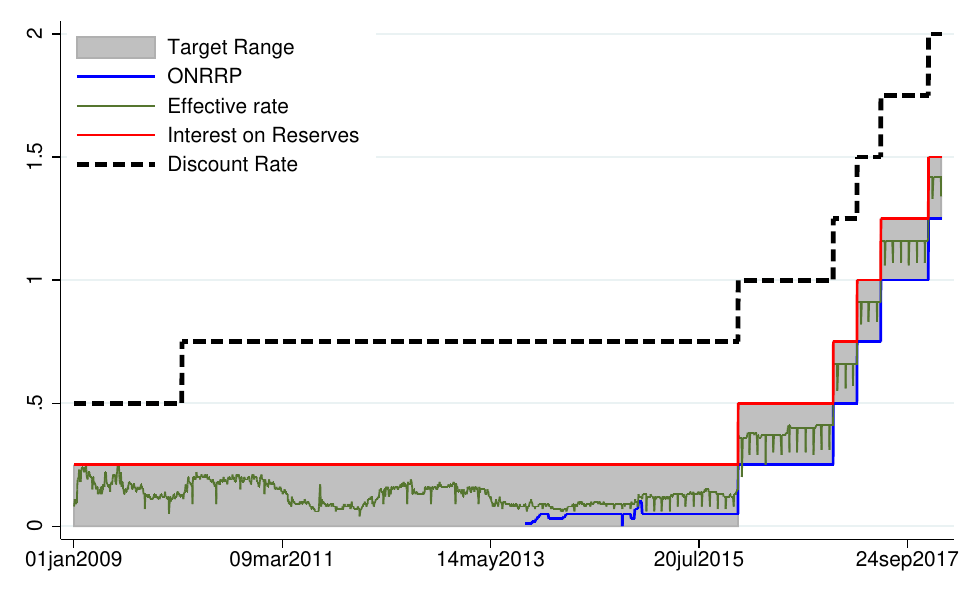}
\caption{Interest on reserves, ON RRP and target range}\label{fig:floor}
\end{figure}

This section discusses details about the interest on reserves and the floor system. The Federal Reserve started paying interest on reserves (IOR) in Oct 2008. In contrast to the idea that paying interest on reserves provides a floor, Figure \ref{fig:floor} shows that interest on reserves has been equal to the upper bound of the Fed's target range. Instead, the interest rate of the Overnight Reverse Repurchase (ON RRP) has been equal to the lower bound of the target rate. To put it simply, the interest on reserves serves as the upper bound of the target range, while the ON RRP rate acts as the lower bound of the target range for most of the period. The short-term policy rate is tightly controlled within this target range.

Given that the federal funds market is a market for unsecured bilateral interbank lending of reserves, a lower federal funds rate than the IOR may seem counterintuitive. This discrepancy is due to the institutional framework. While both depository institutions (DIs) and government-sponsored enterprises (GSEs) can trade in the federal funds market, only DIs are eligible to earn IOR, not GSEs. The GSEs, such as the Federal Home Loan Banks, can earn arbitrage profits by borrowing from the repo market and lending to DIs in the federal funds market. As DIs can earn interest on reserves and the GSEs earn the arbitrage profit, the federal funds rates are usually determined as below IOR. 

\cite{armenter2017excess} present a model of interbank trade in which GSEs' lending to DIs that earn IOR determines the federal funds rates. Since GSEs also participate in the repo market, the ON RRP facility's interest rate serves as an alternative for financial institutions not eligible for the IOR, establishing a lower bound on the federal funds rate.
See \cite{afonso2013sb} and \cite{afonso2013sl} for more discussion on the trade in the federal funds market. 


\section{Retail Sweep}
\label{sec:sweep}
One may think that the structural break in 1992 found in Section \ref{sec:motive} can be attributed to the relaxation of bank deposit regulation in the 1990s that stimulated financial innovations such as retail sweep accounts in the 1990s (e.g., \citealp*{vanhoose2001sweep}, \citealp*{teles2005stable}, \citealp*{lucas2015stability},
\citealp*{berentsen2015financial}). Base on this idea, some previous works have used  an alternative measure of M1 as monetary aggregates, ``M1 Adjusted for Retail Sweeps" (M1S, hereafter).

The rationale for using M1S is that the introduction of the automatic transfer system (ATS) in the early 1990s made highly liquid transaction balances outside M1. The ATS enables convenient transfer from money market deposit account (MMDA) to sweep account. The sweep account is a transaction deposit, whereas the MMDA is a saving deposit that is classified as Non-M1 M2. While the MMDA pays higher deposit rates than other saving deposits, same as other saving deposits, the MMDA had been subject to the restriction on the number of convenient transactions due to Regulation D.\footnote{The MMDA was introduced in the early 1980s after the US Congress permitted its creation as of December 1982. Due to Regulation D, no more than six convenient transactions using the MMDA could be made per statement period. As of March 2020, the Federal Reserve removed this restriction.} Thus, the introduction of ATS may enable people to use MMDA as a liquid deposit. The claim is that this may result in the appearance of  highly liquid transaction balances within instruments outside M1. \cite{cynamon2006redefining} discuss the benefits of using M1S instead of M1 as a measure of money supply.

\begin{figure}[h!]
\centerfloat
\includegraphics[width=11cm,height=5cm]{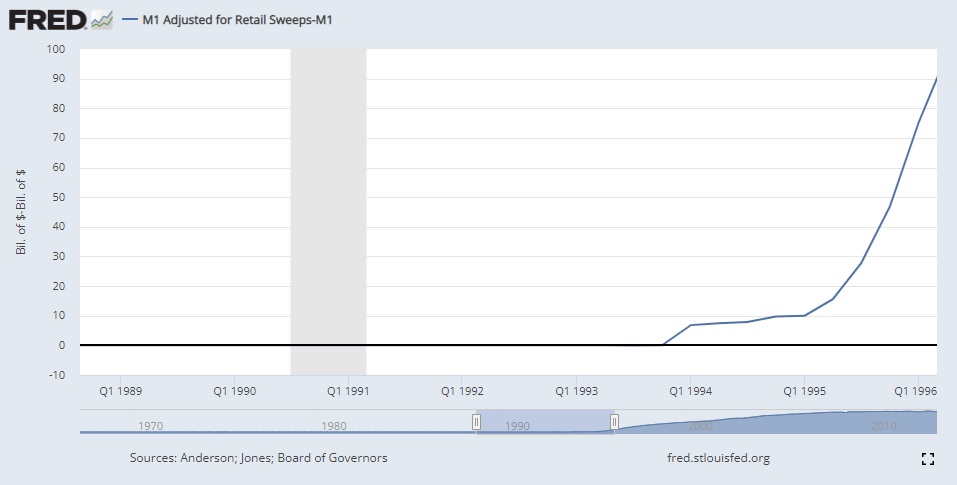}
\caption{Difference between M1 adjusted for retail sweep and M1}\label{fig:m1sweep}
\end{figure}

However, it is worth noting that the ATS was introduced to commercial banks in 1994 while the structural break of M1 from Section \ref{sec:motive} occurred in 1992. Consistent with this, Figure \ref{fig:m1sweep} shows that the difference between M1S and M1 was zero before 1994. Using M1S, \cite{berentsen2015financial} and \cite{kejriwal2022two} also found the structural break of M1 in the early 1990s, suggesting retail sweep consideration cannot explain the break in the early 1990s.

\newpage

\section{Proofs}
\label{sec:proof}
\onehalfspacing
\begin{proof}[Proof of Proposition \ref{prop:pre1}]
Consider the case where the reserve requirement constraint is not binding. Recall equation (\ref{eq:bankfoc_amp2}) and $i_t=i_{s,t+1}$:
$$i_{r,t+1}-i_t = \gamma'(\Tilde{r}_{t+1})$$ 
Given $i_{r,t+1}=0$ and $i_t\geq0$, solving (\ref{eq:bankfoc_amp2}) yields $\Tilde{r}_{t+1}\leq0$ which could not be a equilibrium. Therefore, when the central bank does not pay interest on reserves $i_{r,t+1}=0$ there is no equilibrium with excess reserves $\tilde{r}>\chi\tilde{d}\geq0$.

Next, we move on to the comparative static analysis. Recall (\ref{eq:bank_entry}) and (\ref{eq:bankfoc_amp1})-(\ref{eq:bankfoc_amp3}):

\begin{align}
 \begin{split}
0=G^1&\equiv (1+i_{\ell,t+1})\Tilde{\ell}_{t+1}+(1+i_{r,t+1}) \Tilde{r}_{t+1} \\
 &\quad-(1+i_{t})(\tilde{\ell}_{t+1}+\tilde{r}_{t+1})-\gamma(\Tilde{r}_{t+1}) -\eta(\Tilde{\ell}_{t+1})-\kappa 
  \end{split}\\
0=G^2&\equiv i_{r,t+1}-i_t-\gamma'(\Tilde{r}_{t+1})   \\
0=G^3&\equiv i_{\ell,t+1} -i_t-\eta'(\tilde{\ell}_{t+1}) 
\end{align}
To simplify the notation, drop the time subscripts. Applying the implicit function theorem yields
$$\underbrace{\left[\begin{array}{ccc}
G^1_\ell & G^1_{\tilde{r}} & G^1_{i_{\ell}}    \\
G^2_\ell & G^2_{\tilde{r}} & G^2_{i_{\ell}}    \\
G^3_\ell & G^3_{\tilde{r}} & G^3_{i_{\ell}}    \\
\end{array}\right]}_{\equiv\boldsymbol{G}}\left[\begin{array}{c}
d  \tilde{\ell}    \\
d \tilde{r}   \\
d i_{\ell}     \\
\end{array}\right]=-\left[\begin{array}{c}
G^1_{i_r}d i_r + G^1_{i} d i    \\
G^2_{i_r}d i_r + G^2_{i} d i        \\
G^3_{i_r}d i_r + G^3_{i} d i         \\
\end{array}\right]$$
where
$$\begin{array}{lcl}
G^1_{i_r}=\tilde{r}, &  & G^1_{i}=-(\tilde{\ell}+\tilde{r}), \\
G^2_{i_r}=1,         &  & G^2_{i}=-1, \\
G^3_{i_r}=0,         &  & G^3_{i}=-1.
\end{array}\quad|\boldsymbol{G}|=\begin{vmatrix}
G^1_\ell & G^1_{\tilde{r}} & G^1_{i_{\ell}}    \\
G^2_\ell & G^2_{\tilde{r}} & G^2_{i_{\ell}}    \\
G^3_\ell & G^3_{\tilde{r}} & G^3_{i_{\ell}}    \\
\end{vmatrix}=\begin{vmatrix}
0 &   0       & \tilde{\ell}    \\
0 & -\gamma'' &0    \\
-\eta'' &0 & 1   \\
\end{vmatrix}=-\eta''\gamma''\tilde{\ell}<0$$
By Cramer's rule, we have
$$\frac{\partial \tilde{\ell}}{\partial i}=\frac{1}{|\boldsymbol{G}|}
\begin{vmatrix}
\tilde{\ell}+\tilde{r} & 0       & \tilde{\ell}     \\
1 & -\gamma'' &0    \\
1 &0 & 1    \\
\end{vmatrix}>0, \quad \frac{\partial \tilde{\ell}}{\partial i_r}=\frac{1}{|\boldsymbol{G}|}
\begin{vmatrix}
-\tilde{r} & 0       & \tilde{\ell}    \\
-1 & -\gamma'' &0   \\
0 & 0 & 1   \\
\end{vmatrix}<0, $$
$$\frac{\partial \tilde{r}}{\partial i}=\frac{1}{|\boldsymbol{G}|}
\begin{vmatrix}
0 & \tilde{\ell}+\tilde{r} & \tilde{\ell}    \\
0 & 1 & 0     \\
-\eta'' & 1 & 1    \\ 
\end{vmatrix}<0, \quad \frac{\partial \tilde{r}}{\partial i_r}=\frac{1}{|\boldsymbol{G}|}
\begin{vmatrix}
0 & -\tilde{r}  & \tilde{\ell}     \\
0 & -1 & 0     \\
-\eta'' & 0 & 1     \\
\end{vmatrix}>0, $$
$$\frac{\partial i_\ell}{\partial i}=\frac{1}{|\boldsymbol{G}|}
\begin{vmatrix}
0 & 0 & \tilde{\ell}+\tilde{r}    \\
0 & -\gamma'' & 1  \\
-\eta'' & 0 & 1    \\
\end{vmatrix}>0, \quad \frac{\partial i_\ell}{\partial i_r}=\frac{1}{|\boldsymbol{G}|}
\begin{vmatrix}
0 & 0 & -\tilde{r}      \\
0 & -\gamma'' & -1    \\
-\eta'' & 0 & 0     \\
\end{vmatrix}<0. $$
\end{proof}

\begin{proof}[Proof of Proposition \ref{prop:threshold}] This proof is divided into 2 parts. (i) Existence and uniqueness of the ample-reserve equilibrium; (ii) Existence of the scarce-reserve equilibrium.

\textbf{Part 1} (Existence and uniqueness of the ample-reserve equilibrium)
Consider the ample-reserve equilibrium. The ample-reserve equilibrium satisfies $i_d=i$, $\lambda_{\chi}=0$, and $\tilde{r}>\chi\tilde{d}$. Given $i_d=i$ and $\lambda_{\chi}=0$, the equilibrium solves the following system of equations. 
\begin{align}
0=&  -\gamma'(\tilde{r})  +i_r-i  \label{eq:prop2_pfp1_1} \\ 
0=& -\kappa+\eta'(\tilde{\ell}) \tilde{\ell}+\gamma'(\tilde{r})\tilde{r} -\eta(\tilde{\ell})-\gamma(\tilde{r}) \label{eq:prop2_pfp1_2}  \\
0=& -\frac{1}{\beta}+1-\delta    -\frac{\eta'(\tilde{\ell})}{\beta(1+i)} +f'(k) 
\label{eq:prop2_pfp1_3} \\
0=& -f(k)+f'(k)k+\zeta/U'\big(\{f(k)-\delta k\}N\big) 
\label{eq:prop2_pfp1_4} \\
0=&-(1+i)n\tilde{d}+\max\left\{0,p^*-\bar{b}\right\} 
\label{eq:prop2_pfp1_5} \\
0=& -k N+n\tilde{\ell} \label{eq:prop2_pfp1_6} 
 \\
0=& - i+\nu\sigma_1 \lambda(q) \label{eq:prop2_pfp1_7} 
\end{align}
First, let's check equations (\ref{eq:prop2_pfp1_1})-(\ref{eq:prop2_pfp1_6}) because $q_1$ is already pinned down by $i$. Given $(i, i_r)$ where $i_r\geq i$, equation (\ref{eq:prop2_pfp1_1}) uniquely pins down $\tilde{r}\geq 0$ and $\partial\tilde{r}/\partial i_r>0$ because $\gamma''>0$. When $i_r<i$, there is no solution $\tilde{r}\geq0$ satisfying equation (\ref{eq:prop2_pfp1_1}). Since $\eta''>0$ and $\gamma''>0$, equation (\ref{eq:prop2_pfp1_2}) uniquely pins down $\tilde{\ell}$  as long as  $\bar\Delta+i\geq i_r$ where 
$\bar\Delta=\gamma'(\underline{r})$ and $\underline{r}$ solves $\kappa=\gamma'(\underline{r})\underline{r}-\gamma(\underline{r})$. When $i_r>\bar\Delta+i$, there is no solution for (\ref{eq:prop2_pfp1_2}) given $\tilde{r}\geq 0$ satisfying (\ref{eq:prop2_pfp1_1}). Similarly, since $U''<0$ and $f''<0$, equation (\ref{eq:prop2_pfp1_3}) uniquely pins down $k$. Given $(\tilde{r},\tilde{\ell},k)$ from  (\ref{eq:prop2_pfp1_1})-(\ref{eq:prop2_pfp1_3}), it is straightforward to show that (\ref{eq:prop2_pfp1_4})-(\ref{eq:prop2_pfp1_6}) give   $(n,N,d)$ uniquely. Therefore,  the system of equations (\ref{eq:prop2_pfp1_1})-(\ref{eq:prop2_pfp1_6}) provides a unique solution for $(\tilde{r},\tilde{\ell},k,n,N,d)$ when  $i_r\in(i,\bar\Delta+i)$.

Next, we will check under what condition $\tilde{r}>\chi\tilde{d}$ is satisfied. Consider a case satisfying $\tilde{r}=\chi\tilde{d}$ where $i_d=i$ which is a threshold between ample-reserve case and scarce reserve case.  In this case, we have
$(\tilde{r},\tilde{\ell},k,n,N)=(\bar{r},\bar{\ell},\bar{k},\bar{n},\bar{N})$ where 
$(\bar{r},\bar{\ell},\bar{K},\bar{N},\bar{C},\bar{n})$ solves $\eta'(\bar{\ell})\bar{\ell}+(i_r-i)\bar{r}-\gamma(\bar{r})-\eta(\bar{\ell})=\kappa$, 
$$F_K(\bar{K},\bar{N})=\frac{1}{\beta}-1+\delta+\frac{\eta'(\bar{\ell})}{\beta(1+i)}, \qquad \max\left\{0,\frac{\chi\left\{v(q^*)F_N(\bar{K},\bar{N})/\zeta-\bar{b}\right\}}{\bar{n}}\right\}=(1+i)\bar{r},$$
and  
$\bar{C}+\delta \bar{K}=F(\bar{K},\bar{N})$, $U'(\bar{C})=\zeta/F_N(\bar{K},\bar{N})$, and $\bar{K}=\bar{n}\bar{\ell}$.

Since $\partial \tilde{r}/ \partial i_r>0$, showing $\partial \tilde{d}/ \partial i_r<0$ will    suffice to show (i) part of the 
Proposition \ref{prop:threshold}. Consider the following system of equations. 
\begin{align*}
0=\tilde{G}^1\equiv&  -\gamma'(\tilde{r})  +i_r-i   \\ 
0=\tilde{G}^2\equiv&  -\kappa+\eta'(\tilde{\ell}) \tilde{\ell}+\gamma'(\tilde{r})\tilde{r} -\eta(\tilde{\ell})-\gamma(\tilde{r})  \\
0=\tilde{G}^3\equiv&  -\frac{1}{\beta}+1-\delta    -\frac{\eta'(\tilde{\ell})}{\beta(1+i)} +F_K(n\tilde{\ell},N) \\
0=\tilde{G}^4\equiv&  -F_N(n\tilde{\ell},N)+\frac{\zeta}{U'(F(n\tilde{\ell},N)-\delta n\tilde{\ell})} \\
0=\tilde{G}^5\equiv& -n\tilde{d} + \max\left\{\frac{v(q^*)F_N(n\tilde{\ell},N)/\zeta-\bar{b}}{1+i},0\right\} 
\end{align*}
Applying the implicit function theorem gives the following:
\begin{equation}\label{eq:ample_com_statics}
\underbrace{\left[\begin{array}{ccccc}
0 & \tilde{G}^1_r &  0 & 0  & 0 \\
0 & \tilde{G}^2_r &  \tilde{G}^2_\ell & 0  & 0 \\
\tilde{G}^3_n & 0 &  \tilde{G}^3_\ell  & \tilde{G}^3_N & 0 \\
\tilde{G}^4_n  & 0 &  \tilde{G}^4_\ell & \tilde{G}^4_N  & 0 \\
\tilde{G}^5_n & 0 & \tilde{G}^5_\ell & \tilde{G}^5_N & \tilde{G}^5_d \\
\end{array}\right]}_{=\boldsymbol{\tilde{G}}}\left[\begin{array}{c}
d n     \\
d\tilde{r}  \\
d\tilde{\ell}  \\
d N     \\
d \tilde{d}     \\
\end{array}\right]=-\left[\begin{array}{c}
1     \\
0         \\
0         \\
0          \\
0          \\
\end{array}\right]d i_r-\left[\begin{array}{c}
-1   \\
0       \\
\frac{\eta'}{\beta(1+i)^2}        \\
0     \\
-\frac{v(q^*)F_N}{\zeta(1+i)^2}        \\
\end{array}\right]di-\left[\begin{array}{c}
0   \\
0       \\
0        \\
0     \\
\frac{1}{1+i}       \\
\end{array}\right]d\bar{b}\end{equation}
where
$$\boldsymbol{\tilde{G}}=\left[\begin{array}{ccccc}
0 & -\gamma'' &  0 & 0  & 0 \\
0 & \gamma''\tilde{r} &  \eta''\tilde{\ell} & 0  & 0 \\
F_{KK}\tilde{\ell} & 0 &  -\frac{\eta''(\tilde{\ell})}{\beta(1+i)} +F_{KK}n & F_{KN}  & 0 \\
-\{F_{NK}+\frac{\zeta  U''}{U'^2}(F_K-\delta)\}\tilde{\ell}  & 0 &  -\{F_{NK}+\frac{\zeta  U''}{U'^2}(F_K-\delta)\}n & -F_{NN}-\frac{\zeta F_N U''}{U'^2}  & 0 \\
-\tilde{d}+\frac{v(q^*)F_{KN}\tilde{\ell}}{\zeta(1+i)} & 0 & \frac{v(q^*)F_{KN}n}{\zeta(1+i)} & \frac{v(q^*)F_{NN}}{\zeta(1+i)} & -n \\
\end{array}\right]$$
Using Cramer's rule, we have the following comparative statics:

$$\frac{\partial \tilde{r}}{\partial i_r}=\frac{n\gamma''\eta''\tilde{\ell}^2}{|\boldsymbol{\tilde{G}}|}
\begin{vmatrix}
F_{KK}\tilde{\ell}  &  F_{KN}   \\
-\{F_{NK}+\frac{\zeta  U''}{U'^2}(F_K-\delta)\}\tilde{\ell}   &   -F_{NN}-\frac{\zeta F_N U''}{U'^2}   \\
\end{vmatrix}>0$$

$$\frac{\partial \tilde{d}}{\partial i_r}=\frac{\gamma''\tilde{r}}{|\boldsymbol{\tilde{G}}|}
\begin{vmatrix}
F_{KK}\tilde{\ell}  & -\frac{\eta''}{\beta(1+i)}+F_{KK}n &  F_{KN}   \\
-\{F_{KN}+\frac{\zeta  U''}{U'^2}(F_K-\delta)\}\tilde{\ell} & -\{F_{KN}+\frac{\zeta  U''}{U'^2}(F_K-\delta)\}n  &   -F_{NN}-\frac{\zeta F_N U''}{U'^2}   \\
-\tilde{d}+\frac{v(q^*)F_{KN}\tilde{\ell}}{\zeta(1+i)}  & \frac{v(q^*)F_{KN}n}{\zeta(1+i)} &  \frac{v(q^*)F_{NN}}{\zeta(1+i)}   \\
\end{vmatrix}<0$$
since $(F_{KN})^2=F_{KK}F_{NN}$, $F_{KN}=F_{NK}$ and $F_N>F_{NK}K$. Given $\partial \tilde{r}/\partial i_r>0$  $\partial \tilde{d}/\partial i_r<0$, we can conclude that given $(i, \chi, \bar{b})$, $\exists !$  ample-reserves equilibrium if and only if $i_r\in(\bar{\iota}_r,\bar\Delta+i)$. 

\textbf{Part 2} (Existence of the scarce-reserve equilibrium): 
From Part 1, we already know that there is no scarce-reserve equilibrium when $i_r\in(\bar{\iota}_r,\bar\Delta+i)$. Hence, we only focus on that case where $i_r\in[0,\bar{\iota}_r]$. The scarce-reserve equilibrium solves the following system of equations. 
\begin{align}
0=&-i_{d}+(1-\chi) i_{s}+\chi i_{r}-\chi\gamma'(\tilde{r}) 
\label{eq:prop2_pfp2_1} \\
0=&  -i_{\ell}+ i+\eta'(\tilde{\ell})
\label{eq:prop2_pfp2_2} \\ 
0=& -\kappa+\eta'(\tilde{\ell}) \tilde{\ell}+\gamma'(\tilde{r})\tilde{r} -\eta(\tilde{\ell})-\gamma(\tilde{r}) \label{eq:prop2_pfp2_3} \\ 
0=& -\frac{1}{\beta}+1-\delta    -\frac{\eta'(\tilde{\ell})}{\beta(1+i)} +f'(k) 
\label{eq:prop2_pfp2_4} \\ 
0=& -f(k)+f'(k)k+\zeta/U'\big(\{f(k)-\delta k\}N\big) 
\label{eq:prop2_pfp2_5} \\
0=&-\tilde{r}+\chi\tilde{d}
\label{eq:prop2_pfp2_7} \\
0=& -k N+n\tilde{\ell} 
\label{eq:prop2_pfp2_8} \\
0=& -i+\nu\sigma_1\lambda(q_{1})+\nu\sigma_3\lambda(q_{3}) 
\label{eq:prop2_pfp2_9} \\
0=& -\frac{1+i}{1+i_{d}}+1+ \nu\sigma_2\lambda(q_{2})+\nu\sigma_3\lambda(q_{3})\label{eq:prop2_pfp2_10} \\ 
0=&-v(q_1)+  \min\left\{v(q^*),\frac{\zeta m}{f(k)-f'(k)k}\right\}   
\label{eq:prop2_pfp2_11} \\
0=&-v(q_2)+ 
\min\left\{v(q^*),\frac{\zeta (n\tilde{d}+\bar{b})}{f(k)-f'(k)k}\right\}  
\label{eq:prop2_pfp2_12} \\
0=&-v(q_3)+ 
\min\left\{v(q^*),\frac{\zeta (m+n\tilde{d}+\bar{b})}{f(k)-f'(k)k}\right\} 
\label{eq:prop2_pfp2_13}
\end{align}
The equations (\ref{eq:prop2_pfp2_2})-(\ref{eq:prop2_pfp2_13}) can be simplified as follows:
\begin{align}
0=&\tilde{H}^1\equiv -i+\nu\sigma_1\Lambda\left(\frac{\zeta}{f(k)-f'(k)k} m\right)
\label{eq:prop2_pfp3_f}
\\
&+\nu\sigma_3\Lambda\left(\min\left\{v(q^*), \frac{\zeta}{f(k)-f'(k)k}\left[m+\bar{b}+\frac{[1+i_d]n\tilde{r}}{\chi}\right]\right\}\right) \nonumber  \\ 
0=&\tilde{H}^2\equiv -\frac{1+i}{1+i_d}+1  \\
&+\nu\sigma_2\Lambda\left(\min\left\{v(q^*), \frac{\zeta}{f(k)-f'(k)k}\left[\bar{b}+\frac{[1+i_d]n\tilde{r}}{\chi}\right]\right\}\right) \nonumber  \\ 
&+\nu\sigma_3\Lambda\left(\min\left\{v(q^*), \frac{\zeta}{f(k)-f'(k)k}\left[m+\bar{b}+\frac{[1+i_d]n\tilde{r}}{\chi}\right]\right\}\right) \nonumber  \\ 
0=&\tilde{H}^3\equiv -\kappa+\eta'(\tilde{\ell})\tilde{\ell}-\eta(\tilde{\ell})+\gamma'(\tilde{r})\tilde{r} -\gamma(\tilde{r}) \\
0=&\tilde{H}^4\equiv  -f'(k)+\frac{1}{\beta}-1+\delta+\frac{\eta'(\Tilde{\ell})}{\beta(1+i)}  \\
0=&\tilde{H}^5\equiv  \frac{\zeta}{f(k)-f'(k)k} -U'\left([f(k)-\delta k]N\right)\\
0=&\tilde{H}^6\equiv -k+\frac{n\tilde{\ell}}{N}
\label{eq:prop2_pfp3_l}
\end{align}
\begin{figure}[tp!]
\centerfloat
\input{figures_gen/name.tex}
\caption{RC1 and RC2} 
\label{fig:11}
\end{figure}
Relaxing the equality of (\ref{eq:prop2_pfp2_1}), we obtain $i_{d}\geq (1-\chi) i_{s}+\chi i_{r}-\chi\gamma'(\tilde{r})$. From equation (\ref{eq:prop2_pfp2_1}), it is given that $\tilde{r}$ decreases with $i_d$ for $i_d\in[0,(1-\chi) i_{s}+\chi i_{r}]$. The curve that satisfies equation (\ref{eq:prop2_pfp2_1}) in the $(i_d,\tilde{r})$ space, labeled as RC1, represents the amount of real reserve balances a bank is willing to hold for a given $i_d$. The curve that satisfies equations (\ref{eq:prop2_pfp3_f}) to (\ref{eq:prop2_pfp3_l}) in the $(i_d,\tilde{r})$ space, labeled as RC2, represents the required reserves for transaction deposit demand. This takes into account the values of $i_d$ and the corresponding activities of households and firms. The RC1 curve is a downward-sloping single-valued function, while RC2 can be either a single-valued or a multi-valued curve.

The RC1 curve starts at $(\tilde{r},i_d)=(0,(1-\chi)i+\chi i_r)$, with a continuum of points where $\tilde{r}=0$ and $i\geq i_d\geq(1-\chi)i+\chi i_r$. Along the RC1 curve, $i_d$ weakly decreases with $\tilde{r}$ until it reaches the point $(\tilde{r},i_d)=(\gamma^{-1}\left(\frac{1-\chi}{\chi} i+i_{r}\right),0)$. In equilibrium, since $i\geq i_d$, the right end of the RC2 curve solves (\ref{eq:prop2_pfp3_f})-(\ref{eq:prop2_pfp3_l}) with $i_d=i$. For a given $(i_r,\bar{b})$ with $i_d=i$, this holds when $\tilde{r}>0$ as long as $\bar{b}<p^*$. If $\bar{b}\geq p^*$, then $\tilde{r}=0$. At the left end of the RC2 curve, it solves (\ref{eq:prop2_pfp3_f})-(\ref{eq:prop2_pfp3_l}) where $\tilde{r}=0$ and $i\geq i_d\geq 0$. It is evident that when $\bar{b}\geq p^*$, the RC2 curve becomes a degenerate case with a single point at $(\tilde{r},i_d)=(0,i)$. When $\bar{b}<p^*$, the left end of the RC2 curve satisfies $\tilde{r}>0$, and $i>i_d\geq 0$. Therefore, there is at least one point where the RC1 curve intersects the RC2 curve where the RC1 curve crosses the RC2 curve from below. Consequently, we can conclude that a scarce-reserves equilibrium exists if and only if $0\leq i_r\leq \bar{\iota}_r$.
\end{proof}

\begin{proof}[Proof of Proposition \ref{prop:compastatics1}] 
Recall money multiplier $\xi$:
$$\xi=\frac{m+d}{\phi B}=\frac{m +d}{m+r}=\frac{m +n\tilde{d}}{m+n\tilde{r}}$$
Showing $\partial \tilde{r}/\partial i_r>0$, $\partial \tilde{r}/\partial i<0$, $\partial \tilde{d}/\partial i_r<0$, and $\partial \tilde{d}/\partial i>0$ suffices to show
$\partial\xi/\partial i>0$ and $\partial\xi/\partial i_r<0.$
Recall $\gamma'(\tilde{r})=i_r-i$. It is straightforward to show that $\partial \tilde{r}/\partial i_r>0$ and $\partial \tilde{r}/\partial i<0$. Part 1 of Proposition \ref{prop:threshold} already has shown $\partial \tilde{d}/\partial i_r<0$. To show $\partial \tilde{d}/\partial i>0$, recall 
(\ref{eq:ample_com_statics}). Applying implicit function theorem and comparative statics gives 
$$
\frac{\partial \tilde{d}}{\partial i}=\frac{|\boldsymbol{\tilde{G}_{i,d}}|}{|\boldsymbol{\tilde{G}}|} $$
where 
$$\boldsymbol{\tilde{G}_{i,d}}=\left[\begin{array}{ccccc}
0 & -\gamma'' &  0 & 0  & 1 \\
0 & \gamma''\tilde{r} &  \eta''\tilde{\ell} & 0  & 0 \\
F_{KK}\tilde{\ell} & 0 &  -\frac{\eta''(\tilde{\ell})}{\beta(1+i)} +F_{KK}n & F_{KN}  & -\frac{\eta'}{\beta(1+i)^2} \\
-\{F_{NK}+\frac{\zeta  U''}{U'^2}(F_K-\delta)\}\tilde{\ell}  & 0 &  -\{F_{NK}+\frac{\zeta  U''}{U'^2}(F_K-\delta)\}n & -F_{NN}-\frac{\zeta F_N U''}{U'^2}  & 0 \\
-\tilde{d}+\frac{v(q^*)F_{KN}\tilde{\ell}}{\zeta(1+i)} & 0 & \frac{v(q^*)F_{KN}n}{\zeta(1+i)} & \frac{v(q^*)F_{NN}}{\zeta(1+i)} & \frac{v(q^*)F_N}{\zeta(1+i)^2} \\
\end{array}\right].$$
Solving this yields the following expression:
\begin{align*}
\frac{\partial \tilde{d}}{\partial i}=&\frac{|\boldsymbol{\tilde{G}_{i,d}}|}{|\boldsymbol{\tilde{G}}|} =-\frac{\gamma''\eta''\tilde{\ell}}{|\boldsymbol{\tilde{G}}|}\begin{vmatrix}
F_{KK}\tilde{\ell} & F_{KN} & -\frac{\eta'}{\beta(1+i)^2}  \\
-\{F_{NK}+\frac{\zeta  U''}{U'^2}(F_K-\delta)\}\tilde{\ell}   & -F_{NN}-\frac{\zeta F_N U''}{U'^2} & 0 \\
-\tilde{d}+\frac{v(q^*)F_{KN}\tilde{\ell}}{\zeta(1+i)}  & \frac{v(q^*)F_{NN}}{\zeta(1+i)}  & \frac{v(q^*)F_N}{\zeta(1+i)^2}  \\
\end{vmatrix} \\
&- \frac{\gamma''\tilde{r}}{|\boldsymbol{\tilde{G}}|}\begin{vmatrix}
F_{KK}\tilde{\ell} & -\frac{\eta''(\tilde{\ell})}{\beta(1+i)} +F_{KK}n & F_{KN}  \\
-\{F_{NK}+\frac{\zeta  U''}{U'^2}(F_K-\delta)\}\tilde{\ell}   &  -\{F_{NK}+\frac{\zeta  U''}{U'^2}(F_K-\delta)\}n & -F_{NN}-\frac{\zeta F_N U''}{U'^2} \\
-\tilde{d}+\frac{v(q^*)F_{KN}\tilde{\ell}}{\zeta(1+i)}  & \frac{v(q^*)F_{KN}n}{\zeta(1+i)} & \frac{v(q^*)F_{NN}}{\zeta(1+i)}  \\
\end{vmatrix}>0.
\end{align*}

\end{proof}

\begin{proof}[Proof of Proposition \ref{prop:compastatics2}] This proof is divided into 2 parts: (i) Ample reserves equilibrium and (ii) scarce reserve equilibrium.

\textbf{Ample reserves equilibrium } In the ample reserve equilibrium, since $i_d=i$, it is easy to show that $\partial i_d/\partial \bar{b}=0$.
When $p^*>\bar{b}$  we have 
$$d=\frac{v(q^*)\{f(k)-f'(k)k\}/\zeta-\bar{b}}{1+i}$$ 
in the ample reserves equilibrium. From comparative statics, we have 
$$\frac{\partial \tilde{\ell}}{\partial \bar{b}}=\frac{|\boldsymbol{\tilde{G}_{\bar{b},\tilde{\ell}}}|}{|\boldsymbol{\tilde{G}}|}=0, \quad \frac{\partial n}{\partial \bar{b}}=\frac{|\boldsymbol{\tilde{G}_{\bar{b},n}}|}{|\boldsymbol{\tilde{G}}|}=0, \quad \frac{\partial N}{\partial \bar{b}}=\frac{|\boldsymbol{\tilde{G}_{\bar{b},N}}|}{|\boldsymbol{\tilde{G}}|}=0$$
since
$$|\boldsymbol{\tilde{G}_{\bar{b},\tilde{\ell}}}|=-\frac{1}{1+i}\left|\begin{array}{cccc}
0 &  \tilde{G}^1_r & 0  & 0 \\
0 &  \tilde{G}^2_r & 0  & 0 \\
\tilde{G}^3_n & 0  & \tilde{G}^4_N & 0 \\
\tilde{G}^4_n &  0 & \tilde{G}^4_N  & 0 
\end{array}\right|=0, \quad |\boldsymbol{\tilde{G}_{\bar{b},n}}|=-\frac{1}{1+i}\left|\begin{array}{cccc}
 \tilde{G}^1_r &  0 & 0  & 0 \\
 \tilde{G}^2_r &  \tilde{G}^2_\ell & 0  & 0 \\
 0 &  \tilde{G}^3_\ell  & \tilde{G}^3_N & 0 \\
 0 &  \tilde{G}^4_\ell & \tilde{G}^4_N  & 0 
\end{array}\right|=0$$

$$|\boldsymbol{\tilde{G}_{\bar{b},N}}|=\frac{1}{1+i}\left|\begin{array}{cccc}
 0           &  \tilde{G}^1_r & 0  & 0 \\
 0           &  \tilde{G}^2_r & \tilde{G}^2_\ell  & 0 \\
 \tilde{G}^3_n &  0           & \tilde{G}^3_\ell & 0 \\
 \tilde{G}^4_n &  0           & \tilde{G}^4_\ell  & 0 
\end{array}\right|=0.$$
The above comparative statics implies $\frac{\partial k}{\partial \bar{b}}=0$ since $k=K/N=n\tilde{\ell}/N$.
Then it is straightforward to show $\partial d/\partial \bar{b}<0$ and $\partial \tilde{d}/\partial \bar{b}<0$. 

\textbf{Scarce reserves equilibrium } First, I define $\tilde{z}_j\equiv\frac{\zeta}{f(k)-f'(k)k} \max\{p^*,z_j\}$ to simplify notations. The scarce reserve equilibrium solves the following system of equations:
\begin{align}
0=J^1&\equiv -i+\nu\sigma_1\Lambda\left(\frac{\zeta}{f(k)-f'(k)k} m\right) \\
&+\nu\sigma_3\Lambda\left(\frac{\zeta}{f(k)-f'(k)k} \left\{m+\bar{b}+\frac{[1+i_d]n\tilde{r}}{\chi}\right\}\right) \nonumber \\
0=J^2&\equiv -\frac{1+i}{1+i_d}+1 +\nu\sigma_2\Lambda\left(\frac{\zeta}{f(k)-f'(k)k} \left\{\bar{b}+\frac{[1+i_d]n\tilde{r}}{\chi}\right\}\right) \\
&+\nu\sigma_3\Lambda\left(\min\left\{v(q^*), \frac{\zeta}{f(k)-f'(k)k}\left[m+\bar{b}+\frac{[1+i_d]n\tilde{r}}{\chi}\right]\right\}\right) \nonumber \\
0=J^3&\equiv -\kappa+\eta'(\tilde{\ell})\tilde{\ell}-\eta(\tilde{\ell})+\gamma'(\tilde{r})\tilde{r} -\gamma(\tilde{r}) \\
0=J^4&\equiv  \frac{\zeta}{f(k)-f'(k)k} -U'\left([f(k)-\delta k]N\right)\\
0=J^5&\equiv -k+\frac{n\tilde{\ell}}{N} \\
0=J^6&\equiv -f(k)+\frac{1}{\beta}-1+\delta+\frac{\eta(\tilde{\ell})}{\beta(1+i)}
\end{align}
where $i_d=(1-\chi)i+\chi i_r-\chi \gamma'(\tilde{r})$.

Assuming interior, applying the implicit function theorem yields
\begin{equation}\label{eq:implicit}
\underbrace{\left[\begin{array}{cccccc}
J^1_{m}&J^1_{\tilde{r}}& 0&J^1_{k} & J^1_n & 0 \\
J^2_{m}&J^2_{\tilde{r}}& 0&J^2_{k} & J^2_n & 0 \\
0      &J^3_{\tilde{r}}& J^3_{\tilde{\ell}}& 0 & 0 & 0\\
0      & 0& 0&J^4_{k} & 0 & J^4_N \\
0      & 0& J^5_{\tilde{\ell}} &J^5_{k} & J^5_n & J^5_N \\
0      & 0& J^6_{\tilde{\ell}} & J^6_{k} & 0 & 0 
\end{array}\right]}_{\equiv\boldsymbol{J}}\left[\begin{array}{c}
d m  \\
d\tilde{r}  \\
d\tilde{\ell}  \\
d k  \\
d n   \\
d N 
\end{array}\right]=-\underbrace{\left[\begin{array}{ccc}
J^1_{\bar{b}} & J^1_{i_r} & J^1_{i} \\
J^2_{\bar{b}} & J^2_{i_r} & J^2_{i} \\
0             & 0         & 0      \\
0             & 0         & 0      \\
0             & 0         & 0      \\
0             & 0         & J^6_{i}   
\end{array}\right]}_{\equiv\boldsymbol{\bar{J}}} \left[\begin{array}{c}
d\bar{b} \\
d i_r    \\
di       
\end{array}\right]
\end{equation}
where 

$$\begin{array}{lll}
J^1_{m}=\frac{\nu\zeta}{f(k)-f'(k)k}\{\sigma_1 \Lambda'(\tilde{z}_1)+\sigma_3 \Lambda'(\tilde{z}_3)\}, & J^2_{m}= \frac{\nu\zeta}{f(k)-f'(k)k}\sigma_3 \Lambda'(\tilde{z}_3), & \\
J^1_{\tilde{r}}=\frac{n\nu\zeta \{1+i_d-\tilde{r}\chi \gamma''(\tilde{r})\}}{\chi\{f(k)-f'(k)k\}}  \sigma_3\Lambda'(\tilde{z}_3), & J^2_{r}=-\frac{1+i}{(1+i_d)^2}\chi\gamma''(\tilde{r})+\frac{n\nu\zeta\{1+i_d-\tilde{r}\chi\gamma''(\tilde{r})\} }{\chi\{f(k)-f'(k)k\}}\sum_{j=2}^3\{\sigma_j \Lambda'(\tilde{z}_j)\}, & \\
J^3_{\tilde{r}}= \gamma''(\tilde{r})\tilde{r}, & J^3_{\tilde{\ell}}= \eta''(\tilde{\ell})\tilde{\ell}, &  \\
J^5_{\tilde{\ell}}=\frac{n}{N} , & J^6_{\tilde{\ell}}=\frac{\eta''(
\tilde{\ell})}{\beta(1+i)}, & \\
J^1_{k}=\frac{\nu\zeta f''(k)k}{\{f(k)-f'(k)k\}^2}\{\sigma_1 \Lambda'(\tilde{z}_1)+\sigma_3 \Lambda'(\tilde{z}_3)\}, & J^2_{k}=\frac{\nu\zeta f''(k)k}{\{f(k)-f'(k)k\}^2}\{\sigma_2 \Lambda'(\tilde{z}_2)+\sigma_3 \Lambda'(\tilde{z}_3)\},  & \\
J^4_{k}=\frac{\zeta f''(k)k}{\{f(k)-f'(k)k\}^2}-\{f'(k)-\delta \}N  U''(C), & J^5_{k}=-1 , & \\
J^6_{k}=-f''(k) , & J^1_{n}=\frac{\nu\zeta(1+i_d)\tilde{r} }{f(k)-f'(k)k}\sigma_3 \Lambda'(\tilde{z}_3), & \\
J^2_{n}=\frac{\nu\zeta(1+i_d)\tilde{r}}{f(k)-f'(k)k}\{\sigma_2 \Lambda'(\tilde{z}_2)+\sigma_3 \Lambda'(\tilde{z}_3)\}, & J^5_{n}=\frac{\tilde{\ell}}{N}, & \\
J^4_{N}=-\{f-\delta k\}U''(C), & J^5_{N}= -\frac{n\tilde{\ell}}{N^2}, & \\
J^1_{\bar{b}}=\frac{\nu\zeta}{f(k)-f'(k)k}\sigma_3 \Lambda'(\tilde{z}_3), & J^2_{\bar{b}}=\frac{\nu\zeta}{f(k)-f'(k)k}\{\sigma_2 \Lambda'(\tilde{z}_2)+\sigma_3 \Lambda'(\tilde{z}_3)\}, & \\
J^1_{i_r}=\frac{\sigma_3 n\tilde{r}\nu\zeta}{f(k)-f'(k)k}\Lambda'(\tilde{z}_3), & J^2_{i_r}=\frac{ n\tilde{r}\nu\zeta}{f(k)-f'(k)k}\{\sigma_2 \Lambda'(\tilde{z}_2)+\sigma_3 \Lambda'(\tilde{z}_3)\}, & \\
J^1_{i}=-1+\frac{(1-\chi)d\sigma_3\nu\zeta}{f-f'k}\Lambda(\tilde{z}_3), & J^2_{i}=-\frac{1}{1+i_d}+\frac{(1-\chi)d\nu\zeta}{f-f'k}\{\sigma_2\Lambda(\tilde{z}_2)+\sigma_3\Lambda(\tilde{z}_3)\}+\frac{1+i}{(1+i_d)^2}\chi, & \\
J^6_{i}=-\frac{\eta'(\tilde{\ell})}{\beta(1+i)^2}. &  & 
\end{array}$$

The determinant of $\boldsymbol{J}$ is 
\begin{align*}
|\boldsymbol{J}|=& -U''\gamma''\tilde{r} f''\frac{n}{N}(f-\delta k) \left(\frac{\nu\zeta}{f-f'k} \right)^2 \frac{(1+i_d)d}{n} \\
&\quad  \times\left\{\sigma_1\sigma_2\Lambda'(\tilde{z}_1)\Lambda'(\tilde{z}_2)+\sigma_1\sigma_3\Lambda'(\tilde{z}_1)\Lambda'(\tilde{z}_3)+\sigma_2\sigma_3\Lambda'(\tilde{z}_2)\Lambda'(\tilde{z}_3) \right\}  \\
&  -U''\eta''\tilde{\ell} f''\frac{\ell}{N} (f-\delta k)\frac{\nu\zeta}{f-f'k}\left\{-\frac{1+i}{(1+i_d)^2}\chi\gamma'' \right\} \left\{\sigma_1\Lambda'(\tilde{z}_1)+\sigma_3\Lambda'(\tilde{z}_3) \right\}  \\
&  + U''\gamma''\tilde{r} \frac{\eta''}{\beta(1+i)}\left(\frac{\nu\zeta}{f-f'k} \right)^2\frac{(1+i_d)d}{n}(f-f'k)\\
&\quad  \times\left\{\sigma_1\sigma_2\Lambda'(\tilde{z}_1)\Lambda'(\tilde{z}_2)+\sigma_1\sigma_3\Lambda'(\tilde{z}_1)\Lambda'(\tilde{z}_3)+\sigma_2\sigma_3\Lambda'(\tilde{z}_2)\Lambda'(\tilde{z}_3) \right\}  \\
&-U''\eta''\tilde{\ell}f''\frac{\ell}{N}(f-\delta k)[1+i_d-\tilde{r}\chi\gamma''] \left(\frac{\nu\zeta}{f-f'k} \right)^2\frac{d}{\tilde{r}}\\
&\quad \times \left\{\sigma_1\sigma_2\Lambda'(\tilde{z}_1)\Lambda'(\tilde{z}_2)+\sigma_1\sigma_3\Lambda'(\tilde{z}_1)\Lambda'(\tilde{z}_3)+\sigma_2\sigma_3\Lambda'(\tilde{z}_2)\Lambda'(\tilde{z}_3) \right\}  \\
&+ U'\frac{\tilde{\ell}}{N^2}\gamma''\tilde{r}\frac{\eta''}{\beta(1+i)}\left(\frac{\nu\zeta}{f-f'k} \right)^2\frac{f''k}{f-f'k} \\
&\quad\times \Bigl[(1+i_d)d \left\{\sigma_1\sigma_2\Lambda'(\tilde{z}_1)\Lambda'(\tilde{z}_2)+\sigma_1\sigma_3\Lambda'(\tilde{z}_1)\Lambda'(\tilde{z}_3)+\sigma_2\sigma_3\Lambda'(\tilde{z}_2)\Lambda'(\tilde{z}_3) \right\}       \\
&\qquad\quad   - \left\{z_2\sigma_1\sigma_2\Lambda'(\tilde{z}_1)\Lambda'(\tilde{z}_2)+(z_3-z_1)\sigma_1\sigma_3\Lambda'(\tilde{z}_1)\Lambda'(\tilde{z}_3)+z_2\sigma_2\sigma_3\Lambda'(\tilde{z}_2)\Lambda'(\tilde{z}_3) \right\}\Bigr]
\end{align*}
Since $z_2=z_3-z_1\rightarrow(1+i_d)d$ when $\bar{b}\rightarrow0$, we have 
$$ \lim_{\bar{b}\to0} |\boldsymbol{J}| <0$$
under Assumption 1. One can also show $\lim_{\sigma_3\to0} |\boldsymbol{J}| <0$ as well. Therefore, for sufficiently small $\bar{b}$ or $\sigma_3$, we have $|\boldsymbol{J}| <0$. From comparative statics we have 

\begin{align*}
&\frac{\partial \tilde{r}}{ \partial \bar{b}}=\frac{|\boldsymbol{J_{\bar{b},\tilde{r}}}|}{|\boldsymbol{J}|} \\
&=\frac{1}{|\boldsymbol{J}|}\underbrace{\left(\frac{\nu\zeta}{f-f'k}\right)^2}_{\bigoplus} \underbrace{\{\sigma_1\sigma_2 \Lambda'(\tilde{z}_1)\Lambda'(\tilde{z}_2)+\sigma_1\sigma_3 \Lambda'(\tilde{z}_1)\Lambda'(\tilde{z}_3)+\sigma_3\sigma_2 \Lambda(\tilde{z}_3)\Lambda(\tilde{z}_2)\}}_{\bigoplus}\underbrace{\frac{f''U''\eta''\ell(f-\delta k)}{\beta (1+i)N}}_{\bigoplus}
\end{align*}

\begin{align*}
\frac{\partial n}{ \partial \bar{b}}&=\frac{|\boldsymbol{J_{\bar{b},n}}|}{|\boldsymbol{J}|} \\
&=\frac{1}{|\boldsymbol{J}|}\underbrace{\left(\frac{\nu\zeta}{f-f'k}\right)^2}_{\bigoplus} \underbrace{\{\sigma_1\sigma_2 \Lambda'(\tilde{z}_1)\Lambda'(\tilde{z}_2)+\sigma_1\sigma_3 \Lambda'(\tilde{z}_1)\Lambda'(\tilde{z}_3)+\sigma_3\sigma_2 \Lambda(\tilde{z}_3)\Lambda(\tilde{z}_2)\}}_{\bigoplus} \\
&\quad\times \underbrace{\left[\frac{(f'k-f)U''\gamma''\tilde{r}k \eta''}{\beta(1+i)} +(f-\delta k)U''f''\frac{n}{N}+\frac{-\zeta f''k\eta''k}{\{f-f'k\}^2N\beta(1+i) }  \right]}_{\bigoplus}
\end{align*}
where
$$\boldsymbol{J_{\bar{b},\tilde{r}}}=\left[\begin{array}{cccccc}
J^1_{m}&-J^1_{\bar{b}}& 0&J^1_{k} & J^1_n & 0 \\
J^2_{m}&-J^2_{\bar{b}}& 0&J^2_{k} & J^2_n & 0 \\
0      & 0 & J^3_{\tilde{\ell}}& 0 & 0 & 0\\
0      & 0& 0&J^4_{k} & 0 & J^4_N \\
0      & 0& J^5_{\tilde{\ell}} &J^5_{k} & J^5_n & J^5_N \\
0      & 0& J^6_{\tilde{\ell}} & J^6_{k} & 0 & 0 \\
\end{array}\right], \quad \boldsymbol{J_{\bar{b},n}}=\left[\begin{array}{cccccc}
J^1_{m}&J^1_{\tilde{r}}& 0&J^1_{k} & -J^1_{\bar{b}} & 0 \\
J^2_{m}&J^2_{\tilde{r}}& 0&J^2_{k} & -J^2_{\bar{b}} & 0 \\
0      &J^3_{\tilde{r}}& J^3_{\tilde{\ell}}& 0 & 0 & 0\\
0      & 0& 0&J^4_{k} & 0 & J^4_N \\
0      & 0& J^5_{\tilde{\ell}} &J^5_{k} & 0 & J^5_N \\
0      & 0& J^6_{\tilde{\ell}} & J^6_{k} & 0 & 0 \\
\end{array}\right].$$
Then we have 
$$ \lim_{\bar{b}\to0} \frac{|\boldsymbol{J_{\bar{b},\tilde{r}}}|}{|\boldsymbol{J}|}<0, \quad  \lim_{\bar{b}\to0}  \frac{|\boldsymbol{J_{\bar{b},n}}|}{|\boldsymbol{J}|} <0, \quad \lim_{\sigma_3\to0} \frac{|\boldsymbol{J_{\bar{b},\tilde{r}}}|}{|\boldsymbol{J}|}<0, \quad  \lim_{\sigma_3\to0}  \frac{|\boldsymbol{J_{\bar{b},n}}|}{|\boldsymbol{J}|} <0$$
because 
$|\boldsymbol{J_{\bar{b},\tilde{r}}}|>0$
and $|\boldsymbol{J_{\bar{b},n}}|>0$. Since $d=n\tilde{d}/\chi$, it is easy to show that $\partial d/\partial \bar{b}<0$ for small $\bar{b}$ or $\sigma_3$. Since $i_d=(1-\chi)i+\chi i_r-\chi \gamma'(\tilde{r})$ and $\partial \tilde{r}/\partial\bar{b}<0$ for small $\bar{b}$ or $\sigma_3$, we have $\partial i_d/\partial \bar{b}>0$ for small $\bar{b}$ or $\sigma_3$. 
\end{proof}

\begin{proof}[Proof of Proposition \ref{prop:compastatics4}] 
This proof is divided into 4 parts: 

\textbf{Part 1} (Comparative Statics in the Ample-Reserve Equilibrium): Consider the ample-reserves equilibrium. It can be summarized  as 4 equations 4 unknowns as follows.
\begin{align*}
0=G^1\equiv&  -\gamma'(\tilde{r})  +i_r-i   \\ 
0=G^2\equiv& -\kappa+\{1+i+\eta'(\tilde{\ell}) \}\tilde{\ell}+(1+i_r)\tilde{r}-(1+i)(\tilde{\ell}+\tilde{r})  -\eta(\tilde{\ell})-\gamma(\tilde{r}) \\
0=G^3\equiv& -\frac{1}{\beta}+1-\delta    -\frac{\eta'(\tilde{\ell})}{\beta(1+i)} +f'(k) \\
0=G^4\equiv& -f(k)+f'(k)k+\zeta/U'(\{f(k)-\delta k\}N)  
\end{align*}
where $k=K/H$. Applying the implicit function theorem yields
\begin{equation}
\underbrace{\left[\begin{array}{cccc}
G^1_k& G^1_N & G^1_{\ell} & G^1_{r} \\
G^2_k& G^2_N & G^2_{\ell} & G^2_{r} \\
G^3_k& G^3_N & G^3_{\ell} & G^3_{r} \\
G^4_k& G^4_N & G^4_{\ell} & G^4_{r} \\
\end{array}\right]}_{\equiv\boldsymbol{G}}\left[\begin{array}{c}
d k  \\
d N   \\
d\tilde{\ell}     \\
d\tilde{r}     \\
\end{array}\right]=-\underbrace{\left[\begin{array}{ccc}
G^1_{i_r} & G^1_{i} &  G^1_{b}  \\
G^2_{i_r} & G^2_{i} &  G^2_{b}     \\
G^3_{i_r} & G^3_{i} &  G^3_{b}     \\
G^4_{i_r} & G^4_{i} &  G^4_{b}     \\
\end{array}\right] }_{\equiv\boldsymbol{\bar{G}}}
\left[\begin{array}{c}
di        \\
d i_r     \\
d \bar{b}    \\
\end{array}\right]
\end{equation}
where
\begin{align*}
\boldsymbol{G}=\left[\begin{array}{cccc}
0  & 0 & 0 & -\gamma'' \\
0  & 0 & \eta''\tilde{\ell} & 0 \\
f'' & 0 & -\frac{\eta''}{\beta(1+i)} & 0 \\
f''k-N (f'-\delta)U''\zeta/(U')^2  & -(f-\delta k)U''\zeta/(U')^2 & 0 & 0 \\
\end{array}\right]
\end{align*}
and 
\begin{align*}
\boldsymbol{\bar{G}}=\left[\begin{array}{ccc}
-1   & 1 &   0 \\
-\tilde{r} & \tilde{r} &  0  \\
\frac{\eta'}{\beta(1+i)^2}   & 0 &  0 \\
0 & 0   & 0
\end{array}\right].\end{align*}
By Cramer's rule, we have
$$\frac{\partial \tilde{\ell}}{\partial i}=\frac{1}{|\boldsymbol{G}|}
\begin{vmatrix}
G^1_k & G^1_N & -G^1_i & G^1_{r}  \\
G^2_k & G^2_N & -G^2_i & G^2_{r}  \\
G^3_k & G^3_N & -G^3_i & G^3_{r}  \\
G^4_k & G^4_N & -G^4_i & G^4_{r}  \\
\end{vmatrix} = \frac{-(f-\delta k)U''\frac{\zeta}{(U')^2} f'' \gamma''\tilde{r}}{|\boldsymbol{G}|} >0, $$
$$\frac{\partial \tilde{\ell}}{\partial i_r}=\frac{1}{|\boldsymbol{G}|}
\begin{vmatrix}
G^1_k & G^1_N & -G^1_{i_r} & G^1_{r}  \\
G^2_k & G^2_N & -G^2_{i_r} & G^2_{r}  \\
G^3_k & G^3_N & -G^3_{i_r} & G^3_{r}  \\
G^4_k & G^4_N & -G^4_{i_r} & G^4_{r}  \\
\end{vmatrix} = \frac{(f-\delta k)U''\frac{\zeta}{(U')^2} f'' \gamma''\tilde{r}}{|\boldsymbol{G}|} <0 $$
since $$|\boldsymbol{G}|=-
(f-\delta k) \frac{\gamma''U''}{(U')^2} \zeta f''\eta''\tilde{\ell}<0.$$
Because $\partial\tilde{\ell}/\partial i>0$ and $i_\ell=i+\eta'(\Tilde{\ell})$, we have
$\partial i_\ell /\partial i=1+\eta''(\Tilde{\ell})\frac{\partial \Tilde{\ell}}{\partial i}>0$.  Similarly, because $\partial\tilde{\ell}/\partial i_r<0$ and $i_\ell=i+\eta'(\Tilde{\ell})$,  $\partial i_\ell /\partial i_r=\eta''(\Tilde{\ell})\frac{\partial \Tilde{\ell}}{\partial i_r}<0$. The real lending rate can be written as 
$\rho=1/\beta-1+\frac{\eta'(\tilde{\ell})}{\beta(1+i)}$. It is straightforward to show
$\partial\rho/\partial i_r=\frac{\partial \tilde{\ell}}{\partial i_r}\frac{\eta''(\tilde{\ell})}{\beta(1+i)} <0$ and 
$\partial\rho/\partial i=\frac{\partial \tilde{\ell}}{\partial i}\frac{\eta''(\tilde{\ell})}{\beta(1+i)} - \frac{\eta'(\tilde{\ell})}{\beta(1+i)^2} \lessgtr 0.$ Since $i_d=i$, $\partial i_d/\partial i=1>0$ and $\partial i_d/\partial i_r=0$.

\textbf{Part 2} (Comparative Statics in the Scarce-Reserve Equilibrium with respect to $i_r$): 

$$\boldsymbol{J_{i_r,\tilde{r}}}=\left[\begin{array}{cccccc}
J^1_{m}&-J^1_{i_r}& 0&J^1_{k} & J^1_n & 0 \\
J^2_{m}&-J^2_{i_r}& 0&J^2_{k} & J^2_n & 0 \\
0      & 0 & J^3_{\tilde{\ell}}& 0 & 0 & 0\\
0      & 0& 0&J^4_{k} & 0 & J^4_N \\
0      & 0& J^5_{\tilde{\ell}} &J^5_{k} & J^5_n & J^5_N \\
0      & 0& J^6_{\tilde{\ell}} & J^6_{k} & 0 & 0 \\
\end{array}\right].$$

$$|\boldsymbol{J_{i_r,\tilde{r}}}|=(J^1_m J^2_{i_r}-J^2_m J^1_{i_r})\underbrace{J^3_{\tilde{\ell}}J^6_{k}J^4_{N}J^5_{n}}_{\bigoplus}$$
where 
\begin{align*}
&J^1_m J^2_{i_r}-J^2_m J^1_{i_r}= \\
& \left(\frac{\nu\zeta}{f-f'k}\right)
\{\sigma_1 \Lambda'(\tilde{z}_1)+\sigma_3\Lambda'(\tilde{z}_3)\} \chi\left[\frac{1+i}{(1+i_d)^2}+ \left\{\frac{(n\tilde{r}/\chi)\nu\zeta}{f-f'k}\right\}\{\sigma_2\Lambda'(\tilde{z}_2)+\sigma_3\Lambda'(\tilde{z}_3)\} \right]  \\
& -n\tilde{r}\left(\frac{\nu\zeta}{f-f'k}\right)^2 \left\{\sigma_3\Lambda'(\tilde{z}_3)\right\}^2.
\end{align*}
Therefore, as long as 
\begin{equation}\label{eq:com_condition}
J^1_m J^2_{i_r}-J^2_m J^1_{i_r}<0
\end{equation}
holds (which we will check later), we have $\partial \tilde{r}/\partial i_r>0$ for small $\bar{b}$ which also implies $\partial \tilde{\ell}/\partial i_r<0$ for small $\bar{b}$. Immediate results of above results are $\partial i_{\ell}/\partial i_r<0$ and $\partial\rho/\partial i_r<0$. The next part (Part 3) verifies the condition  (\ref{eq:com_condition}). 

\textbf{Part 3} (Confirming $J^1_m J^2_{i_r}-J^2_m J^1_{i_r}<0$ holds when scarce-reserve equilibrium is unique):
To confirm $J^1_m J^2_{i_r}-J^2_m J^1_{i_r}<0$, recall (\ref{eq:prop2_pfp3_f})-(\ref{eq:prop2_pfp3_l}) and apply implicit function theorem:
$$\underbrace{\left[\begin{array}{cccccc}
\tilde{H}^1_{m}&\tilde{H}^1_{\tilde{r}}& 0&\tilde{H}^1_{k} & \tilde{H}^1_n & 0 \\
\tilde{H}^2_{m}&\tilde{H}^2_{\tilde{r}}& 0&\tilde{H}^2_{k} & \tilde{H}^2_n & 0 \\
0      &\tilde{H}^3_{\tilde{r}}& \tilde{H}^3_{\tilde{\ell}}& 0 & 0 & 0\\
0      & 0& \tilde{H}^4_{\tilde{\ell}} & \tilde{H}^4_{k} & 0 & 0 \\
0      & 0& 0&\tilde{H}^5_{k} & 0 & \tilde{H}^5_N \\
0      & 0& \tilde{H}^6_{\tilde{\ell}} &\tilde{H}^6_{k} & \tilde{H}^6_n & \tilde{H}^6_N \\
\end{array}\right]}_{\equiv\boldsymbol{\tilde{H}}}\left[\begin{array}{c}
d m  \\
d\tilde{r}  \\
d\tilde{\ell}  \\
d k  \\
d n   \\
d N 
\end{array}\right]= - \left[\begin{array}{c}
(n\tilde{r}/\chi)\nu\sigma_3 \Lambda'(\tilde{z}_3)  \\
\frac{1+i}{(1+i_d)^2}+(n\tilde{r}/\chi)\nu \left[\sigma_2 \Lambda'(\tilde{z}_2)+\sigma_3 \Lambda'(\tilde{z}_3)\right]  \\
0  \\
0  \\
0   \\
0
\end{array}\right]di_d$$
where
$$\begin{array}{lll}
H^1_{m}=\frac{\nu\zeta}{f(k)-f'(k)k}\{\sigma_1 \Lambda'(\tilde{z}_1)+\sigma_3 \Lambda'(\tilde{z}_3)\}, & H^2_{m}= \frac{\nu\zeta}{f(k)-f'(k)k}\sigma_3 \Lambda'(\tilde{z}_3), & \\
H^1_{\tilde{r}}=\frac{n\nu\zeta \{1+i_d-\tilde{r}\chi \gamma''(\tilde{r})\}}{\chi\{f(k)-f'(k)k\}}  \sigma_3\Lambda'(\tilde{z}_3), & H^2_{r}=-\frac{1+i}{(1+i_d)^2}+\frac{n\nu\zeta(1+i_d) }{\chi\{f(k)-f'(k)k\}}\sum_{j=2}^3\{\sigma_j \Lambda'(\tilde{z}_j)\}, & \\
H^3_{\tilde{r}}= J^3_{\tilde{r}}, & H^3_{\tilde{\ell}}= J^3_{\tilde{\ell}}, &  \\
H^5_{\tilde{\ell}}=J^5_{\tilde{\ell}}, & H^6_{\tilde{\ell}}=J^6_{\tilde{\ell}}, & \\
H^1_{k}=J^1_{k}, & H^2_{k}=J^2_{k}, & \\
H^4_{k}=J^4_{k}, & H^5_{k}=J^5_{k}, & \\
H^6_{k}=J^6_{k}, & H^1_{n}=J^1_{n}, & \\
H^2_{n}=J^2_{n}, & H^5_{n}=J^5_{n}, & \\
H^4_{N}=J^4_{N}, & H^5_{N}=J^5_{N}, & 
\end{array}$$

From Part 2 of  Proposition \ref{prop:threshold}, there exists at least one point where RC1 crosses RC2 curve from below. If the equilibrium is unique, the equilibrium exists where RC1 crosses RC2 curve from below which implies that the equilibrium satisfies $\partial\tilde{r}/ \partial i_d>0$. The uniqueness of scarce reserve implies $\partial \tilde{r} / \partial i_d>0$ which also means $\frac{|\boldsymbol{\tilde{H}_{i_d,\tilde{r}}}|}{|\boldsymbol{\tilde{H}}|}>0$. 
This holds if and only if  
\begin{align*}
&\tilde{H}^2_{m}\tilde{H}^1_{i_d} -\tilde{H}^1_{m}\tilde{H}^2_{i_d} \\
&=  (n\tilde{r}/\chi)\left(\frac{\nu\zeta}{f-f'k}\right)^2\left\{\sigma_3\Lambda'(\tilde{z}_3) \right\}^2 \\
&\quad-\frac{\nu\zeta}{f-f'k}\{\sigma_1\Lambda'(\tilde{z}_1)+\sigma_3\Lambda'(\tilde{z}_3)\}\left[\frac{1+i}{(1+i_d)^2} +\frac{(n\tilde{r}/\chi)\nu\zeta}{f-f'k}\{ \sigma_1\Lambda'(\tilde{z}_1)+\sigma_3\Lambda'(\tilde{z}_3)\} \right]>0
\end{align*}
because \begin{align*}
|\boldsymbol{\tilde{H}}|=(\tilde{H}^1_m \tilde{H}^2_r-\tilde{H}^1_r\tilde{H}^2_m)\tilde{H}^3_{\ell}\tilde{H}^4_{k}(-\tilde{H}^5_{N}\tilde{H}^6_{n})+(\tilde{H}^1_m 
\tilde{H}^3_r\tilde{H}^2_k - \tilde{H}^2_m \tilde{H}^3_r\tilde{H}^1_k )\tilde{H}^3_{\tilde{\ell}} (-\tilde{H}^6_{n}\tilde{H}^5_{N})>0
\end{align*}
and $\tilde{H}^2_{m}\tilde{H}^1_{i_d} -\tilde{H}^1_{m}\tilde{H}^2_{i_d}=-(J^1_m J^2_{i_r}-J^2_m J^1_{i_r})\chi$, we have $J^1_m J^2_{i_r}-J^2_m J^1_{i_r}<0$ when the equilibrium is unique. 

\textbf{Part 4} (Comparative Statics in the Scarce-Reserve Equilibrium with respect to $i$):
Recall (\ref{eq:implicit}). Using Cramer's rule, we have $\frac{\partial\tilde{r}}{\partial i}=\frac{|\boldsymbol{J_{i,\tilde{r}}}|}{|\boldsymbol{J}|}$ where
$$\boldsymbol{J_{i,\tilde{r}}}=\left[\begin{array}{cccccc}
J^1_{m}&-J^1_{i}& 0&J^1_{k} & J^1_n & 0 \\
J^2_{m}&-J^2_{i}& 0&J^2_{k} & J^2_n & 0 \\
0      & 0 & J^3_{\tilde{\ell}}& 0 & 0 & 0\\
0      & 0& 0&J^4_{k} & 0 & J^4_N \\
0      & 0& J^5_{\tilde{\ell}} &J^5_{k} & J^5_n & J^5_N \\
0      & -J^6_{i}& J^6_{\tilde{\ell}} & J^6_{k} & 0 & 0 \\
\end{array}\right]$$
and
\begin{align*}
|\boldsymbol{J_{i,\tilde{r}}}|=&(J^1_{m}J^2_{i}-J^2_{m}J^1_{i})J^3_{\tilde{\ell}} J^6_{k} J^5_nJ^4_N -J^6_i J^3_\ell J^4_k (J^1_m J^2_n J^5_N - J^2_m J^1_n J^5_N)   \\
&-J^6_i J^3_\ell J^4_N \{J^1_m(J^2_k J^5_n -J^2_n J^5_k)-J^2_m (J^1_k J^5_n -J^1_n J^5_k )\}.
\end{align*}
Given that, one can show that 
$$ \lim_{\sigma_3\rightarrow0} \frac{|\boldsymbol{J_{i,\tilde{r}}}|}{|\boldsymbol{J}|}<0$$
which implies  $\frac{\partial\tilde{r}}{\partial i}<0$ and $\frac{\partial\tilde{\ell}}{\partial i}>0$. Given these results, it is straightforward to show that $\partial i_{\ell}/\partial i >0$ and $\partial i_{d}/\partial i >0$ for small $\sigma_3$.
\end{proof}

\begin{proof}[Proof of Proposition \ref{prop:dmt}]
Recall equation (\ref{eq:DM_trade})-(\ref{eq:DM_trade2}). 
\begin{align*}  
\frac{i_t}{\nu}&=\sigma_1\lambda(q_{1})+\sigma_3\lambda(q_{3})  \\
\left\{ \frac{1+i}{1+i_d}-1\right\}\frac{1}{\nu}&=\sigma_2\lambda(q_{2})+\sigma_3\lambda(q_{3})  
\end{align*}
where $v(q_1)= \zeta m/\{f(k)-f'(k)k\}$, $v(q_2)=\zeta(d+\bar{b})/\{f(k)-f'(k)k\}$, and $v(q_3)=\zeta(m+d+\bar{b})/\{f(k)-f'(k)k\}$. When $i_r\geq \bar{\iota}_r$, we have $i_d=i$ which gives $\lambda(q_2)=\lambda(q_3)$. Therefore, it is straightforward to show $q_2=q_3=q^*$. 
\end{proof}

\begin{proof}[Proof of Proposition \ref{prop:dmt2}]
Since $\partial\tilde{r}/\partial \bar{b}<0$ in the ample reserve equilibrium and $\bar{\iota}_r=\gamma'(\bar{r})+i$, $\bar{\iota}_r$ is decreasing in $\bar{b}$.
When the credit limit is sufficiently large $\bar{b}>p^*$,  $z_2$,$z_3>p^*$ which results in  $q_2=q_3=q^*$. Since $\bar{b}>p^*$ the households do not have any incentive to hold transaction deposits, $d=0$ which implies $\chi d= 0$. Therefore, the reserve requirement constraint does not bind. In this case, each bank's reserve balance is determined by
$\gamma'(\Tilde{r})=i_{r}- i$. The bank holds excess reserves as long as $i_{r}>i$. 
\end{proof}

\section{Additional Results}
\label{sec:append}
\subsection{Chow Test}
\label{sec:append_1}
Figure \ref{fig:multip1} includes the Chow test for structural breaks. The test result reported in the bottom-left panel of Figure \ref{fig:multip1} is implemented by estimating following regression. 
\begin{align*}
\text{Money multiplier}_t=&\beta_0+\beta_1\text{(RequiredReserves/Deposit)}_t\\
+&\textbf{1}_{t\geq1992Q2}[\gamma_0+\gamma_1\text{(RequiredReserves/Deposit)}_t]\\
+&\textbf{1}_{t\geq2008Q4}[\delta_0+\delta_1\text{(RequiredReserves/Deposit)}_t]+\epsilon_t
\end{align*}
Table \ref{tab:regrra} reports $F$-statistics which are obtained by testing  $\gamma_0=\gamma_1=\delta_0=\delta_1=0$.
The Chow test in the bottom-right panel of Figure \ref{fig:multip1} is implemented by estimating following regression. 
\begin{align*}
\text{Money multiplier}_t=&\beta_0+\beta_1\text{(Currency/Deposit)}_t\\
+&\textbf{1}_{t\geq2008Q4}[\delta_0+\delta_1\text{(Currency/Deposit)}_t]+\epsilon_t
\end{align*}
Table \ref{tab:regcda}  reports $F$-statistics is obtained by testing  $\delta_0=\delta_1=0$. The regression estimates and the Chow test results are summarized at Table \ref{tab:chow}.
\begin{table}[hp!]
\footnotesize 
\caption{Chow test for structural breaks}
\label{tab:chow}
\begin{subtable}{.5\linewidth}
\centering
\caption{Require reserve ratio}
\label{tab:regrra}
\begin{tabular}{lD{.}{.}{6}} 
\hline
\multicolumn{2}{l}{Dependent Variable: Money Multiplier}     \\ 
&        \\ \hline
RR     & -0.601  \\ 
                           & (0.365)           \\
$\text{RR}\times\textbf{1}_{t\geq1992Q2}$ 
                           & 132.279^{***}   \\ 
                           & (0.031)             \\
$\text{RR}\times\textbf{1}_{t\geq2008Q4}$ 
                           & -147.943^{***}  \\ 
                           & (8.574)            \\
$\textbf{1}_{t\geq1992Q2}$ & 9.091^{***}     \\
                           & (0.557)            \\ 
$\textbf{1}_{t\geq2008Q4}$ & 0.074^{***}   \\
                           & (0.611)            \\ 
Constant                   & 2.813^{***}   \\
                           & (0.053)              \\ \hline
Obs. & \multicolumn{1}{c}{228}  \\
$R^2$ & \multicolumn{1}{c}{0.963}  \\
DF for numerator & \multicolumn{1}{c}{4}  \\
DF for denominator  & \multicolumn{1}{c}{222}  \\
$F$ Statistic for Chow test & \multicolumn{1}{c}{1711.32}  \\
$F$ Statistic for 1\% sig. level & \multicolumn{1}{c}{3.40}  \\
$F$ Statistic for 0.1\% sig. level & \multicolumn{1}{c}{4.79}  \\
\hline \bottomrule
\end{tabular}
\end{subtable} 
\begin{subtable}{.5\linewidth}
\centering
\caption{Currency deposit ratio}
\label{tab:regcda}
\begin{tabular}{lD{.}{.}{6}} 
\hline
\multicolumn{2}{l}{Dependent Variable: Money Multiplier}      \\
& \\  \hline
CD     & -1.301^{***}  \\ 
                           & (0.027)           \\
$\text{CD}\times\textbf{1}_{t\geq2008Q4}$ 
                           & -52.018^{***}  \\ 
                           & (4.995)            \\
$\textbf{1}_{t\geq2008Q4}$ & 3.061^{***}   \\
                           & (0.409)            \\ 
Constant                   & 3.159^{***}   \\
                           & (0.015)              \\ 
     &              \\ 
     &              \\ 
     &              \\ 
     &              \\ \hline
Obs. & \multicolumn{1}{c}{228}  \\
$R^2$ & \multicolumn{1}{c}{0.974}  \\
DF for numerator & \multicolumn{1}{c}{2}  \\
DF for denominator  & \multicolumn{1}{c}{224}  \\
$F$ Statistic for Chow test & \multicolumn{1}{c}{1245.69}  \\
$F$ Statistic for 1\% sig. level & \multicolumn{1}{c}{4.70}  \\
$F$ Statistic for 0.1\% sig. level & \multicolumn{1}{c}{7.13}  \\
\hline \bottomrule
\end{tabular}
\end{subtable} 
\begin{center}
\begin{minipage}{0.9\textwidth} 
{\footnotesize Notes: Newy-West standard errors are in parentheses. ***, **, and * denote significance at the 1, 5, and 10 percent levels, respectively. Degree of freedom is denoted by DF.   \par}
\end{minipage} 
\end{center}
\end{table}

\subsection{Unit Root Test}\onehalfspacing
\label{sec:append_2}
Columns (3) and (6)  in Table \ref{tab:motive_reg} includes the canonical cointegrating regression estimates and the cointegration tests. This section reports unit root tests for the series used in Columns (3) and (6). For all four variables, the unit root tests fail to reject the null hypothesis of non-stationarity while their first difference rejects the null hypothesis of non-stationarity at 1\% significance level. All series are demeaned before implementing the unit root test following to \cite{elliott2006minimizing} and \cite{harvey2009unit}, because the magnitude of the initial value can be problematic. Let ***, **, and * denote significance at the 1, 5, and 10 percent levels, respectively.  The data are quarterly from 1980Q1 to 2007Q4.
\begin{table}[hp!]
\centering
\caption{Unit root test}
\label{tab:unit_1}
\begin{tabular}{lD{.}{.}{6}D{.}{.}{6}}
\toprule \hline
& \multicolumn{2}{c}{Phillips-Perron test}   \\ \hline
& \multicolumn{1}{c}{$Z(\rho)$} & \multicolumn{1}{c}{$Z(t)$}  \\ \hline
$ln(m)$  & 0.567   & 0.297  \\ 
$ln(d)$  & 1.275   & 1.054  \\ 
$ln(uc)$ & -1.114  & -1.710 \\ 
$r$      & -7.721  &-2.471  \\ \hline
$\Delta ln(m)$ & -46.623^{***}   & -5.335^{***}  \\ 
$\Delta ln(d)$ & -42.267^{***}   & -5.060^{***}  \\ 
$\Delta ln(uc)$ & -41.998^{***}  & -5.107^{***} \\ 
$\Delta r$ &-94.183^{***}   &-9.263^{***}  \\
\hline\bottomrule 
\end{tabular}
\end{table}

\clearpage

\section{Data Sources and Variable Definitions}
\singlespacing
The quantitative analysis uses the annual average of the below series. The empirical analysis uses data from the same sources.
\begin{itemize}
\item Federal funds rates: ``Effective Federal Funds Rate" (FRED series FEDFUNDS).
\item Interest on reserves: ``Interest Rate on Excess Reserves" (FRED series IOER) and  ``Interest Rate on Required Reserves" (FRED series IORR). 
\item 3-month treasury rate: ``3-Month Treasury Bill: Secondary Market Rate" (FRED series TB3MS).
\item Deposit (Total checkable deposits): ``Total Checkable Deposits"  (FRED series TCDSL). 
\item Excess reserves: ``Excess Reserves of Depository Institutions" (FRED series EXCRESNS) and (FRED series EXCSRESNS).
\item Excess reserve ratio: $\frac{\text{Excess reserves}}{\text{Total checkable deposits}}$.
\item Required reserves:
``Required Reserves of Depository Institutions" (FRED series REQRESNS).
\item Required reserves ratio: $\frac{\text{Required reserves}}{\text{Total checkable deposits}}$.
\item Reserves: ``Total Reserves of Depository Institutions" (FRED series TOTRESNS).
\item M1: ``M1 Money Stock" (FRED series M1SL).
\item M2: ``M2 Money Stock" (FRED series M2SL).
\item Monetary base: ``Monetary Base; Total" (FRED series BOGMBASE).
\item M1 money multiplier: ``M1 Money Multiplier" (FRED series MULT).
\item Currency: ``Currency Component of M1" (FRED series CURRSL).
\item Deposit Component of M1: M1$-$Currency.
\item Unsecured credit: ``Revolving Consumer Credit Owned and Securitized"  (FRED series REVOLSL) 
\item GDP:  ``Gross Domestic Product" (FRED series GDP), quarterly and ``Gross Domestic Product" (FRED series GDPA), annual.
\end{itemize}
\end{appendices}

\end{document}

%% file: figures_gen/name.tex
\tikzset{every picture/.style={line width=0.75pt}} 

\begin{tikzpicture}[x=0.75pt,y=0.75pt,yscale=-1,xscale=1]

\draw [color={rgb, 255:red, 74; green, 144; blue, 226 }  ,draw opacity=1 ][line width=1.5]    (45,137.17) .. controls (69.5,98.92) and (133.35,50.21) .. (180.3,43.93) ;
\draw  [dash pattern={on 0.84pt off 2.51pt}]  (180.3,43.93) -- (133.35,44.13) -- (46.9,44.52) ;
\draw [line width=1.5]    (45.35,158.35) -- (201.8,158.18) ;
\draw [shift={(204.8,158.17)}, rotate = 179.94] [color={rgb, 255:red, 0; green, 0; blue, 0 }  ][line width=1.5]    (14.21,-4.28) .. controls (9.04,-1.82) and (4.3,-0.39) .. (0,0) .. controls (4.3,0.39) and (9.04,1.82) .. (14.21,4.28)   ;
\draw [line width=1.5]    (45.35,158.35) -- (45.32,3.54) ;
\draw [shift={(45.32,0.54)}, rotate = 89.99] [color={rgb, 255:red, 0; green, 0; blue, 0 }  ][line width=1.5]    (14.21,-4.28) .. controls (9.04,-1.82) and (4.3,-0.39) .. (0,0) .. controls (4.3,0.39) and (9.04,1.82) .. (14.21,4.28)   ;
\draw  [dash pattern={on 0.84pt off 2.51pt}]  (420.27,43.93) -- (373.32,44.13) -- (286.87,44.52) ;
\draw [line width=1.5]    (285.32,158.35) -- (441.78,158.18) ;
\draw [shift={(444.78,158.17)}, rotate = 179.94] [color={rgb, 255:red, 0; green, 0; blue, 0 }  ][line width=1.5]    (14.21,-4.28) .. controls (9.04,-1.82) and (4.3,-0.39) .. (0,0) .. controls (4.3,0.39) and (9.04,1.82) .. (14.21,4.28)   ;
\draw [line width=1.5]    (285.32,158.35) -- (285.29,3.54) ;
\draw [shift={(285.29,0.54)}, rotate = 89.99] [color={rgb, 255:red, 0; green, 0; blue, 0 }  ][line width=1.5]    (14.21,-4.28) .. controls (9.04,-1.82) and (4.3,-0.39) .. (0,0) .. controls (4.3,0.39) and (9.04,1.82) .. (14.21,4.28)   ;
\draw [color={rgb, 255:red, 208; green, 2; blue, 27 }  ,draw opacity=1 ][line width=1.5]    (45.4,44.37) -- (45.4,73.97) ;
\draw [color={rgb, 255:red, 208; green, 2; blue, 27 }  ,draw opacity=1 ][line width=1.5]    (45,72.92) .. controls (69.6,143.42) and (135.81,150.92) .. (174.33,157.83) ;
\draw  [color={rgb, 255:red, 74; green, 144; blue, 226 }  ,draw opacity=1 ][fill={rgb, 255:red, 74; green, 144; blue, 226 }  ,fill opacity=1 ] (282.19,42.88) .. controls (282.19,41.34) and (283.71,40.1) .. (285.6,40.1) .. controls (287.48,40.1) and (289,41.34) .. (289,42.88) .. controls (289,44.42) and (287.48,45.67) .. (285.6,45.67) .. controls (283.71,45.67) and (282.19,44.42) .. (282.19,42.88) -- cycle ;
\draw [color={rgb, 255:red, 208; green, 2; blue, 27 }  ,draw opacity=1 ][line width=1.5]    (285.5,42.79) -- (285.5,77.42) ;
\draw [color={rgb, 255:red, 208; green, 2; blue, 27 }  ,draw opacity=1 ][line width=1.5]    (285.5,76.42) .. controls (310.1,146.92) and (374.75,150.14) .. (413.27,157.06) ;

\draw (-20.97,71.21) node [anchor=north west][inner sep=0.75pt]  [font=\tiny]  {$( 1-\chi ) i+\chi i_{r}$};
\draw (115.3,165.77) node [anchor=north west][inner sep=0.75pt]  [font=\tiny]  {$\gamma ^{\ ^{-1}}\left(\frac{1-\chi }{\chi } i+i_{r}\right)$};
\draw (34.46,163.23) node [anchor=north west][inner sep=0.75pt]  [font=\tiny]  {$0$};
\draw (34.12,37.82) node [anchor=north west][inner sep=0.75pt]  [font=\tiny]  {$i$};
\draw (220.5,72.87) node [anchor=north west][inner sep=0.75pt]  [font=\tiny]  {$( 1-\chi ) i+\chi i_{r}$};
\draw (351.6,165.5) node [anchor=north west][inner sep=0.75pt]  [font=\tiny]  {$\gamma ^{\ ^{-1}}\left(\frac{1-\chi }{\chi } i+i_{r}\right)$};
\draw (274.43,163.23) node [anchor=north west][inner sep=0.75pt]  [font=\tiny]  {$0$};
\draw (274.09,37.82) node [anchor=north west][inner sep=0.75pt]  [font=\tiny]  {$i$};
\draw (26.92,2.62) node [anchor=north west][inner sep=0.75pt]  [font=\scriptsize]  {$i_{d}$};
\draw (206.8,161.57) node [anchor=north west][inner sep=0.75pt]  [font=\scriptsize]  {$\tilde{r}$};
\draw (266.25,2.07) node [anchor=north west][inner sep=0.75pt]  [font=\scriptsize]  {$i_{d}$};
\draw (72.67,201) node [anchor=north west][inner sep=0.75pt]   [align=left] {(a) Case 1 };
\draw (446.78,161.57) node [anchor=north west][inner sep=0.75pt]  [font=\scriptsize]  {$\tilde{r}$};
\draw (153.14,198.16) node [anchor=north west][inner sep=0.75pt]    {$\overline{b} < p^{*} \ $};
\draw (283.71,200.9) node [anchor=north west][inner sep=0.75pt]   [align=left] {(a) Case 2 };
\draw (363.9,198.78) node [anchor=north west][inner sep=0.75pt]    {$\overline{b} \geq p^{*} \ $};
\draw (143.67,134) node [anchor=north west][inner sep=0.75pt]  [font=\small,color={rgb, 255:red, 208; green, 2; blue, 27 }  ,opacity=1 ] [align=left] {{\small RC1} };
\draw (148.33,55.33) node [anchor=north west][inner sep=0.75pt]  [font=\small,color={rgb, 255:red, 74; green, 144; blue, 226 }  ,opacity=1 ] [align=left] {{\small RC2} };
\draw (383.33,135) node [anchor=north west][inner sep=0.75pt]  [font=\small,color={rgb, 255:red, 208; green, 2; blue, 27 }  ,opacity=1 ] [align=left] {{\small RC1} };
\draw (291,25) node [anchor=north west][inner sep=0.75pt]  [font=\small,color={rgb, 255:red, 74; green, 144; blue, 226 }  ,opacity=1 ] [align=left] {{\small RC2} };

\end{tikzpicture}